\documentclass[%
 reprint,
 amsmath,amssymb,
 aps,onecolumn,pra
]{revtex4-1}

\usepackage{graphicx}
\usepackage{hyperref}
\usepackage{multirow}

\newcommand{\eqn}[1]{\begin{eqnarray} #1 \end{eqnarray}}
\newcommand{\tit}[1]{\textit{#1}}
\newcommand{\tbf}[1]{\textbf{#1}}
\newcommand{\trm}[1]{\textrm{#1}}
\newcommand{\brm}[1]{\textrm{\textbf{#1}}}

\newcommand{\pa}[1]{\textbf{pa}(#1)}
\newcommand{\ch}[1]{\textbf{ch}(#1)}

\newcommand{\tr}[1]{  \textrm{Tr}\left[ #1 \right]  }
\newcommand{\trc}[2]{  \textrm{Tr}_{#1}\left[ #2 \right]  }
\newcommand{\zum}[2]{\displaystyle\sum_{#1}^{#2}}
\newcommand{\bra}[1]{\langle #1 |}
\newcommand{\ket}[1]{| #1 \rangle}

\begin{document}

\title{Quantum causal models via QBism}

\author{Jacques Pienaar}
\altaffiliation[Currently at: ]{QBism group, University of Massachusetts Boston, 100 Morrissey Boulevard, Boston MA 02125, USA.}
\affiliation{
 International Institute of Physics, Universidade Federal do Rio Grande do Norte, Campus Universitario, Lagoa Nova, Natal-RN 59078-970, Brazil.
}

\date{\today}



\begin{abstract}
This paper presents a framework for Quantum causal modeling based on the interpretation of causality as a relation between an observer's probability assignments to hypothetical or counterfactual experiments. The framework is based on the principle of `causal sufficiency': that it should be possible to make inferences about interventions using only the probabilities from a single `reference experiment' plus causal structure in the form of a DAG. This leads to several interesting results: we find that quantum measurements deserve a special status distinct from interventions, and that a special rule is needed for making inferences about what would happen if they are not performed (`un-measurements'). One natural candidate for this rule is found to be an equation of importance to the QBist interpretation of quantum mechanics. We find that the causal structure of quantum systems must have a `layered' structure, and that the model can naturally be made symmetric under reversal of the causal arrows.
\end{abstract}

\maketitle


\section{Introduction}

The advent of quantum information theory has brought with it the idea that there is a limited sense in which physical systems -- not necessarily human or conscious -- might be said to perform observations. For instance, the reduction in visibility of quantum interference phenomena traditionally attributed to `observation' of which-path information does not require observation \tit{per se}, but only that the relevant information could be obtained \tit{in principle} from some extraneous physical system. Thus, any system capable of obtaining and storing information about another system through a physical interaction is capable of `observation' in the broader information-theoretic sense. Parallel to these developments, the quantum physics community has also recently expanded the formal notion of \tit{causality} beyond deterministic physics to encompass probabilistic causality, following seminal developments in statistical modeling of causation \cite{PEARL,SGS}. A key part of this new probabilistic notion of causality is the concept of manipulation by an external agent. The `agent' is usually assumed to be a human experimenter, but the term may be extended to encompass physical systems in general, provided the notion of a `manipulation' by such a system can be meaningfully defined \cite{WOODW}. Both these recent developments blur the line between the notion of `observer' and `physical system', and invite us to re-examine the meaning of the intertwined concepts of observation and causation in contemporary quantum physics.

The present work is part of a growing field of research on \tit{quantum causal modeling} \cite{COSHRAP,ALLEN,RIED,RIEDPHD,FRITZ,HLP,PIEBRUK,BARRETTQCM,GIARM,LEIF08,MILZ17}, which aims to consolidate quantum information theory and probabilistic causation into a single framework. A major initial stimulus for these efforts was the work of Wood \& Spekkens \cite{WOOD}, who showed that classical causal models could not explain quantum correlations without `unnatural fine-tuning'. Much of the subsequent literature on the topic can be seen as a concerted effort to show that quantum correlations can be explained without such fine-tuning by suitably generalizing the notion of a `causal model'. On this account the research program has been successful: most of the works just cited are more than capable of performing all practical functions of causal modeling for both quantum and classical systems, and are able to do so without fine-tuning, or invoking causal pathways that run counter to classical intuition. This literature strongly suggests that although quantum theory may not be local in the sense originally defined by Bell \cite{BELL76}, it may nevertheless be called a `causal theory' in the sense of probabilistic causation. This compelling idea is an invitation to examine the relevance of quantum causal modeling to quantum foundations.

However, in their emphasis carrying over the operational functionality of the classical causal models into the quantum domain, these approaches have so far skirted around foundational issues regarding the meaning of causality itself, and how it might be revised in light of quantum theory. That these matters have been overlooked is easy to verify: nearly all of the proposed models are compatible with any interpretation of quantum mechanics, and nearly all of them adopt without much critical reflection the basic definition of causality from the classical framework. None of them question the basic idea that causality is \tit{only} about manipulations; it is instead tacitly assumed that the sole task of a causal model is to produce probabilities for measurement outcomes in response to manipulations \tit{and nothing more}. Most of the proposed frameworks can therefore be transformed into each other with relative ease, as there are only so many ways to define a generalized quantum process that maps a set of inputs to a set of outputs. One therefore finds that the differences between these models are invariably of a mainly technical nature and not a matter of deep principle.

The present work has in common with these other works the commitment to a notion of causality that is probabilistic and manipulationist. We differ from them in that we define causality as a predominantly \tit{counterfactual} concept, rather than being entirely restricted to manipulations. Whereas an emphasis on manipulation would regard counterfactuals as being about the different possible manipulations that could be performed, we will instead consider manipulations to be just one of a broader class of counterfactual experiments that one could perform. We will argue that, apart from manipulations, it is also important to consider counterfactual experiments in which certain measurements are \tit{not} performed. Just as there is a rule for determining the result of a manipulation, our framework demands a rule for inferring the result of such an `un-measurement'. The rule in the classical case is trivial, which allows us to overlook it; but in the quantum case the rule takes on a fascinating mathematical form that is known in quantum foundations research in connection with the QBist interpretation of quantum theory. Our approach therefore immediately manifests a connection between causality and quantum foundations.

A related advantage of our counterfactual approach is that it emphasizes the observer's involvement in the process of causal discovery. The counterfactuals represent mutually exclusive alternative experimental situations, which by implication are referred to some observer who has the power to bring about one or the other. Consequently, we view causality as a relation that holds between the probabilities that the observer assigns to different experimental contexts. Which contexts? This brings us to an old puzzle in the foundations of causality. 

Imagine that a fire burns down an apartment, and the investigation reveals that the tenant had left the gas stove on. The landlord accuses the tenant of causing the fire, since if he had remembered to switch off the stove, the fire would not have occurred. The tenant, who happens to be a philosopher, counters that if there had been no oxygen in the apartment, the fire also would not have occurred. To see what is wrong with this argument, one has to recognize that causal relations can only be established with reference to the `status quo' as to what is and what is not considered reasonably possible. The presence of oxygen in the apartment should not be regarded as a possible cause because nobody would reasonably expect it to be absent. If, on the other hand, the defendant lived inside a vacuum chamber that could be readily evacuated by the push of a button, the case might turn out differently.

When dealing with counterfactual causality it is therefore quite natural to introduce the idea of a `reference' experiment relative to which the observer is contemplating possible deviations. In this work we formalize this idea with the help of a control variable that indicates the different alternative `contexts' an observer can bring about. By contrast, an emphasis on manipulation would tend to presume the existence of some independent physical structure that conveys the input to the output, which hides the fact that what is considered a `mechanism' is itself observer-dependent. The mechanistic viewpoint thus tends to reify an observer's possibilities relative to a system into absolute properties of the system itself. A tractor is thought to be a `mechanism' even when there is nobody around who is trained to operate it. What use, then, in insisting that it is a mechanism? We can only make sense of such a claim by appealing to counterfactuals, for instance, by supposing what \tit{would} happen if there \tit{were} somebody there who knew how to drive the tractor -- but this brings us back to the dependence on an observer (or perhaps one should say `user'), which was hidden when one was thinking in terms of mechanisms.

Thus on a counterfactual account we are constantly reminded of the observer's role in establishing a causal connection. For on this account causality it is a statement about how an \tit{observer}'s probability assignments should be related between two counterfactual experiments that the \tit{observer} can concievably perform. This way of understanding causal relations is quite unheard of in fundamental physics, where the overwhelming preference is to appeal to mechanisms (deterministic or otherwise). We will argue, however, that it is the more fruitful way to think about causality when the systems in question are quantum.

We close the introduction with some remarks about how the observer-dependence of causality manifests itself in the present work. First, it leads us to make a general demand that all the fundamental rules of the framework should be expressed as equations that relate the probabilities of events in one experimental context to the probabilities of events in another context. In particular, although we make extensive use of standard formal devices like linear operators on Hilbert spaces, our aim is always to elimiate any reference to such objects from the basic rules of our framework. This is because probabilities are things that we may take to refer to the direct experience of the observer making the experiment, whereas Hilbert space operators are far removed from this experience and only make themselves felt in the way that they guide our probability assignments to events. This view is closely allied with the QBist interpretation of quantum theory, which takes probabilities to be subjective judgements of the observer. We note, however, that while we take QBism as our motivation in this regard, our framework is entirely compatible with an objective interpretation of probability. Secondly, in the course of applying our framework to quantum systems, we find it natural to postulate a fundamental symmetry with respect to the reversal of the direction of causality. This suggests that causal relations might be non-directional at the fundamental level, and that the direction may depend in some way on the observer's capacities relative to the system. 

\tit{Remark:} Of course, this is not to imply that it is within the observer's powers to change the causal direction at will. A detailed discussion of this issue is beyond the scope of this paper, but it seems plausible that the observer's powers over a system are constrained by their own thermodynamic arrow of time, in which case reversing the direction of causality might be no easier than un-scrambling an egg, which is to say impossible for all practical purposes.

The paper is structured into three main sections. Section \ref{sec:defs} outlines a general framework for causal modeling of general physical systems (classical, quantum or otherwise) based on an emphasis on counterfactual inference. A key idea is the principle of \tit{causal sufficiency}, which asserts that the only formal mathematical structure needed for inferring counterfactuals, beyond the probabilities, is a graph of the causal structure. This is a driving force behind the whole work, which can be alternatively seen as a purely mathematical exercise in exorcising Hilbert spaces from causal modeling of quantum systems. (We mention that this general framework is in fact conceptually posterior to the causal modeling of quantum systems -- which comes later in Sec. \ref{sec:qbism} -- for it was the unique problems faced in causal modeling of quantum systems that motivated the counterfactual framework). Section \ref{sec:ccms} applies the framework to classical causal models, comparing and contrasting it to the way that these models are usually formulated. Section \ref{sec:qbism} then applies it to quantum systems, resulting in the definition of a quantum causal model. Along the way, we make contact with some of the formal mathematical apparatus of QBism, which we generalize to suit our purposes. The result is a model that is manifestly symmetric with respect to the reversal of the causal arrows. We discuss the meaning of this result and address other questions about the framework at the end in Sec. \ref{sec:discuss}.

\section{Causal modeling as counterfactual inference \label{sec:defs}}

In this section, we describe a general framework for causal modelling that applies to any class of physical systems, which emphasizes a counterfactual definition of causality.

\subsection{Preliminaries and notation \label{sec:notation}}
It is generally agreed (among non-philosophers) that causation is transitive: if $A$ causes $B$ and $B$ causes $C$, then $A$ must cause $C$ (in this case $A$ is called an `indirect' cause of $C$). Furthermore, it stands to reason that no cause can be its own effect, hence there can be no chain of causes leading from $A$ back to itself. Based on these axioms, the causal relations among a set of propositions $A,B,C,\dots$ can be represented schematically by a directed acyclic graph (DAG). In this representation, a variable $A$ is a \tit{cause} of $B$ iff there is at least one path leading from $A$ to $B$ following the directions of the arrows. It is also standard to use the additional terminology that $A$ is a \tit{direct cause} of $B$ if there is a single arrow from $A$ to $B$, and an \tit{indirect cause} if there is a path consisting of two or more arrows from $A$ to $B$. These special cases do not exclude each other, thus $A$ can be both a \tit{direct} and \tit{indirect} cause of $B$. (Note that while \tit{cause} is transitive, the more refined notion of \tit{direct cause} is not). 

It is conventional to use `family tree' terminology to describe relationships in a DAG, eg. in addition to the \tit{parents} of $X$, one can define its \tit{children}, \tit{ancestors} and \tit{descendants} in an analogous way. In terms of the family tree nomenclature, $A$ is a cause of $B$ whenever $A$ is an \tit{ancestor} of $B$, irrespective of whether it is also a \tit{parent} of $B$.

In this work, $P(X)$ represents a normalized discrete probability function on the domain $\trm{dom}(X)$ of the random variable $X$. When evaluating probabilities for specific values of $X$, we adopt the shorthand $P(x) := P(X=x)$. Thus, for example, $P(X,y,w|Z)$ should be understood as a function of just two variables, $X,Z$, equivalent to the function $P(X,Y,W|Z)$ evaluated at the specific values $Y=y,W=w$. For sets of random variables, we often replace the `$\cup$' with just a comma (or a space) when taking the union, eg. $A,B,C = A \, B \, C := A \cup B \cup C$. Sets of random variables can be used to define composite random variables, which we denote by bold letters, eg. the composite variable $\tbf{X} := A,B$ takes values that are tuples $\tbf{x}:=(a,b)$ with $a \in \trm{dom}(A),\, b \in \trm{dom}(B)$.

\subsection{Causality and measurements \label{sec:measurements}}

The setting of \tit{causal modeling} is a collection of \tit{localized measurements} that are performed on some external arrangement of physical matter evolving in time for the duration of an experiment; this material is referred to simply as the \tit{system}. These measurements are assumed to be fixed in advance of the experiment and cannot be dynamically changed as the experiment progresses. We now specify more carefully what we mean by `measurements':\\ 

\tbf{Localized Measurement:} A localized measurement is a physical interaction of an observer with a system that takes place within a region of space-time that is \tit{localized} relative to the experiment, and produces an outcome (i.e. a piece of classical data). Here, \tit{localized} means that the space-time extent of each region is small compared to the sizes of the distances and times between the measurements in the experiment under consideration. (This is a slight generalization of the similar concept of a space-time random variable introduced in Ref. \cite{CBRenner}, where the regions were assumed to be point-like). \\

\tit{Remark:} The space-time co-ordinates of each measurement region are only given relative to the experimental instance or run. Thus, if we consider the whole ensemble of repetitions of the experiment (whether parallel in space or sequential in time), then strictly speaking each measurement actually corresponds to a whole set of disjoint space-time regions, each one corresponding to a unique experimental run. Conventionally it is useful to re-set the space-time co-ordinates at the beginning of each experimental run so as to identify multiple repetitions of the `same' measurement by giving it the same space-time co-ordinates in each run.
 
Each measurement is associated with a random variable $X_i$, $i \in \{1,2,\dots,N \}$ whose values $x_i \in \trm{dom}X_i$ correspond to the possible outcomes of the measurement, with $P(X_i = x_i)$ the probability of obtaining the outcome $x_i$. It is conventional to choose the labelling such that $i<i'$ whenever $X_i$ is a cause of $X_{i'}$. Note that under this definition, all random variables represent the `outcomes' of measurements, even in cases where the value of a variable is completely determined. For instance, the action of a scientist turning a dial to some chosen value is still considered a `localized physical action on a system that produces an outcome'. We thus avoid making any fundamental distinction between `settings' and `outcomes' as is typical elsewhere in the literature. 

In this work, we adopt the usual manipulationist definition of causality, but carefully re-worded for our own purposes:\\

\tbf{MC. Manipulationist Causation:} \\
Let $A,B$ be correlated random variables. Then the statement `$A$ causes $B$' means that $A$ and $B$ \tit{remain} correlated in an experiment in which we perform a \tit{manipulation} of $A$ (and only $A$). Formally, we introduce a \tit{control variable} $C^{A}$ (the superscript indicates that it controls only how $A$ is measured), whose values $c \in \trm{dom}C$ represent different possible \tit{ways} of measuring $A$. In particular, let $C^A=\brm{do}$ represent a manipulation of $A$; then `$A$ causes $B$' means that:

\eqn{ \label{eqn:manipulationist}
P(AB| C^{A}=\brm{do} \, ) \neq P(A| C^{A}=\brm{do} \, ) P(B| C^{A}=\brm{do} \, )  \, .
}
In this instance, we call $A$ the \tit{cause} and $B$ the \tit{effect}.\\

The terminology `\tbf{do}' will be used to refer to an \tit{intervention}, which is a specific type of manipulation, but the definition above is intended to hold for manipulations more generally. The constraints on what defines a manipulation will be discussed in more depth in Sec. \ref{sec:activemanips}.

The above definition is rather different to what one usually finds in textbooks. Elsewhere, it is usually emphasized that $A$ can \tit{signal} to $B$, which in the present notation means there are two manipulations $C^A=\brm{do}(A=a')$ and $C^A=\brm{do}(A=a'')$ such that 
\eqn{
P(B|C^A=\brm{do}(A=a')) \neq P(B|C^A=\brm{do}(A=a'')) \, . 
}

However, this way of defining causality obscures the essential point that what matters is not the particular manipulation of $A$, but the bare \tit{fact} of manipulation. If $A$ and $B$ are correlated and I am controlling $A$, this is already sufficient to establish qualitatively that $A$ is the cause and $B$ is the effect, without needing to say precisely what I am doing to $A$. To be sure, we can always refine our notion of manipulation to so as to describe different settings of the lever as distinct manipulations, but this is incidental. Of course, for the purposes of predicting the \tit{quantitative} consequences of interventions, we will associate some distribution $P'(A)$ to the intervention on $A$; see Sec. \ref{sec:passactclass} .

The above definition draws our attention to a formal device that we will use throughout this work, whereby the \tit{physical method of measurement} of a variable $X$ is itself specified by the special variable $C^X$. Even for cases where the mode of measurement is in some sense `passive' or `neutral', we will reserve a special symbol `$\oslash$', which represents the default mode of measurement whenever no particular value of $C^X$ is specified; thus $P(X)$ is equivalent to $P(X|C^X=\oslash)$. This notation means that $P(X)$ and $P(X|C^X=\brm{do})$ represent the probabilities of \tit{distinct} events within the same sample space. Specifically, `$X=x|C^X=\oslash$' refers to the event that the outcome $X=x$ is measured without manipulating it, while `$X=x|C^X=\brm{do}$' refers to the event that $X=x$ is measured by intervention. The random variable $C^X$ here plays a special role of toggling between different subsets of the sample space, which represent different physical modes of observation of $X$. For this reason, we will generally not bother to assign probabilities to the values of $C^X$ and will always treat its value as conditioned upon. Since it is interpreted as defining part of the experimental context, we will not include it as a variable within the system and will not represent it by a node in causal diagrams. However, from a formal mathematical point of view, it can be regarded as just another random variable.

Some further clarification is needed regarding the case of a `non-manipulation' of $X$ that we express as $C^X=\oslash$. In this work, there are two special instances of a `non-manipulation' that we will consider. The first type represents an observation that is made expressly for the purposes of causal inference on the system, and is taken as the conventional laboratory standard of measurement:\\

\tbf{Reference measurement:} The notation $C^X=\oslash$ indicates a \tit{reference measurement} of $X$, which means that $X$ is the outcome of a measurement on the system that has the following essential properties:\\
(i) The value of $X$ is maximally informative about the system, in a technical sense that will be elaborated shortly at the end of Sec. \ref{sec:experiments};\\
(ii) The measurement of $X$ does not break causal chains, meaning, if $X$ appears in a causal chain $A \rightarrow X \rightarrow B$, then it is still the case that $A$ causes $B$, i.e. that $A,B$ remain correlated (ignoring the value of $X$) under manipulations of $A$.\\

The second type witnesses the actual \tit{absence} of a measurement of $X$:\\

\tbf{Un-measurement:} The notation $C^X=\brm{undo}$ (sometimes shortened to $\brm{un}(X)$) indicates an \tit{un-measurement} of $X$. This is characterized by the following properties:\\
(i) It represents the enforcement of the physical \tit{absence} of a measuring device in the designated space-time region of $X$ (eg. by removing a measuring device that was previously present there);\\
(ii) The un-measurement of $X$ does not break causal chains, meaning, if $X$ appears in a causal chain $A \rightarrow X \rightarrow B$, then it is still the case that $A$ causes $B$ after the un-measurement of $X$, i.e. that $A,B$ remain correlated (conditional on $C^X=\brm{undo}$) under manipulations of $A$.\\

Note that we do not insist that the reference measurements be `passive' or `non-disturbing' in any sense. For notational convenience, we will adopt the convention that all variables are measured as per the reference measurement scheme unless otherwise stated, and the conditionals of the form $C^A=\oslash$ will accordingly be omitted unless they are needed for clarity. 

\tit{Remark:} The designation of what is an `un-measurement' is of course ambiguous. Consider a Young's double-slit experiment with single photons. Here, an un-measurement could mean the removal of any photon-counters from the location of one of the slits; but then it may be asked whether this un-measuremet also requires us to remove tiny particles of dust from the air around the slit, since the scattering of light by these particles would make available (in principle) the information about which slit the photon passed though, even if our apparatuses do not collect this information. This ambiguity is to be settled by choosing a convention. If the dust particles are imagined to constitute a kind of `measurement device' then an un-measurement would indeed require that they be physically removed; but if they are designated as part of what we call the system's `environment', then the process of un-measurement does not mandate their removal. The only difference lies in how we tell the story, say, to explain the loss of interference due to the dust particles' presence; in the first case it would be attributed to the presence of a `measuring device' at the slit, while in the second case it would be attributed to the system `interacting with its environment'. Evidently no inconsistency will arise, provided we are careful to state what is considered a `measuring device' for the purposes of a given experiment. 

\subsection{Experiments and counterfactual inference \label{sec:experiments}}

In this work, we will be concerned with three special categories of experiment, classified according to how the non-exogenous variables in the system are measured. In a \tit{reference observational scheme} (also called the \tit{reference experiment}) all measurements are reference measurements. In a \tit{passive observational scheme} all the non-exogenous measurements are \tit{non-disturbing} (to be defined shortly in  Sec. \ref{sec:passactclass} ). Finally, in a \tit{manipulationist scheme} some of the measurements in the system are \tit{manipulations} (and when these are interventions, it is called an \tit{interventionist scheme}). Any \tit{experiment} proceeds in two stages. In the preparation stage, a system of the desired type is identified, isolated inside the laboratory, and `cleaned up' to meet certain quality standards; in the measurement stage the set of localized measurements $\tbf{X}:=\{ X_i : i = 1,2,\dots,N \}$ are performed, always in the same way and with each measurement occurring within its designated region of space-time. From the collected data-set of outcomes, the observer may infer a joint probability distribution $P(\tbf{X})$; this will be called the system's \tit{behaviour} relative to the given experiment.

\tit{Remark:} Although we do often make reference to the `system', it is the concept of an \tit{experiment} that is more fundamental, because it is only through experimentation that the properties of the system become known to us. In fact, the system may be regarded entirely as a conceptual abstraction that serves to represent the object to which our (the observer's) measurements are supposed to be directed.

An experiment is assumed to be carefully designed to as to have the following features:\\

\noindent \tbf{(i).} Distinct measured variables should refer to logically distinct properties (eg. `number of ravens' $R$ and `number of birds' $B$ are not logically distinct, and so must not both be used in the same experiment);\\
\tbf{(ii).} If two variables $X$ and $Y$ in the system are judged to have a common cause that is in the system, this common cause must be measured and included as a variable (or, if it is impractical to measure, included as an unmeasured \tit{latent variable} -- see Sec. \ref{sec:physMarkov});\\
\tbf{(iii).} If the \tit{experiment} has a chance of failure, care must be taken not to introduce \tit{spurious correlations} in $P(\tbf{X})$ when post-selecting on the successful runs. (One method to avoid this is simply to enlarge the scope of the experiment to explicitly include its `failure modes' as a special class of possible outcomes).\\
\tbf{(iv).} The \tit{exogenous} variables -- those whose causes are judged to lie outside of the system of interest -- must be initialized or selected so as to be statistically independent of one another, i.e. $P(E_1,E_2)=P(E_1)P(E_2)$ whenever $E_1,E_2$ are exogenous. Note that this can be achieved by doing independent manipulations of the exogenous variables.\\


Causal relations are relevant to experiments because they enable us to make inferences about counterfactuals. Specifically, we define \tit{counterfactual inference} as the procedure of taking a certain experiment as a reference (hence defining its measurements as the \tit{reference measurements}) and then using the system's behaviour in the reference experiment to deduce what would happen in a variety of \tit{counterfactual experiments}, which are defined by allowing $C^{X_i}$ to take values other than $\oslash$ for some of the $X_i \in \tbf{X}$. The rules for making such counterfactual deductions from the reference experiment are called \tit{inference rules}.

Causal inference refers to any form of counterfactual inference whose inference rules depend on \tit{causal structure}, i.e. the causal relations between all of the measurements in the system. More importantly, it is assumed that the inference rules don't depend on anything beyond the bare causal structure. To formalize these ideas, we define a causal model:\\ 

\tbf{CM. Causal model:} Let $P(\tbf{X})$ represent data about a system obtained in the reference experiment, and let $G(\tbf{X})$ be a DAG that represents the (actual or hypothesized) \tit{causal structure} of the system. Then the pair $\{ P(\tbf{X}),G(\tbf{X})\}$ defines a \tit{causal model} for the system;\\

and we supplement this definition with the postulate:\\

\tbf{CS. Causal sufficiency:} Causal structure is sufficient for counterfactual inference. That is, given a causal model $\{ P(\tbf{X}),G(\tbf{X})\}$ for a system in an experiment, there exist \tit{inference rules} that can be used to deduce from this model what would happen under \tit{manipulations} of the variables in the system.\\

\tit{Remark:} Counterfactual inference is, in general, a one-way operation. An \tit{intervention} gives a concrete example of the one-way nature of inference: since intervening on $W \in \tbf{X}$ effectively removes causal influences on $W$ that previously existed, a specification of $P(\tbf{X}|C^W=\brm{do})$ and the causal structure of the intervened-upon system will not contain enough information to deduce what would have happened if the intervention had not been performed. This might be rephrased as saying that there is no inference rule for `un-interventions'. This induces a natural (possibly partial) ordering on the set of counterfactual experiments, such that the higher ranking members in the order are those with the power to make counterfactual inferences about the lower-ranking members, and not vice-versa. It is therefore most natural to choose the \tit{reference experiment} to be one of the experiments that ranks at the top of this natural hierarchy, because this guarantees that it can be used to make inferences about the largest possible set of counterfactual experiments under consideration. This is what we meant in the previous section by calling the reference measurements `maximally informative'. It also underlies our assumption that reference measurements do not break causal chains (i.e. because if they did we would have the same difficulty un-measuring them as we do with interventions).

The above definition \tbf{CM} and postulate \tbf{CS} are the core of our counterfactual approach to causal modeling. However, as stated, they are still very vague. There remain two details that must be specified in order to obtain a rigorous framework for causal modeling. The first is to specify exactly what conditions the pair $\{ P(\tbf{X}),G(\tbf{X})\}$ needs to satisfy in order that we can say that $G(\tbf{X})$ represents the ``actual or hypothesized causal structure" of the system; these conditions are called the \tit{physical Markov conditions} and will be discussed in the next section. The second important detail is what precisely are the \tit{inference rules} referred to in \tbf{CS}. The inference rules for un-measurements and interventions on classical systems will be discussed in Sec. \ref{sec:passactclass}, and their quantum counterparts deferred to Sec. \ref{sec:qinterf}.   

\subsection{Physical Markov conditions \label{sec:physMarkov}}

Once we have fixed a convention for the reference experiment, it is useful for the purposes of inference to restrict our attention to some particular class of physical systems, which can be broadly or narrowly defined. For example one could restrict attention to the narrow class of classical pendulums, or to the broader class of relativistic $N$-body mechanical systems, and so forth. Suppose that we gather statistics from a reference experiment performed on many different systems all belonging to the particular class of physical systems of interest. Now, depending on the particular features of this class, one will find that certain conditions always hold between the variables in the reference experiment that depend explicitly on the causal structure $G(\tbf{X})$. 

One condition that is natural and commonplace for many classes of physical systems is the condition that \tit{causation implies correlation}, i.e. that if the variables $A,B$ are found to be uncorrelated in the reference experiment then neither will be found to be a cause of the other under manipulations of either variable. In fact, if one subscribes to our definition \tbf{MC} (recall Sec. \ref{sec:measurements}), then causation presupposes correlation, so we cannot escape this rule. However, on a broader conception of causality, this need not be the case. One example is the class of cryptographic ``one-time pads", whereby the system is a string of bits called the `message' $M$ that is added modulo 2 to a random bit string called the `key' $K$ to produce a coded message called the `cipher' $C$. On eg. a mechanistic account of causality it seems natural to say that both $M$ and $K$ are causes of $C$, yet manipulations of either one (ignoring the values of the other) remain uncorrelated with $C$.

The principle that causation implies correlation is just one example of what we will call a \tit{physical Markov condition}, which describes any rule that relates the causal structure of a system (belonging to a particular class) to its anticipated behaviour in the reference experiment. To see why \tit{physical Markov conditions} play a central role in causal modeling, consider the class of \tit{deterministic systems}, defined by the property that the value of each variable $X_i$ is fully determined by the values of its parents $\pa{X_i}$. For such systems, the rule \tit{causation implies correlation} holds, as does another more interesting condition \cite{PEARL,SGS}:\\

\noindent \tbf{FCC. Factorization on common causes:} \\
Suppose neither of $X_1,X_2$ is a cause of the other and $\tbf{C}$ is the complete set of their shared ancestors, i.e. $\tbf{C}$ contains all `common causes' (see Fig. \ref{fig:threedags}(a)); then $P(X_1 X_2|\tbf{C})=P(X_1|\tbf{C})P(X_2 |\tbf{C})$ .\\  

The \tbf{FCC} is a physical Markov condition that has an important consequence: if variables $A,B,C$ are found to be correlated in the reference experiment, and $A,B$ remain correlated conditional on the value of $C$, then it may be concluded that one of the pair $A,B$ must be a cause of the other, so long as we hold firm in our belief that the system is of the deterministic class. This provides a simple illustration of \tit{inference of causal structure}, whereby one leverages knowledge about the class of physical systems plus the behaviour in the reference experiment to infer facts about the causal structure. Thus, physical Markov conditions allow one to reduce the number of interventions needed to establish causal claims.

The most commonly considered class of systems in the literature on causal modeling is the class of \tit{classical stochastic systems}. This class can be defined as a slight generalization from the class of deterministic systems, as follows:\\

\noindent \tbf{Classical stochastic systems:}\\
These are systems defined by the requirement that, for every variable $X_i$ relevant to the system, it is possible to introduce a hypothetical auxiliary exogenous variable $A_i$ that has $X_i$ as its only child, such that the value of $X_i$ is fully determined by the values of $\pa{X_i}$ and $A_i$.\\

Informally, a \tit{classical stochastic system} is observationally equivalent to a deterministic system in which there are hidden sources of noise (represented by the $A_i$) independently affecting each node. This class of systems is particularly interesting because it has been found to be powerful enough to describe a wide range of real physical systems, including biological, ecological, and mechanical systems, and they form the basis for the standard textbooks on causal inference \cite{PEARL,SGS}. Let $P(\tbf{X})$ be the observed statistics of the reference experiment for any classical stochastic system with causal structure $G(\tbf{X})$. Then the physical Markov conditions for this class of systems may be summarized by the constraint:\\

\tbf{CMC. The Causal Markov Condition:}\\
$P(\tbf{X})$ factorizes according to:
\eqn{ \label{eqn:cmc}
P(\tbf{X})=\prod_i \, P(X_i|\pa{X_i})
}
where $\pa{X_i}$ are the parents of $X_i$ in $G(\tbf{X})$.\\

Given a class of physical systems, such as the classical stochastic systems, there are two distinct ways to establish the physical Markov conditions for this class. The first way is to start with an abstract formal definition of the class of systems (eg. by postulating a `mechanism') and then derive the \tbf{CMC} as a logical consequence ( see eg. Pearl \cite{PEARL}).

The second route is more empirical and involves the iteration of two fundamental steps. First we restrict attention to experiments in which the causal structure takes one of several simple forms (specifically, the common-cause, causal chain and common-effect structures shown in Fig. \ref{fig:threedags} of Sec. \ref{sec:cmc}). At this stage, the `class of systems' merely refers to some set of laboratory preparation procedures in which we are interested. Under these restrictions, we observe the behavior of the class of systems over many trials. Given the causal structure, any statistical independence that is found to hold for \tit{all} systems in the chosen class (or at least is true for any `typical' member of the class) is then declared to be a physical Markov condition for that class. In the second step, we extrapolate these empirically derived conditions to arbitrary causal structures. This extrapolation is \tit{postulated}, rather than derived, and serves to \tit{define} the class of systems in more general causal structures (the method of extrapolation is discussed in Sec. \ref{sec:cmc} ). This then enables us to infer causal structure by fixing the class of physical systems and comparing the observations to different candidate causal graphs.

The advantage of this second approach is that it emphasizes that causal structure and the properties of material systems are inextricably interwoven. First we make an assumption about the causal structure and use it to establish the behaviour of a class of systems, then we fix the class of systems and use it to deduce further causal relations. This way of thinking about physical Markov conditions has the advantage that it can easily accommodate the experimental evidence that quantum systems exhibit different behaviour than classical systems in the same causal structure. Whereas Bell's theorem makes this difference appear dramatic and even paradoxical, on the present account it is interpreted as displaying a simple empirical truth: that quantum systems interact with causal structure in a way that is fundamentally different to the way classical systems do. It also leads to a much more intuitive understanding of the condition of \tit{no fine-tuning}, which we will discuss next.

\subsection{Fine-tuning and latent variables}

The principle of \tit{no-fine-tuning} may be stated as follows:\\

\tbf{NFT. No Fine-tuning:}\\
Let $P(\tbf{X})$ be the behaviour of a typical member of a given class of physical systems, in an experiment where the causal structure is $G(\tbf{X})$. Then there are no statistical independences in $P(\tbf{X})$ beyond those that are implied by the Causal Markov Condition and $G(\tbf{X})$ for that class of systems.\\

The assumption \tbf{NFT} can be motivated from the considerations of the previous section. First consider the case that $G(\tbf{X})$ is one of the three `simple cases' (common-cause, causal chain, or common effect). Any conditional independences that hold in $G(\tbf{X})$ for all (or for `typical') members of the given class of systems must then be implied by the Causal Markov Condition, because the \tbf{CMC} has effectively been \tit{defined} so as to include them. In these cases, therefore, \tbf{NFT} holds as a matter of definition. For general causal structures, one essentially postulates that \tbf{NFT} continues to hold, hence that the \tbf{CMC} continues to capture all of the typical features of the class of systems in these more general experiments. This postulate is useful for it serves as a powerful aid to causal inference: it allows one to eliminate any causal structures that don't explain (via the \tbf{CMC}) all of the independences in the observed behaviour $P(\tbf{X})$.
 
\tit{Remark:} The reader may be uneasy about the vague usage of the word `typical' in the above. One way to formalize this notion is to imagine selecting a system at random from the given class, using some measure defined on the space of possible systems within the class. Then \tbf{NFT} can be read as saying that the subset of systems exhibiting extra independences beyond \tbf{CMC} has measure zero within the class. To maintain greater generality, however, we prefer to leave `typical' a flexible notion to be decided as a matter of practice.

So far we have talked about cases in which all relevant variables of the system are measured in the reference experiment. Frequently, it is not practical to measure all of the relevant variables. In such cases, the causal structure includes \tit{latent variables} $\tbf{L}$ that do not appear in the observed behaviour $P(\tbf{X})$. This defines the strictly larger class of \tit{classical stochastic systems with latent variables} (in which we can recover the classical stochastic systems by setting $\tbf{L}=\emptyset$). The physical Markov conditions for this class are given by the following (strictly weaker) conditions:\\    

\tbf{CMC2. Causal Markov Condition (with latent variables):}\\
There exists an extended distribution $P(\tbf{X},\tbf{L})$, such that $P(\tbf{X},\tbf{L})$ satisfies the \tbf{CMC} for the causal structure of the system $G(\tbf{X},\tbf{L})$, and $P(\tbf{X})$ is obtained from $P(\tbf{X},\tbf{L})$ by marginalizing over the latent variables, i.e.
\eqn{
P(\tbf{X}) = \zum{\tbf{l}}{} \, P(\tbf{X},\tbf{l}) \, .
}

For the sake of simplicity, we will assume from here onwards (unless stated otherwise in the text) that there are no \tit{latent variables} in the systems of interest.

\subsection{Manipulations and un-measurements \label{sec:activemanips}}

In this work, we consider manipulations as modes of measurement that break causal connections, hence they do not include either reference measurements $C^X=\oslash$ or un-measurements $C^X=\brm{undo}$. We avoid making strong commitments as to whether `manipulations' must be effected by agents, and if so whether these should be conscious, etc, but propose only some minimal properties that manipulations ought to satisfy, of which the first is:\\

\tbf{Externality.} A manipulation represents the physical influence on the system by an \tit{external} entity, so as to exclude all \tit{causes} within the system from affecting the manipulated variable. We assume that the system's response to the manipulation does not depend on the nature of this external entity, i.e. whether it is a conscious agent, a physical system, an artificial intelligence, an environment, God, and so on.\\

Note that this leads us into a circularity, since manipulations depend on the definition of a \tit{cause} (i.e. by asserting that the influencing entity has no \tit{causes} in the system), but our proposed definition of manipulationist causation \tbf{MC} is itself based on the concept of a manipulation! Fortunately, this is not a vicious circle, as any realistic situation always involves some causal relations that may be postulated \tit{a priori}. For instance, it is generally accepted that the experimenter has the ability to freely choose which buttons to push on the apparatus, independently of the variables within the system. Having specified such originating causes, we can then deduce \tit{other} causal relations that hold \tit{within} the system.\\

Note that externality is necessary but not sufficient property of any manipulation. Hence any variable whose causes lie entirely outside the system is a \tit{candidate} for being a manipulation, but whether or not it \tit{is} a manipulation may depend on other considerations beyond the scope of our analysis (that we leave to philosophers). Thus, while the convention is to write $C^Z=\oslash$ for an exogenous variable $Z$ in the reference experiment -- thereby declaring it not to be a manipulation -- the property of externality suggests that our reasoning would be unaffected if we were to regard them as manipulations. 

In fact, this allows us to infer what would happen if an exogenous variable were to be manipulated, since (according to externality) nothing about the system would change. We can formalize this as a special inference rule for manipulations of exogenous variables: \\

\tbf{Exogenous indifference:} Let $Z$ be an exogenous variable in $G(\tbf{X},Z)$. Then the externality of manipulations implies that the system's behaviour under manipulations of $Z$ is the same as in the reference experiment:\\ 
\eqn{ \label{eqn:exogextern}
P(\tbf{X}=\tbf{x},Z=z| C^Z=\brm{do}(Z=z)) = P(\tbf{X}=\tbf{x},Z=z| C^Z=\oslash) \qquad  \, \forall \, z \in \trm{dom}(Z)
}

Besides externality, manipulations satisfy the following principle, which also applies to measurements more generally:\\

\tbf{CNS. Counterfactual no-signalling:} If one doesn't condition on the descendants of $Z$, then different ways of measuring $Z$ cannot affect the causal non-descendants of $Z$. Formally, let $C^Z=\{ c,c' \}$ toggle between different ways of measuring $Z$ (that need not be confined to manipulations) in a system whose causal relations are described by a DAG $G(\tbf{A} \tbf{D} \tbf{R} Z)$. Here, $\tbf{A}$ are the causal ancestors of $Z$, $\tbf{D}$ are the descendants of $Z$, and $\tbf{R}$ are the remainder. Then:
\eqn{ \label{eqn:extracontext}
P(\tbf{A} \, \tbf{R}|\,C^Z=c)=P(\tbf{A} \, \tbf{R}|\,C^Z=c') \, \qquad \forall c,c' \in \trm{dom}(C)\,.
}

It is important to note that \tbf{CNS} is conceptually distinct from the principle of \tit{no-signalling} found elsewhere in the literature, which states that, within the reference experiment, an exogenous variable $Z$ (often called a `measurement setting') can only be correlated with its causal descendants. Formally, it can be expressed as:\\

\tbf{NS. No-signalling:}\\
For an exogenous variable $Z$ with non-descendants $\tbf{R}$ we have: 
\eqn{
P(\tbf{R}|Z=z) = P(\tbf{R}|Z=z') \, , \qquad \forall z \in \trm{dom}(Z) \, ,
}
or, more prosaically, $P(\tbf{R}|Z)=P(\tbf{R})$.\\

Although they are conceptually distinct, \tbf{CNS} and \tbf{NS} can be linked by the following rationale. Since an exogenous variable $Z$ has no causes within the system (i.e. $\tbf{A}=\emptyset$) we can, according to externality, equally imagine that it has an external cause given by a set of manipulations of the form $\{ C^Z=\brm{do}(Z=z) : \, z \in \trm{dom}(Z) \}$, and this should make no difference to the statistics, i.e.
\eqn{
P(\tbf{R}|C^Z=\brm{do}(Z=z)) = P(\tbf{R}|Z=z) \, , \qquad \forall z \in \trm{dom}(Z) \, .
}
By applying \tbf{CNS} to that equation, we recover rule \tbf{NS}, which can therefore be thought of as the special case of \tbf{CNS} applied to exogenous variables. This is significant because \tbf{CNS} is a general principle that is expected to hold regardless of the class of systems one is working with. Hence, within the general framework discussed here, all classes of physical systems are assumed to obey \tbf{CNS} and hence also no-signalling, regardless of whether they are classical, quantum, or something else.

The principles of externality and counterfactual no-signalling are assumed to apply to all manipulations, but specific types of manipulations may also have additional defining properties. 

In classical causal modeling it is customary to restrict attention to interventions, but in this work we will include inferences about un-measurements. We add to the properties mentioned in Sec. \ref{sec:measurements} of un-measurements the condition that, so long as a variable does not depend on the value of $Z$, it also should not depend on whether or not $Z$ is measured. More precisely:\\

\tbf{CSO. Counterfactual screening-off:} Suppose that for some disjoint $\tbf{A},\tbf{B},Z$ we have that $\tbf{A}$ is independent of $Z$ conditional on $\tbf{B}$ in the reference behaviour. Then $\tbf{A}$ is also independent of \tit{whether or not Z is measured} conditional on $\tbf{B}$. Formally,
\eqn{
P(\tbf{A}|\tbf{B} Z , \, C^Z=\oslash) &=& P(\tbf{A}|\tbf{B} , \,C^Z=\oslash) \nonumber \\ \Rightarrow P(\tbf{A}|\tbf{B} , \, C^Z=\brm{undo} ) &=& P(\tbf{A}|\tbf{B} , \, C^Z=\oslash) \, .
}


\section{Counterfactual classical causal models \label{sec:ccms}}

In this section, we restrict attention to causal modeling with the class of classical stochastic systems, assuming no latent variables. We discuss the origin and characteristics of the physical Markov conditions that hold for these systems, and then we introduce \tit{non-disturbing measurements} and \tit{interventions} and discuss their corresponding counterfactual inference rules.

\subsection{The Causal Markov Condition \label{sec:cmc}}

\tit{Classical causal models} refer to causal models of classical stochastic systems. Hence the physical Markov conditions are those entailed in \tbf{CMC} and these tell us how the causal relations of the system constrain the allowed behaviour in the reference experiment under the assumption of \tbf{NFT}. We then have:\\

\tbf{Classical Causal Model:}\\
A Classical Causal Model consists of a pair $\{P(\tbf{X}), G(\tbf{X}) \}$ where $P(\tbf{X})$ satisfies the \tit{Causal Markov Condition} and no fine-tuning for the DAG $G(\tbf{X})$.\\

As discussed earlier in Sec. \ref{sec:physMarkov}, the \tbf{CMC} can be extrapolated to general causal structures from three special cases. We will now discuss the details of how this extrapolation is carried out. The first special case is already familiar; we repeat it here for convenience:\\

\begin{figure}[!htb]
\centering\includegraphics[width=0.7\linewidth]{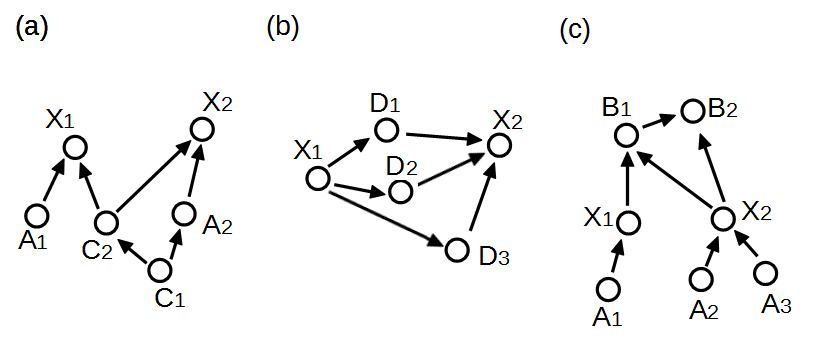}
\caption{Three special cases in which the Causal Markov Condition reduces to simpler principles: (a) variables $X_1,X_2$ only connected through common causes, where \tbf{CMC} reduces to \tbf{FCC}; (b) $X_1,X_2$ only connected through intermediate causes, where \tbf{CMC} reduces to \tbf{SSO}; (c) $X_1,X_2$ connected only through common effects, where \tbf{CMC} reduces to \tbf{BK}.}
\label{fig:threedags}
\end{figure}

\noindent \tbf{FCC. Factorization on common causes:} \\
Suppose neither of $X_1,X_2$ is a cause of the other and $\tbf{C}$ is the complete set of their shared ancestors, i.e. $\tbf{C}$ contains all `common causes' (see Fig. \ref{fig:threedags}(a)); then $P(X_1 X_2|\tbf{C})=P(X_1|\tbf{C})P(X_2 |\tbf{C})$ \footnote{Perhaps contrary to one's first intuition, it is not sufficient to condition only on the set of variables that are \tit{parents} of both $X_1,X_2$. A trivial counterexample is $X_1 \leftarrow A_1 \leftarrow A_3 \rightarrow A_2 \rightarrow X_2$.}.\\  

Taken as a postulate, the principle \tbf{FCC} was historically conceived as just one part of a more general postulate proposed by Hans Reichenbach \cite{RPCC}. The \tbf{FCC} is sometimes called the \tit{quantitative} part of Reichenbach's Principle to distinguish it from the \tit{qualitative} component \cite{CAVLAL}, which we will here simply refer to as `Reichenbach's Principle':\\

\noindent \tbf{RP. Reichenbach's Principle:}\\
If neither of $X_1,X_2$ is a cause of the other and they have no shared ancestors, then they are statistically independent: $P(X_1 X_2)=P(X_1)P(X_2)$. (Note: Following Ref. \cite{ALLEN} we have presented it in the contrapositive of its more common form: `if two variables are correlated, one must cause the other or they must have a common cause, or both').\\ 

Since \tbf{RP} can be obtained from \tbf{FCC} by setting $\tbf{C}=\emptyset$, it is a strictly weaker principle. \tbf{RP} captures the intuitive fact that two systems with independent causal histories should be initially uncorrelated. Unlike \tbf{FCC}, whose application to quantum systems is controversial, \tbf{RP} is widely accepted to hold for quantum systems. This will be discussed further in Sec. \ref{sec:qmc}). In addition to \tbf{RP}, it is usually assumed that not only are physical systems independent prior to interaction, but that they are correlated afterwards. In Ref. \cite{PRICEBOOK}, Price calls this the `principle of independence' and summarized it by the slogan `innocence precedes experience'. However, it must be emphasized that the principle is composed of two conceptually distinct components: first, that systems are independent before they interact (\tbf{RP} in the present framework), and second, that they are typically correlated after they interact. We define this latter requirement as:\\

\noindent \tbf{PE. The Principle of Experience:}\\
If neither of $X_1,X_2$ is a cause of the other and they do have shared ancestors, then one generally expects them to be correlated: $P(X_1 X_2) \neq P(X_1)P(X_2)$. \\ 

Price's `principle of independence' in our framework is then the conjunction of \tbf{RP} and \tbf{PE}, which is a manifestly asymmetric combination. Price argues (and we agree) that this asymmetric combination, while intuitive in the macroscopic classical world, does not extend to microscopic systems, and hence that quantum systems should satisfy a more symmetric principle. Later on in the present work we will advocate retaining \tbf{RP} and dropping \tbf{PE} for quantum systems. For the moment we are discussing classical systems, and so will retain \tbf{PE}. The second simple case on which the \tbf{CMC} is based is that of the causal chain, for which the following is assumed to hold:\\

\noindent \tbf{SSO. Sequential screening-off:}\\
Suppose $X_1$ causes $X_2$ and every path connecting $X_1$ to $X_2$ is intercepted by a variable in $\tbf{D}$, i.e. contains a chain $X_1 \rightarrow D \rightarrow X_2$, where the $D$ are not causes of one another (see Fig. \ref{fig:threedags}(b)); then $P(X_1 X_2|\tbf{D})=P(X_1|\tbf{D})P(X_2 |\tbf{D})$ .\\ 

The principle \tbf{SSO} says that conditioning on $\tbf{D}$ `screens off' the future measurement $X_2$ from the past, because knowing $\tbf{D}$ makes the information $X_1$ redundant. The third simple case on which the \tbf{CMC} is based is:\\

\noindent \tbf{BK. Berkson's rule:}\\
Suppose neither of $X_1,X_2$ is a cause of the other and they have no shared ancestors, and suppose $B$ is a common descendant of $X_1,X_2$ (see Fig. \ref{fig:threedags}(c)); then one generally expects $X_1,X_2$ to be correlated conditional on $B$, i.e. $P(X_1X_2|B) \neq P(X_1|B)P(X_2|B)$.\\

The principle \tbf{BK} derives from a well-known result in statistics called `Berkson's Paradox', after the medical statistician Joseph Berkson \cite{BERK}. Despite not really being a paradox, newcomers to statistics often find it counter-intuitive.

\tit{Remark:} Whereas \tbf{FCC} and \tbf{SSO} give conditions under which statistical independence is \tit{necessary}, the principle \tbf{BK} gives conditions under which correlations are `typical' but not necessary. In fact it is a convention of causal modeling that physical Markov conditions either assert the \tit{necessity} of independence or the \tit{possibility} of correlation, but never assert the \tit{necessity} of correlation or the \tit{possibility} of independence. That is because it is not possible to encode all four types of statements within a single graph. If one adheres to the first two types of statements, the graph is called an `independence map' of correlations; if one opts for the latter two types, it is a `dependence map'.

Strictly speaking, the aforementioned conditions \tbf{FCC}, \tbf{RP}, \tbf{PE}, \tbf{SSO}, \tbf{BK} can only be applied to causal graphs having the form of one of the three special cases displayed in Fig. \ref{fig:threedags}. Ideally, we would like to have an empirical postulate that could apply to a system with arbitrary causal structure. One way to achieve this is to convert the conditions into corresponding \tit{graphical} rules that allow their consequences to be deduced by direct inspection of the causal structure. The graphical rules can then be jointly applied to any causal structure. The full details of how one obtains graphical rules from principles are given in Appendix \ref{app:graphrules}. The result is the following alternative graphical formulation of the \tbf{CMC} \cite{PEARL,SGS}:\\

\tbf{CMC3. Causal Markov Condition (graphical version)}:\\
Let $\tbf{U}$,$\tbf{V}$ and $\tbf{W}$ be disjoint sets of nodes in a DAG $G(\tbf{X})$. A path from $X_1$ to $X_2$ in $G(\tbf{X})$ is said to be \tit{blocked} by $\tbf{W}$ iff at least one of the following graphical conditions holds:\\

\noindent \tbf{g-FCC.} There is a fork $X_1 \leftarrow C \rightarrow X_2$ on the path where $C$ is in $\tbf{W}$;\\
\tbf{g-SSO.} There is a chain $X_1 \rightarrow C \rightarrow X_2$ on the path where $C$ is in $\tbf{W}$;\\
\tbf{g-BK.} There is a collider $X_1 \rightarrow C \leftarrow X_2$ on the path where $C$ is \tit{not} in $\tbf{W}$ and has no descendants in $\tbf{W}$.\\
If \tit{all} paths between $\tbf{U}$,$\tbf{V}$ are blocked by $\tbf{W}$, then the \tit{d-separation theorem} \cite{PEARL} states that $\tbf{U}$ and $\tbf{V}$ are independent conditional on $\tbf{W}$ in any distribution $P(\tbf{X})$ that satisfies the \tbf{CMC} (in the form of Eq. \eqref{eqn:cmc}) for the DAG $G(\tbf{X})$. \\

Note that the principles \tbf{PE} and \tbf{RP} are implicit in the graphical rules, in the following way. If \tbf{PE} did not hold, then there would be another way in which two variables could be independent, namely, by sharing a common cause that is not conditioned upon. The absence of such a rule in the above list is the result of enforcing \tbf{PE}. The principle \tbf{RP} is implicit in the rule \tbf{g-BK}. If \tbf{RP} were false, then the mere presence of a common descendant might enable two variables to be correlated, which would mean that \tbf{g-BK} could not be a graphical rule. 

Thus by interpreting the conditions \tbf{FCC}, \tbf{RP}, \tbf{PE}, \tbf{SSO}, \tbf{BK} as \tit{graphical} criteria as described above, one can derive any and all constraints implied by the \tbf{CMC}. In this way, although the three conditions \tbf{FCC},\tbf{SSO}, \tbf{BK} individually apply to only limited classes of causal structures, they can be combined via the graphical representation to obtain a condition that applies to arbitrary causal structures, and this condition is the \tbf{CMC}. 

\subsection{Observation and intervention in classical causal models \label{sec:passactclass}}

Our restriction to the class of classical stochastic systems allows us to further refine the measurements involved in the reference experiment. For causal modeling, we need to consider only two kinds of measurements: non-disturbing measurements and interventions.

A non-disturbing measurement is a measurement that can be performed without disturbing the system in any way. To formalize this idea, we need to clarify what is meant by a `disturbance of the system'. In the present framework we are concerned only with what can be detected at the level of probabilities, and so whether the presence or absence of the measurement affects the probabilities of other variables in the system.

More precisely, we partition the measurements on a system in the reference experiment into two sets $\tbf{X} \cup \tbf{Z}$, and write the system's behaviour as $P(\tbf{X} \tbf{Z}|C^Z=\oslash)$ where the conditional $C^Z=\oslash$ is to remind us that `the reference measurements $\tbf{Z}$ are actually performed in this experiment'. Conversely, let $P(\tbf{X}|C^Z=\brm{undo})$ indicate the behaviour of the system in a counterfactual experiment in which the measurements $\tbf{Z}$ are not performed. Then we define:\\

\tbf{ND. Non-disturbing measurements:} \\
The set of measurements $\tbf{Z}$ is called \tit{non-disturbing relative to the set of variables} $\tbf{S} \subseteq \tbf{X}$ if:
\eqn{ \label{eqn:nondisturb}
P(\tbf{S}|C^Z=\brm{undo})  &=& P(\tbf{S}|C^Z=\oslash) \, \nonumber \\
 &=& \zum{\tbf{z}}{} \, P(\tbf{S},\tbf{z}|C^Z=\oslash) \, .
}
In the special case that $\tbf{Z}$ is non-disturbing relative to all other variables in the system (i.e. $\tbf{S} = \tbf{X}$) we will simply call $\tbf{Z}$ \tit{non-disturbing}.\\

We can summarize equation \eqref{eqn:nondisturb} as saying that $\tbf{Z}$ are non-disturbing iff not performing them is equivalent to performing them and summing over their outcomes (which resembles, but is conceptually distinct from, the `law of total probability'). Note that this equation is an example of counterfactual inference rule, although in this instance it does not depend on the causal structure of the system.

\tit{Remark:} Strictly speaking the above definition \tbf{ND} is ambiguous when applied to exogenous variables; since the causes of exogenous variables (if any) lie outside the system and are not subject to analysis, we cannot say what would have happened if an exogenous measurement were not performed. Indeed, since they play a role in setting the very conditions that define the system, we cannot `un-measure' them without enlarging the scope of analysis to include a larger encompassing system, but the new exogenous variables of the larger system will again necessarily be ambiguous.

In the case of classical causal models, we restrict attention to experiments in which all variables represent either non-disturbing measurements or interventions. In this case an experiment will be at the top of the hierarchy of counterfactual experiments (cf Sec. \ref{sec:experiments} ) if and only if all non-exogenous variables are non-disturbing. Thus, in classical causal models, the reference experiment is typically also a passive observational scheme. However, this need not be true in general, as we will see later with quantum systems.

In contrast to non-disturbing measurements, an \tit{intervention} is a type of manipulation that disturbs the system in a precise way that targets a specific variable. An intervention can usefully be thought of as equivalent to introducing a randomized control into an experiment. Physically, it means forcing a target variable $W$ to take a particular value in a manner that is independent of its causes $\pa{W}$ within the system. One example is actively controlling the temperature of a system instead of passively measuring its temperature under ambient conditions. Another is assigning patients in a drug trial randomly to the treatment or control groups, instead of allowing them to choose whether to take the treatment themselves. 
 
An intervention $C^{W} = \brm{do}$ results in a new set of probabilities $P(\tbf{X}|C^{W}=\brm{do})$ that describes the behavior of the system in the counterfactual experiment in which the variable $W$ is intervened upon. If we wish to be more specific, we may use $C^{W} = \brm{do}(W=w)$ to mean that the intervention intends to fix the value of $W$ to $w$. 

\tit{Remark:} We must be careful to distinguish our notation from that of the \tit{do-conditional} notation used in the literature, eg. $P(\tbf{X}|\trm{do}(W=w))$ \cite{PEARL,SGS}. The key difference is that our notation treats separately the fact that $W$ is intervened upon, as represented by $C^{W} = \brm{do}$ or $C^{W} = \brm{do}(W=w)$, from the fact that it takes the value $W=w$, which is expressed just by the value of $W$. Thus, for instance, we can assign a non-zero probability to the event $P(W=w'|C^W=\brm{do}(W=w''))$, $w' \neq w''$, which we interpret as the probability that an intervention whose aim is to fix $W$ to $w''$ results in the undesired outcome $W=w'$, as might occur if there were some noise or errors in the physical implementation of the intervention. By constrast, the `do conditional' expression $P(W=w'|\trm{do}(W=w''))$ is either undefined or defined to be zero. Our notation incorporates the standard do-conditional as a special case that obtains when the intervention perfectly achieves its aim, that is, when $P(W=w'|C^W=\brm{do}(W=w'')) = \delta(w',w'')$. Under that assumption, we may identify our expressions of the form $P(\tbf{X}|C^W=\brm{do}(W=w))$ with standard do-conditionals of the form $P(\tbf{X}|\trm{do}(W=w))$. For convenience, in the remainder of this work we will assume that this is the case.

\begin{figure}[!htb]
\centering\includegraphics[width=0.7\linewidth]{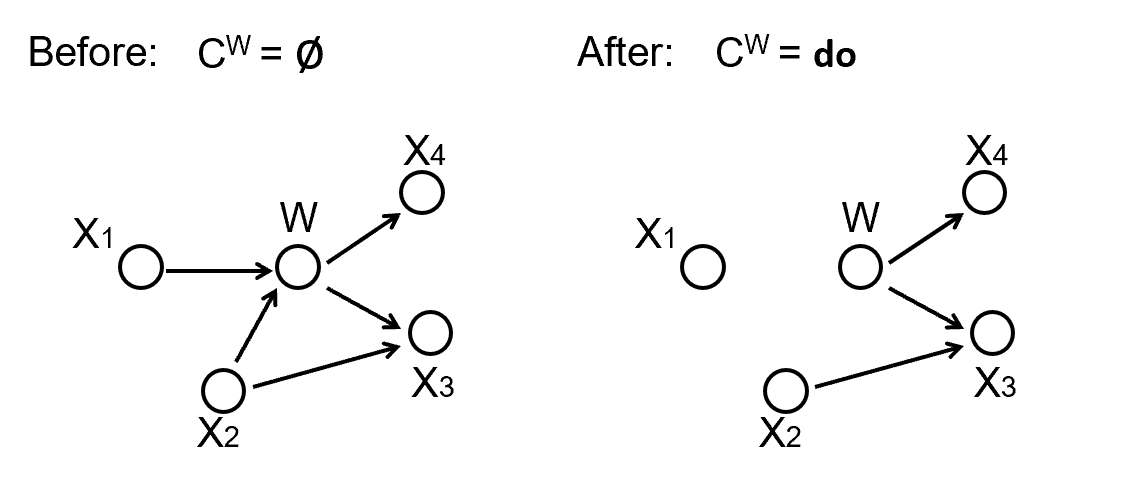}
\caption{A causal graph before and after a manipulation (or an intervention) of $W$.}
\label{fig:intervention}
\end{figure}

Like all manipulations, interventions satisfy externality, which suggests that the causal structure after the intervention, $G(\tbf{X}|C^{W}=\brm{do})$, should be obtained from $G(\tbf{X})$ in the reference experiment by deleting all incoming arrows to $W$ (see Fig. \ref{fig:intervention}).\\

Beyond this basic rule, interventions are assumed to precisely target $W$, which means that any effect the intervention has on other variables must be mediated through the intervention's effect on $W$ itself. More specifically, if we are contemplating performing interventions on any of the causal parents of some variable $D$, then the probabilities of $D$ should be insensitive to which, if any, of its parents are intervened-upon. Practically speaking, this rule amounts to the elimination of `placebo effects' in the experimental design, hence we define it as:\\

\tbf{NPE. No placebo effect:}\\
Let $\tbf{W} \subseteq \pa{X}$ be any subset of the parents of a variable $X$. Then, conditional on all of its parents, $X$ should be insensitive to whether $\tbf{W}$ is intervened-upon:
\eqn{ \label{eqn:NPE}
P(X|\pa{X},C^{\tbf{W}}=\brm{do}) = P(X|\pa{X}) \, .
}

For a single variable $X$, the intuition behind \tbf{NPE} is straightforward. Intervening on a parent of $X$ can only affect $X$ through two avenues: either through $X$'s direct dependence on the values of its parents, or through the indirect effect of deleting the causes incoming to $\pa{X}$; \tbf{NPE} states that only the former route is legitimate. The latter route can only directly affect the (non-parental) ancestors of $X$, since the induced correlations among these may depend on the causal links to $\pa{X}$ which are disrupted by interventions. For these effects to plausibly `propagate' to $X$ would require the existence of an unblocked path from these ancestors to $X$, however, all such paths are blocked: either the path contains a member of $\pa{X}$ and so is blocked by \tbf{g-SSO}, or else it contains an unconditioned collider and so is blocked by rule \tbf{g-BK}. Hence deleting causal arrows incoming to $\pa{\tbf{X}}$ should have no means within the causal structure of affecting the conditional probabilities $P(X|\pa{\tbf{X}})$; this is what \tbf{NPE} asserts.

In the present work, we will need to generalize this principle to more variables. It is not clear how to do this in the most general way, because the parents of one variable $X_1$ might be children of another variable $X_2$, so some variables might reasonably be sensitive to interventions on the parents of other variables. However, it suffices here to make a restricted generalization of the principle as follows:\\

\tbf{NPE2. Generalized no placebo effect:}\\
Let $\tbf{X}$ be a set of variables, let $\pa{\tbf{X}}$ be the union of all their parents, and let $\tbf{W} \subseteq \pa{\tbf{X}}$ be any subset of these. Finally, let $\tbf{D}$ be all descendants of $\tbf{X}$ such that all directed paths to $\tbf{D}$ from the ancestors of $\tbf{X}$ pass through $\tbf{X}$ itself. Under the assumption that the members of $\pa{\tbf{X}}$ are not causes of one another (direct or indirect), then, conditional on all $\pa{\tbf{X}}$, we expect both $\tbf{X},\tbf{D}$ to be insensitive to whether $\tbf{W}$ is intervened-upon:
\eqn{ \label{eqn:NPE2}
P(\tbf{X}\tbf{D}|\pa{\tbf{X}},C^{\tbf{W}}=\brm{do}) = P(\tbf{X}\tbf{D}|\pa{\tbf{X}}) \, .
}

As with \tbf{NPE}, this assumption is motivated on the grounds that, conditional on the values of $\pa{\tbf{X}}$, the mere fact of intervening on $\tbf{W} \subseteq \pa{\tbf{X}}$ can only directly affect the ancestors of $\pa{\tbf{X}}$. As these have no unblocked paths connecting them to $\tbf{X}$ or $\tbf{D}$, we expect $P(\tbf{X}\tbf{D}|\pa{\tbf{X}})$ to be insensitive to such effects, and this is what \tbf{NPE2} asserts. (Note: since the $\pa{\tbf{X}}$ are not causes of one another, it is again true that any path from the unconditioned ancestors of $\tbf{X}$ leading to $X,\tbf{D}$ must either go through $\pa{\tbf{X}}$ and so be blocked by \tbf{g-SSO}, or else contain an unconditioned collider and so be blocked by \tbf{g-BK}).

We are now ready to derive the counterfactual inference rule for interventions. It is enough to posit that the post-intervention probabilities $P(\tbf{X}|C^W=\brm{do})$ should satisfy the $\tbf{CMC}$ relative to the new DAG $G(\tbf{X}|C^{W}=\brm{do})$. This is a useful requirement, because it means that the post-intervention pair $\{P(\tbf{X}|C^{W}=\brm{do}), G(\tbf{X}|C^{W}=\brm{do})\}$ is again a valid causal model and can therefore be used as the starting point for further counterfactual inferences, such as additional interventions. Given the structure of $G(\tbf{X}|C^{W}=\brm{do})$, the \tbf{CMC} implies that $P(\tbf{X}|C^{W}=\brm{do})$ factorizes into a product of the general form:
\eqn{ \label{eqn:infruleprimer}
P(\tbf{X}|C^{W}=\brm{do})=P(W|\pa{W} , \, C^{W}=\brm{do}) \, \prod_i \, P(X_i|\pa{X_i} , \, C^{W}=\brm{do}) \, ,
}
where $\pa{W}$ refers to the variables that were parents of $W$ in the pre-intervention graph. By externality, we expect $W$ to be independent of its former parents after the intervention, so $P(W|\pa{W}, \, C^{W}=\brm{do})=P(W|C^{W}=\brm{do})$. If we wish to be more precise and use fine-grained ideal interventions, we can reduce this to $P(W|\pa{W}, \, C^{W}=\brm{do}(W=w'))=\delta(w,w')$. 

The remaining terms $P(X_i|\pa{X_i} C^{W}=\brm{do})$ must be treated differently depending on whether $X_i$ is a descendant of $W$ or not. In both cases we obtain the same rule,\\
\eqn{ \label{eqn:infrule2}
P(X_i|\pa{X_i} C^{W}=\brm{do}) = P(X_i|\pa{X_i} C^{W}=\oslash) \, ,
}
but the justification differs in each case. For the non-descendants of $W$, \eqref{eqn:infrule2} follows from \tbf{CNS}, whereas for the descendants of $W$, it follows from \tbf{NPE}. Putting these together in \eqref{eqn:infruleprimer}, we finally obtain the inference rule for interventions:\\
                
\tbf{IR. Inference of interventions:} An observer's probability assignments for a counterfactual experiment where an intervention is performed on $W$ are given by:
\eqn{  \label{eqn:intervaxiom}
P(\tbf{X}|C^{W}=\brm{do}) = P'(W) \, \prod_i \, P(X_i|\pa{X_i} , \, C^{W}=\oslash) \, ,
}
where $P'(W):=P(W|C^{W}=\brm{do})$ is a distribution of values that characterizes the particular intervention. For the case of fine-grained interventions,

\eqn{  \label{eqn:intervaxiom2}
P(\tbf{X}|C^{W}=\brm{do}(W=w')) = \delta(w,w') \, \prod_i \, P(X_i|\pa{X_i} , \, C^{W}=\oslash) \, ,
}

Note that in textbooks the above rule is usually stipulated as an axiom, rather than derived from principles as we have done here. To infer the result of interventions on multiple variables $\tbf{W} \subseteq \tbf{X}$, the procedure for intervening on one variable can simply be iterated; it can easily be proven that the order of interventions does not affect the final result, i.e. sequential interventions on different variables commute.

 \section{Counterfactual causation of quantum systems \label{sec:qbism}}

\subsection{Problems with defining quantum causal models \label{sec:qm}}

Following the general pattern that was established in the classical case, there are three main questions that need to be answered when attempting to define a quantum causal model. First, what should be used as the reference experiment, and what characterizes the reference measurements used in it? Second, what are the relevant physical Markov conditions for the class of quantum systems, and in what ways do these deviate from the classical ones (i.e. the \tbf{CMC})? Third, what are the inference rules that tell us how to compute the probabilities for interventions ($C^W=\tbf{do}$) and un-measurements ($C^W=\tbf{undo}$), using only the causal model consisting of the reference behaviour $P(\tbf{X})$ and the causal structure $G(\tbf{X})$? In order to answer these questions, we must first address two well-known obstacles to quantum causal modeling: the fact that quantum measurements are disturbing, and the fact that common-causes don't factorize (Bell's Theorem). These are the topics of the next two subsections.
 
\subsubsection{Screening-off and measurement disturbance \label{sec:soff}}

In quantum mechanics, the most general way to describe a quantum measurement associated with a random variable $Y$ is by a \tit{quantum instrument}:\\

\tbf{QI. Quantum instrument:} Given a random variable $Y$, a \tit{quantum instrument} assigns a completely positive (CP) linear map $\mathcal{M}_y : \mathcal{H}_{\trm{in}} \mapsto \mathcal{H}_{\trm{out}}$ to each outcome $y \in \trm{dom}(Y)$, subject to the conditions:\\
(i) The outcome probabilities can be expressed as $P(Y)=\tr{\mathcal{M}_y(\rho)}$ where $\rho$ is a density operator representing the \tit{input state} to $Y$;\\
(ii) The induced map $\mathcal{C}(\cdot) := \zum{y}{} \mathcal{M}_y (\cdot)$ defined by summing over the outcomes $Y$ is a valid \tit{quantum channel}, i.e. $\mathcal{C}$ is completely positive and trace-preserving (CPTP) and hence maps density operators on $\mathcal{H}_{\trm{in}}$ to density operators on $\mathcal{H}_{\trm{out}}$.\\
(iii) $\mathcal{H}_{\trm{in}}$ (resp. $\mathcal{H}_{\trm{out}}$) represents the Hilbert space of the system immediately prior to (resp. after) the measurement $Y$. Note: it is natural to assume that the measurement process preserves the dimension, so we will adopt the convention that dim($\mathcal{H}_{\trm{in}}$)=dim($\mathcal{H}_{\trm{out}}$):=$d_Y$, and will sometimes use the notation $\mathcal{H}_Y$ to refer to any Hilbert space of dimension $d_Y$.\\

A measurement described by a quantum instrument $\{ \mathcal{M}_Y \}$ is non-disturbing in the sense of \tbf{ND} only if it describes a channel that does not change the quantum state, i.e. only if $\mathcal{C}(\rho)=\rho$ for all relevant input states $\rho$. The proof is by counterexample: if $\mathcal{C}(\rho) \neq \rho$, then the probabilities  $P(Z)$ of an immediately subsequent measurement $Z$ would suffice to probabilistically distinguish the state $\mathcal{C}(\rho)$ from $\rho$, and hence to distinguish an experiment in which $Y$ is performed prior to $Z$ (and its outcome disregarded) from an experiment in which $Y$ is not performed prior to $Z$, hence $Y$ is disturbing relative to $Z$.

A problem then arises because of the well-known fact that for quantum systems there is `no information without disturbance' \cite{BUSCH}. Formally, this is expressed by the mathematical theorem that the only quantum instrument that can represent a non-disturbing measurement is the \tit{trivial instrument}, whose elements $\mathcal{M}_y$ are all equal to some constant $c_y$ times the identity operator. Since $P(Y=y)=c_y$ is evidently independent of the input state, the outcome of such a measurement provides no information about the measured system.   

\tit{Remark:} it is instructive to point out why non-trivial non-disturbing measurements can exist for classical systems. We may interpret the \tit{classical limit} as referring to the special case in which all relevant states $\rho$ are confined to a subset of \tit{classical states} that are defined to be diagonal in a particular basis of Hilbert space (whose selection may be justified, for instance, on the grounds of environmental decoherence). Classical measurements can then be modeled using the quantum formalism as a special case of quantum instruments that map the classical subspace to itself. An instrument is then non-disturbing only if it preserves the states within the classical subspace, which amounts to a much weaker constraint than requiring that all quantum states be preserved. For instance, a projective measurement in the classical basis using the L\"{u}der's rule to define the outgoing state is non-disturbing relative to the classical subspace, and yet it provides sufficient information to reconstruct the input state, i.e. it is \tit{informationally complete} relative to the classical subspace. 

The fact of no information without disturbance implies that quantum reference measurements must be disturbing. Recalling from Sec. \ref{sec:cmc} that the screening-off condition \tbf{SSO} requires that the measurement produce enough information about the state to render its previous history redundant, this seems to put \tbf{SSO} at odds with the desire for measurements to be minimally disturbing. This has led some authors to propose that the \tbf{SSO} should be relaxed for quantum systems, either by introducing `quantum nodes' that cannot be conditioned upon as in Ref. \cite{HLP}, or by dropping the requirement altogether, as in Refs. \cite{PIEBRUK,FRITZ}. At the opposite extreme, one might choose to allow quantum measurements to be arbitrarily disturbing, even to the point of breaking the causal link between input and output. On the latter view, quantum measurements are a generalization of classical manipulation \cite{COSHRAP,ALLEN}. In these frameworks \tbf{SSO} can trivially be upheld for interventions in which the post-measurement state is simply discarded and a new state prepared in its place independently of the measurement outcome. 

From the perspective of the present work, neither of these options is appealing. As one of the physical Markov conditions, \tbf{SSO} is supposed to tell us something fundamental about the nature of possible measurements on physical systems, namely, that it is possible to measure a system in such a way that the acquired information renders the information from previous measurements \tit{redundant} for future measurements (recall Sec. \ref{sec:cmc}). Achieving this by manipulations or interventions is too heavy-handed, for in that case the past information is not merely made redundant by the measurement outcome, but is actually \tit{destroyed} along with the causal link between input and output. Better would be to find a middle ground in which \tbf{SSO} can be retained for quantum reference measurements while avoiding the destructiveness of interventions. 

To see how this can be done, first consider the simplest case of \tbf{SSO} involving three sequential measurements $A,B,C$, whose causal relations are assumed to be given by the causal chain $A \rightarrow B \rightarrow C$. Let the measurement of $B$ be represented by a quantum instrument $\{\mathcal{M}_B \}$. The state after preparing a state $\rho_A$ as the input to $B$ and obtaining the outcome $B=b$ may be written as:
\eqn{
\rho_b(A) := \frac{\mathcal{M}_b(\rho_A)}{\tr{\mathcal{M}_b(\rho_A)}} \, .
}
According to \tbf{SSO}, we must have that $A$ and $C$ become uncorrelated conditional on the value of $B$, i.e. that $P(A,C|B)=P(A|B)P(C|B)$. Since the input state to $C$ conditional on $B=b$ is $\rho_b(A)$, it will always be possible to find some measurement $C$ whose outcome is correlated with $A$, so long as $\rho_b(A)$ depends explicitly on the value of $A$. The only way to avoid such correlations for any $\rho_A$ is therefore to demand that $\mathcal{M}_b$ has the form:
\eqn{ 
\mathcal{M}_b(\rho_A) := \tr{ \mathcal{M}_b \rho_A } \, \rho(b)
} 
for all $b \in \trm{dom}(B)$, where $\rho(b)$ is a density matrix that can only depend on the outcome $b$. The quantum channel produced by such measurements after summing over $B$ has the form:
\eqn{ \label{eqn:Holevo}
\mathcal{C}(\cdot):= \zum{b}{} \, \tr{ \mathcal{M}_b (\cdot) } \, \rho(b) \, ,
}
which characterizes a class of channels first studied by Holevo \cite{HOLEVO}. These have the interesting property of being equivalent to the class of \tit{entanglement breaking} channels \cite{HORODECKI}, which are defined by the property that the output state $\mathcal{C}(\rho_A)$ cannot be entangled to any other systems, regardless of the input. The form \eqref{eqn:Holevo} shows that it is possible to maintain \tbf{SSO} while at the same time preserving a causal link between the input and output of the measurement $B$. This can be seen in a number of ways, but it is sufficient to note that the output state $\mathcal{C}(\rho_A)$ after summing over $B$ may be written in `ensemble' form as:
\eqn{
\mathcal{C}(\rho_A) = \zum{b}{} \, P(B=b|A) \, \rho(b) \, ,
} 
from which it can clearly be seen that the output state depends on $A$ not through the individual states $\rho(b)$, but through their relative weights $P(B=b|A)$ in the ensemble. Therefore, so long as $\rho(b)$ maintains an explicit dependence on $B$, and so long as $\mathcal{M}_b$ is non-trivial (i.e. to ensure that the weights $P(B=b|A)$ do depend on $A$), the instruments of this class do not break the causal link.

Having identified the general form of the quantum reference measurements, we might ask whether there is some sense in which they are analogous to the classical passive observations. One benefit of defining an appropriate quantum analog of a classical passive observational scheme is that this would enable us to ask whether quantum systems are better resources for causal inference than classical systems, under the constraint of passive observation (or its analog), see eg. \cite{KUEBLER, RIED, RIEDPHD}. We will discuss this later in Sec. \ref{sec:discuss}; for the time being we take the conservative view that quantum reference measurements belong to their own special class, distinct from both manipulations and passive observations.

\subsubsection{Common causes and entanglement \label{sec:comm}}

In the previous sections, we explained how \tbf{SSO} could be maintained for quantum reference measurements, which were found to be necessarily disturbing measurements. In this section we turn to another of the classical Markov conditions, \tbf{FCC}, and review how it fails for quantum systems. Consider a quantum experiment consisting of three measurements $A,B,C$ represented by the common cause graph $B \leftarrow A \rightarrow C$. In this case the \tbf{FCC} (if applicable to quantum systems) would imply $P(A,B,C)=P(B|A)P(C|A)P(A)$.

To see how this condition can fail, consider a particular implementation of this experiment in which $A$ measures a system with Hilbert space $\mathcal{H}_A := \mathcal{H}_B  \otimes \mathcal{H}_C $, and $B,C$ are subsequently performed on the parts of the system that are represented by the respective sub-spaces $\mathcal{H}_B$ and $\mathcal{H}_C $. If the state $\rho_a$ produced by the event $A=a$ is entangled between the partitions corresponding to $B,C$, then it is possible to violate the factorization condition \tbf{FCC}, even under ideal experimental conditions. When dealing with classical systems, a natural response would be to guess that there must be additional \tit{latent variables} $\tbf{L}$ serving as additional common causes of $B,C$, which, if conditioned on together with $A$, would eliminate the correlations (i.e. that the extended principle \tbf{CMC2} should still hold). 

There are two strong reasons why this explanation does not work in the quantum implementation just described. The first is the observation that there are no variables in the standard quantum formalism to play the role of $\tbf{L}$, and so the standard quantum formalism would have to be regarded as incomplete, and the missing variables sought after experimentally. Yet despite much effort (and notwithstanding philosophical arguments for their inclusion) direct experimental evidence for such variables remains elusive.  

\begin{figure}[!htb]
\centering\includegraphics[width=0.3\linewidth]{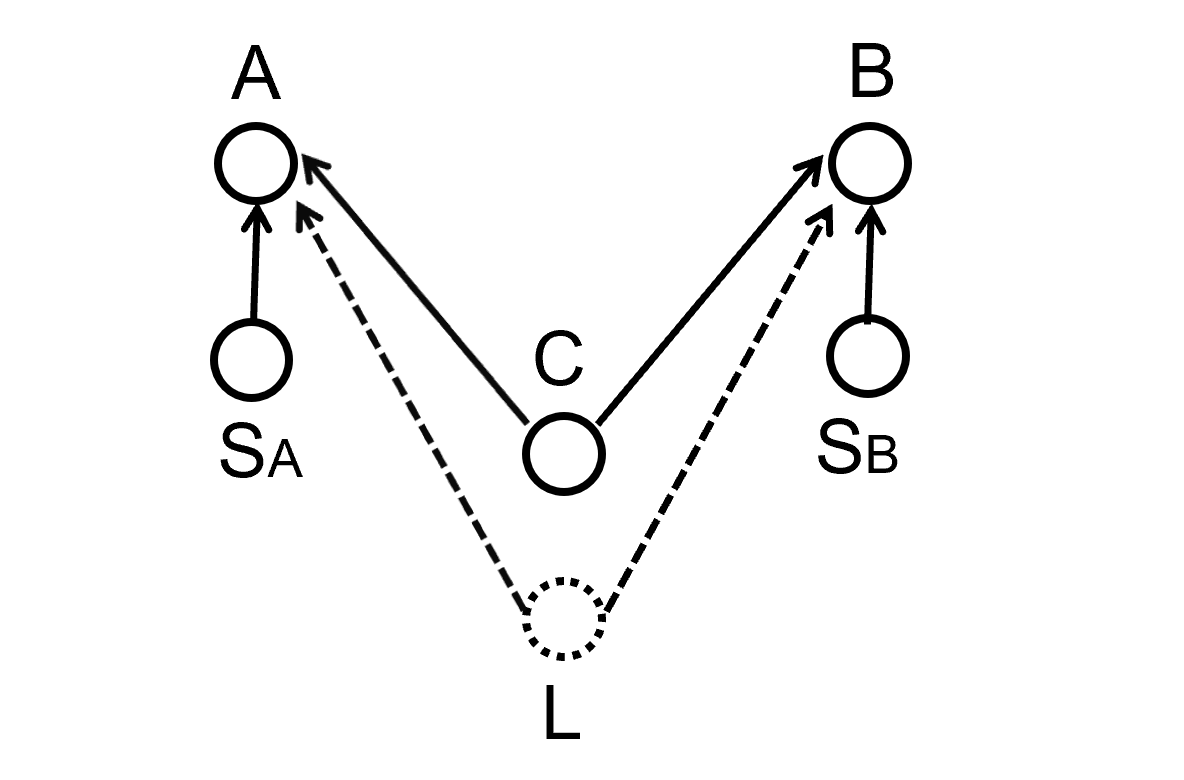}
\caption{The Bell scenario, \tit{aka} the common-cause scenario with variable measurement settings. Assuming no fine-tuning, this causal hypothesis is ruled out as an explanation for quantum systems whose statistics violate Bell inequalities.}
\label{fig:Bell}
\end{figure}

Perhaps the most compelling argument against the existence of the hypothesized latent variables is Bell's theorem \cite{BELL76}, which is widely regarded as showing that such hidden variables, if they exist, must possess some highly counter-intuitive properties. Bell's theorem requires that we introduce new exogenous variables $S_A,S_B$ corresponding to the respective measurement settings of $A,B$. In this experiment the corresponding causal graph is assumed to be the common-cause scenario as shown in Fig. \ref{fig:Bell} and the \tbf{CMC2} (allowing for latent variables \tbf{L}) implies the constraint:
\eqn{
P(A,B,C,S_A,S_B)= \zum{\tbf{l}}{} P(A|S_A,C,\tbf{l})P(B|S_B,C,\tbf{l})P(S_A)P(S_B)P(C,\tbf{l}) \, .
}
This constraint can be proven to imply mathematical inequalities on the marginal distribution $P(A,B,C,S_A,S_B)$, which have been found to be violated in experiments using entangled states, in agreement with quantum theory, ruling out any reasonable explanation in terms of latent common causes.

A landmark paper by Wood and Spekkens \cite{WOOD} showed that Bell's theorem can be alternatively expressed as the impossibility of explaining quantum correlations using \tit{any} classical causal model, under the assumptions of \tbf{CMC2} and \tit{no fine-tuning}. This way of formulating Bell's theorem is very powerful. Since it refers to causal structure, it may be generalized to contextuality scenarios in which space-like separation is not important \cite{CAVNFT}. For the same reason, it applies even to latent variables that defy known physics by travelling faster than light or backwards in time. While this has led some authors to questioned whether the assumption of \tbf{NFT} is always reasonable (see eg. Ref. \cite{ALMADA} ), we cannot give it up in our framework because as we discussed in Sec. \ref{sec:physMarkov}, \tbf{NFT} is here taken as a fundamental assumption by which physical Markov conditions such as \tbf{CMC2} come to be established. Instead, our approach forces us to simply reject the classical Markov conditions \tbf{CMC} and \tbf{CMC2}, and replace them with something else that is better suited to quantum systems. In particular, in light of Bell's theorem, the quantum Markov conditions should not include \tbf{FCC}. 

\subsection{Quantum reference measurements and counterfactual inference \label{sec:born}} 

We are now in a position to tackle the first key question of causal modeling: what are the reference measurements that define a reference observational scheme for quantum systems? In Sec. \ref{sec:soff} it was determined that the quantum reference measurements are distinct from either non-disturbing measurements or interventions. In this section we will propose, as a matter of convention, a precise form for the quantum reference measurements that will provide us with particularly elegant mathematical expressions. Along the way, we unexpectedly make contact with an approach to quantum foundations known as QBism.

For simplicity, let us again take as our reference experiment the example of the `causal chain', in which three measurements $X,Y,Z$ are performed in succession (that is, in time-like separated space-time regions) and whose causal relations are represented by the causal graph $X \rightarrow Y \rightarrow Z$. (Note: we use $X,Y,Z$ here instead of $A,B,C$ to avoid confusing $C$ with the control variable). Here, $Y$ is the quantum measurement whose properties will be investigated. Now suppose that the detector or measuring apparatus that is responsible for measuring $Y$ in its designated space-time region is to be removed from the experiment, or deactivated, as indicated by the conditonal $C^Y=\brm{undo}$.

Recall that in the special case where $Z$ is non-disturbing relative to $\tbf{X}$, the appropriate inference rule is (by definition) that given by Eq. \eqref{eqn:nondisturb} in Sec. \ref{sec:passactclass}. For quantum reference measurements, however, a different rule is required. Returning to our example of the causal chain, we begin by asking what is the inference rule to obtain $P(X,Z|C^Y=\brm{undo})$. In fact, this rule has been worked out elsewhere in the literature, for quite different reasons: in the ``QBist" approach to quantum theory (eg. \cite{QBCOH,QPLEX} and references therein) it appears in the guise of the Born rule expressed probabilistically (with no direct reference to Hilbert space operators). Due to its importance in QBism it is there named the \tit{Urgleichung}. We now review how the rule is derived.

First note that if we could fully reconstruct the state $\rho(x)$ (which represents the input to the measurement $Y$ conditioned on $X=x$) using only the probabilities $P(Y|X)$, and also fully reconstruct the POVM elements $\{ E_{z} : z \in \trm{dom}Z \}$ from the probabilities $P(Z|Y)$, then the inference rule for obtaining the probabilities $P(X,Z|C^Y=\brm{undo})$ would just be the Born rule itself:
\eqn{ \label{eqn:Born}
P(X,Z|\brm{undo}) = \tr{\rho(X) E_{Z} } \, ,
}
since then the RHS could then be expanded into some function of the reference probabilities $P(Y|X),P(Z|Y)$. Evidently the full reconstruction of an arbitrary $\rho(X)$ from the probabilities $P(Y|X)$ is possible if and only if $Y$ is an \tit{informationally complete} (IC) instrument:\\

\tbf{ICM. Informationally complete instrument:} An instrument $\{ \mathcal{M}_Y \}$ represents an \tit{informationally complete instrument} if its elements can be decomposed as:
\eqn{
\mathcal{M}_y(\cdot) = \sqrt{F_{y} \vphantom{F^{\dagger}_{y}}} \, (\cdot) \, \sqrt{F^{\dagger}_{y}} \, , 
}
such that $\{ F_y : y \in \trm{dom}(Y) \}$ spans the space of linear operators on $\mathcal{H}_Y$ and $\zum{y}{} F_{y} = \mathbb{I}_{Y}$, where $\mathbb{I}_{Y}$ is the identity matrix in $\mathcal{H}_{Y}$. (Note that a random variable $Y$ can only be associated with an informationally complete instrument on a system if $Y$ has at least as many outcomes as the square of the system's dimension, $d^2_Y$, since otherwise there won't be enough $F_y$'s to span the space of linear operators on $\mathcal{H}_{Y}$).\\  

For the purposes of obtaining simple and elegant expressions, we choose $Y$ to be a \tit{symmetric} informationally complete instrument (called a \tit{SIC-instrument}): \\

\tbf{SIC. Symmetric informationally complete instrument:} An instrument $\{ \mathcal{M}_Y \}$ represents a \tit{SIC}-measurement if dom$(Y)=d^2_Y$ and
\eqn{
\mathcal{M}_y(\cdot) = \frac{1}{d_Y} \Pi_{y} \, (\cdot) \, \Pi_{y} \, , 
}
where $\{ \Pi_y : y \in \trm{dom}Y \}$ have the special property:
\eqn{
\tr{\Pi_y \Pi_{y'}}=\frac{d \, \delta(y,y')+1}{d+1} \, \, , \qquad \forall \, y,y' \in \trm{dom}Y \, ,
}
and $\{ \frac{1}{d_Y} \Pi_Y \} := \{ \frac{1}{d_Y} \Pi_y : y \in  \trm{dom}Y \}$ defines a POVM called a \tit{SIC-POVM}.\\

\tit{Remark:} The post-measurement state corresponding to the outcome $Y=y$ is equal to the pure state projector $\Pi_y$. This may be regarded as an `unsharp' generalization of the L\"{u}ders rule for updating the state after measurement \cite{BUSCH2}, as the $\Pi_y$ are necessarily not quite orthogonal. 

Since the elements of the SIC-POVM define a basis for the space of linear operators, we can expand $\rho(X)$ and the POVM $E_Z$ as:
\eqn{ \label{eqn:rhoE}
\rho(X) &=& \zum{y}{} \, \alpha_y(X) \, \frac{1}{d_Y} \Pi_y \nonumber \\
E_{Z} &=& \zum{y}{} \, \beta_y(Z) \, \frac{1}{d_Y} \Pi_y \, .
}
The coefficients $ \alpha_Y(X), \, \beta_Y(Z)$ are related to the measurement probabilities according to \cite{QBCOH}:
\eqn{ \label{eqn:coeffab}
\alpha_Y(X) &=& d_Y(d_Y+1)P(Y|X )-1 \nonumber \\  
\beta_Y(Z) &=& (d_Y+1)P(Z|Y) - \frac{1}{d_Y} \zum{y}{} \, P(Z|y) \, .
}
By substituting \eqref{eqn:rhoE},\eqref{eqn:coeffab} into the right hand side of the Born Rule \eqref{eqn:Born}, one can then establish the QBist re-formulation of the Born rule, called the \tit{Urgleichung} \cite{QBCOH}: 
\eqn{ \label{eqn:urg}
P(Z|X, C^Y=\brm{undo})= \zum{y}{} \, P(Z|y) \left[ (1+d_Y) P(y|X)-\frac{1}{d_Y} \right] \, .
}
(Note that the probabilities on the RHS of this equation refer to the reference measurement scheme; we have suppressed the conditioning on $C^Y=\oslash$). The \tit{Urgleichung} gives us $P(Z|X, C^Y=\brm{undo})$, but we require the full distribution $P(X,Z| C^Y=\brm{undo})$. Using elementary probability theory, we can decompose this as:
\eqn{ \label{eqn:preurg}
P(X,Z|C^Y=\brm{undo}) = P(Z|X,C^Y=\brm{undo})P(X|C^Y=\brm{undo}) \, .
}
To proceed, we can make use of the fact that un-measurements satisfy counterfactual no-signalling \tbf{CNS}. Hence the value of $C^Y$ (whether or not $Y$ is measured) should not affect $X$ (a causal ancestor of $Y$), i.e. 
\eqn{ \label{eqn:abminus}
P(X|C^Y=\brm{undo})=P(X|C^Y=\oslash) \, .
}
Substituting \eqref{eqn:abminus} and the Urgleichung \eqref{eqn:urg} into \eqref{eqn:preurg} we finally obtain the sought-after inference rule:
\eqn{ \label{eqn:causalurg}
P(X,Z|C^Y=\brm{undo}) &=& P(Z|X,C^Y=\brm{undo})P(X) \, \nonumber \\
&=& \zum{y}{} \, P(Z|y) \left[ (1+d_Y) P(y|X)-\frac{1}{d_Y} \right] P(X) \, .
}
Therefore, in the context of causal modeling, the Urgleichung gives us the foundation for an inference rule for un-measurements. Note that the rule depends on the causal structure. To see this explicitly, consider what would happen if the accompanying causal structure had instead been $G(X,Y,Z):=X \leftarrow Y \leftarrow Z$. In that case the condition \eqref{eqn:abminus} would not follow from \tbf{CNS}, since $X$ is now in the causal future of $Y$. Instead, \tbf{CNS} implies $P(Z|C^Y=\brm{undo})=P(Z|C^Y=\oslash)$, and we obtain a different form of the rule: 
\eqn{ \label{eqn:causalurgbackwards}
P(X,Z|C^Y=\brm{undo}) &=& P(X|Z,C^Y=\brm{undo})P(Z|C^Y=\brm{undo}) \, \nonumber \\
&=& P(X|Z,C^Y=\brm{undo})P(Z) \, \nonumber \\
&=& \zum{y}{} \, P(X|y) \left[ (1+d_Y) P(y|Z)-\frac{1}{d_Y} \right] P(Z) \, ,
}
which in general is not equivalent to the constraint \eqref{eqn:causalurg} (more precisely, neither of \eqref{eqn:causalurg},\eqref{eqn:causalurgbackwards} implies the other, but nor do they exclude each other, i.e. both can be satisfied simultaneously). What is important is that the causal structure is sufficient to determine the particular form of the inference rule, and hence the principle of causal sufficiency \tbf{CS} is maintained for un-measurements on quantum systems (cf Sec. \ref{sec:experiments} ). Of course, this \tit{inference rule} still needs to be generalized to more interesting causal structures; this will be done in Sec. \ref{sec:qinterf}. We conclude this section with our final definition of the quantum reference measurements:\\

\tbf{QRM. Quantum reference measurements:} The \tit{quantum reference measurements} are quantum instruments that are \tit{informationally complete} and whose corresponding channels are of the Holevo form \eqref{eqn:Holevo}, i.e. they are entanglement-breaking. By convention, we take them to be SIC-instruments.\\

Accordingly, a \tit{reference experiment} on a quantum system is an experiment (as per Sec. \ref{sec:experiments}) in which the non-exogenous variables represent SIC-instruments. The terminal nodes (i.e. those that have no effects in the causal graph) may be represented by SIC-POVMS, while the exogenous variables (those without causes) will be assumed to have the maximally mixed state as input. The justification for this convention will be given in the next section.

\tit{Remark:} It is currently not known whether SIC-POVMs actually exist in all Hilbert space dimensions, but that does not present a problem to our program, which requires only the informational completeness of the measurements and not necessarily their symmetry or minimality; the latter are adopted on purely aesthetic grounds. Nevertheless, it remains an intriguing idea to ask what theory results from elevating the Urgleichung to the level of a postulate that holds prior to the existence of any Hilbert space representation; the QBists explore this idea in Ref.\cite{QPLEX}.

\subsection{Quantum Markov Conditions \label{sec:qmc}} 

Now that the essential details of a quantum \tit{reference experiment} have been identified, we turn to the second key problem, namely, that of finding a set of \tit{physical Markov conditions} for quantum systems based upon how they would behave in the reference experiment. We begin by considering the general properties of quantum systems for the three special cases considered in Sec. \ref{sec:cmc}: common causes, causal chains, and common effects. We will discover that there is an opportunity for the physical Markov conditions of quantum systems to be \tit{causally symmetric}, that is, invariant under a reversal of the directions of all arrows in the causal graph. This is because, besides the rejection of \tbf{FCC}, quantum systems also satify a condition \tbf{BK*} that is equivalent to the causal inverse of Berkson's rule. To achieve full symmetry, we will further impose the restriction as a matter of convention that the marginal probabilities of the exogenous nodes are uniformly distributed over their outcomes (equivalently, that the inputs to the exogenous nodes are maximally mixed states) and that these are preserved by the dynamics. This constraint enforces an additional physical Markov condition that we call \tbf{RP*}, which is the causal inverse of Reichenbach's Principle, and whose inclusion makes the whole set of quantum Markov conditions causally symmetric. In order to extend these conditions to arbitrary causal structures, we postulate a causally symmetric graphical criterion based on them, which we call the Quantum Markov Condition (QMC). 

To begin with, we will argue that quantum systems continue to satisfy the physical Markov conditions \tbf{SSO}, \tbf{RP} and \tbf{BK} which also hold classically. 

The argument for retaining \tbf{SSO} has already been given in Sec. \ref{sec:soff} for the case of a simple causal chain. We only need to show that \tbf{SSO} extends also to the more general `multi-chain' case shown in Fig. \ref{fig:threedags} (b) of Sec. \ref{sec:cmc}. This can be done by noting that the tensor product of a set of quantum instruments is again a quantum instrument, namely $\{ \mathcal{M}_{\vec{D}} \} := \{\mathcal{M}_{D_1} \otimes \dots \otimes \mathcal{M}_{D_N} \}$. The principle \tbf{SSO} will be upheld in this scenario if it is upheld for the total instrument $\{ \mathcal{M}_{\vec{D}} \}$ applied to an arbitrary input state $\rho_{X_1}$ defined on the tensor product Hilbert space of the individual measurements $\mathcal{H}:= \mathcal{H}_{D_1} \otimes \dots \otimes \mathcal{H}_{D_N}$. To see that it is upheld, it is enough to note that the tensor product of a set of entanglement breaking channels is also entanglement breaking, and so the total instrument is of the Holevo form \eqref{eqn:Holevo} and we can apply the same reasoning as in Sec. \ref{sec:soff} to conclude that it respects \tbf{SSO}, i.e. that knowledge of all the outcomes $D_1,\dots,D_N$ renders $X_1$ and $X_2$ uncorrelated: $P(X_1 X_2|D_1,\dots,D_N)=P(X_1|D_1,\dots,D_N)P(X_2|D_1,\dots,D_N)$. (It should be noted that the total instrument $\{ \mathcal{M}_{\vec{D}} \}$ is also informationally-complete, a fact that will become important in Sec. \ref{sec:qinterf}).

Turning now to \tbf{RP}, let us recall its definition: if neither of $X_1,X_2$ is a cause of the other and they have no shared ancestors, then they are statistically independent: $P(X_1 X_2)=P(X_1)P(X_2)$. To see how this can be arranged to hold for quantum systems in a natural manner, let $\tbf{E}_1$ be the set of ancestors of $X_1$ that are exogenous, and similarly let $\tbf{E}_2$ be the exogenous ancestors of $X_2$. By assumption, $\tbf{E}_1,\tbf{E}_2$ have no members in common, and are independent $P(\tbf{E}_1,\tbf{E}_2)=P(\tbf{E}_1)P(\tbf{E}_2)$ (condition (iv), Sec. \ref{sec:experiments} ). These measurements can be regarded as preparing a quantum state with the form $\rho_{\tbf{E}_1} \otimes \rho_{\tbf{E}_2}$ on a Hilbert space $\mathcal{H}_{\tbf{E}_1} \otimes \mathcal{H}_{\tbf{E}_2}$, which is mapped by some channel $\mathcal{T}$ to the input spaces $\mathcal{H}_{X_1} \otimes \mathcal{H}_{X_2}$ of the measurements $X_1,X_2$. Since by assumption the causal structure contains no causal pathways from $\tbf{E}_1$ to $X_2$ or from $\tbf{E}_2$ to $X_1$, it is reasonable to posit that the channel $\mathcal{T}$ does not generate correlations, that is, to postulate that $\mathcal{T} = \mathcal{T}_1 \otimes \mathcal{T}_2$ where $\mathcal{T}_1: \mathcal{H}_{\tbf{E}_1} \mapsto \mathcal{H}_{X_1}$ and $\mathcal{T}_2: \mathcal{H}_{\tbf{E}_2} \mapsto \mathcal{H}_{X_2}$. This shows that we can enforce \tbf{RP} for quantum systems without difficulty. 

\tit{Remark:} The fact that \tbf{RP} can be retained despite the loss of \tbf{FCC} is one of the main motivations for thinking that quantum correlations could be explained by a suitably defined causal model without fine-tuning. For example, \tbf{RP} is identified and claimed to hold for quantum systems in many of the early works on the topic \cite{FRITZ,PIEBRUK,CAVLAL}.

Next we recall the definition of \tbf{BK}: If neither of $X_1,X_2$ is a cause of the other and they have no shared ancestors, and $B$ is a common descendant of them, then they are typically correlated conditional on $B$. This principle holds for quantum systems for essentially the same reasons as it did the classical case: conditioning on common effects of independent variables typically renders them correlated. The only new feature that appears in the quantum case is that the induced `spurious correlations' can be stronger than classical, i.e. they can exhibit entanglement. This effect has already been studied under the name of the ``quantum Berkson effect" in Refs. \cite{SPEKBERK,RIEDPHD}. 

Beyond the above three conditions, there are also some additional physical Markov conditions that are special to quantum systems. The case of \tbf{FCC} has already been partially dealt with in Sec. \ref{sec:comm}, where we pointed out that entangled quantum systems exhibit counterexamples to it. However, this leaves open the possibility that \tbf{FCC} might nevertheless hold for any `typical' quantum system, in which case we might have grounds to retain it as a physical Markov condition, albeit in weaker form. But does it typically hold?

To make this more precise, consider the case of a single common cause, $X_1 \leftarrow C \rightarrow X_2$. Conditioning on $C=c$ effectively results in the preparation of an outgoing post-measurement state $\rho_c$ on the Hilbert space $\mathcal{H}_{C}$. The causal arrows mean that manipulations of this state can signal to the measurements at $X_1$ and $X_2$, which implies that the system's dynamics can be expressed as a quantum channel $\mathcal{T}: \mathcal{H}_{C} \mapsto \mathcal{H}_{X_1} \otimes \mathcal{H}_{X_2}$ which conveys information about $\rho_C$ to each of $X_1$ and $X_2$, but is otherwise unconstrained. The result is that the conditioned probabilities can be expressed as:\\
\eqn{ \label{eqn:antiberk}
P(X_1,X_2|C=c) = \frac{1}{d_1d_2} \, \tr{ \Pi_{x_1} \otimes \Pi_{x_2} \cdot \rho'_c} \nonumber \, ,
}
where $\rho'_c := \mathcal{T}(\rho_c)$ may be assumed to be an arbitrary state on $\mathcal{H}_{X_1} \otimes \mathcal{H}_{X_2}$. For any reasonable measure on the set of density matrices, the ones that do not exhibit any correlations between the $\mathcal{H}_{X_1}$ and $\mathcal{H}_{X_2}$ subspaces will be a set of measure zero. 

\tit{Remark:} It is tempting to attribute the typicality of correlations in this case to entanglement. However, depending on the measure one uses, the correlations will generally be separable and will not necessarily exhibit entanglement even in most cases \cite{ZYCZ}). Notwithstanding this observation, the fact that the entangled density operators have full measure in the space of all density operators is enough to rule out any hope of rescuing \tbf{FCC} by appealing to latent common causes and typicality arguments. 

Thus we are led not only to reject \tbf{FCC} but to introduce a \tit{new} physical Markov condition that asserts the contrary, namely the typicality of correlations conditional on common causes:\\

\noindent \tbf{BK*. Non-factorization on common causes (a.k.a. the causal inverse of Berkson's rule):}\\
Suppose neither of $X_1,X_2$ is a cause of the other and they have no shared descendants, and suppose $C$ is a common ancestor of $X_1,X_2$; then one generally expects them to be correlated conditional on $C$, i.e. that $P(X_1X_2|C) \neq P(X_1|C)P(X_2|C)$. (This condition is related to \tbf{BK} by switching the roles of `ancestors' and `descendants' in its definition, which is why we have labelled it the causal inverse of \tbf{BK}). \\

\tit{Remark:} the restriction to variables $X_1,X_2$ that have no shared descendants might seem arbitrary here, but it is included so as to make \tbf{BK*} perfectly symmetric with \tbf{BK}. To remove this clause would be to assert something extra, namely, that variables are not correlated by the mere fact of interaction in their common future. The latter assertion is in fact a corollary of the principle \tbf{RP}, so to assert it here would be redundant.

\tbf{BK*} marks an interesting departure from classical stochastic systems. In the \tbf{CMC} there was a marked asymmetry in the physical Markov conditions, due to the simultaneous presence of \tbf{FCC} and \tbf{BK}, which together asserted that variables are correlated conditioned on common effects but not on common causes. What is remarkable is that not only do quantum systems reject \tbf{FCC}, but they actually replace it with \tbf{BK*}, which as we have noted perfectly restores the symmetry with \tbf{BK}. This points to the intriguing possibility that the quantum Markov conditions for quantum systems might have the property of being fully causally symmetric. 

In order to achieve this, there is still another asymmetry that needs to be dealt with, present in the opposition between \tbf{RP} and \tbf{PE}. Note that these principles refer to what is implied when there are common effects (i.e. `future interactions') or common causes (`past interactions'), respectively, whose outcomes are \tit{not} conditioned upon. The point at stake in these principles is whether the \tit{mere occurrence} of a common future or common past measurement can imply correlations or independence between two variables. As discussed in Sec. \ref{sec:cmc} the asymmetric pairing of these two principles for macroscopic classical systems is commonplace, where it is summarized by the principle that systems are uncorrelated before interaction and typically correlated afterwards (or as Price put it, `innocence precedes experience' \cite{PRICEBOOK} ). 

Whereas Price prefers to restore symmetry by rejecting \tbf{RP} for quantum systems, thereby asserting that quantum states may be correlated due to the mere existence of a future interaction, we are inclined, given the marked importance of \tbf{RP} in quantum causal modeling, to take the opposite route and restore symmetry by upholding \tbf{RP} and rejecting \tbf{PE}. This leads us to the counter-intuitive proposition that quantum systems can remain independent of one another even after interaction. On further exploration, however, we find that this idea is more sensible than it first appears.

While it is true that in the laboratory it is commonplace to see independent systems becoming correlated after interaction, this typically occurs only when the systems have been carefully prepared \tit{in known initial states}. In our framework, that means only when we are conditioning on the values of the exogenous variables. But in that case, as we have just pointed out above, \tbf{BK*} would lead us to expect correlations. The point is that a rejection of \tbf{PE} only mandates independence when the common past is \tit{not conditioned upon}, and this represents a situation that is rarely encountered in practice. In any normal laboratory setting, we are not ignorant of the values of the exogenous variables. For instance, one usually does not begin an optics experiment until one has verified that one's sources are producing photons. Conditional that these are working, one then gathers statistics, and one then has the choice whether to post-select on the functioning of the detectors or keep all of the statistics including the cases where the \tit{detectors} failed to detect photons. If we now wish to imagine a scenario in which the exogenous variables are \tit{not} conditioned upon, we would have to gather statistics for the whole experiment even in cases where the photon \tit{sources} failed to work. This runs quite counter to intuition and efficiency: why would anybody go ahead with an experiment in which they knew their sources were not working? It would be beyond the scope of this work to account for this asymmetry in the way we conduct experiments (Price's book \cite{PRICEBOOK} does a good job); the main point is that \tit{if we were} to take statistics even when our sources were not working, thus \tit{not conditioning} on the exogenous variables in the system, we might well find that variables with a common source remain uncorrelated (or, in a phrase, `garbage in, garbage out'). The rejection of \tbf{PE}, then, is not so unnatural as it first appears. We can formalize this idea by postulating the causal inverse of \tbf{RP}, namely:\\

\noindent \tbf{RP*. Causal inverse of Reichenbach's Principle:}\\
If neither of $X_1,X_2$ is a cause of the other and they have no shared descendants, then they are statistically independent: $P(X_1 X_2)=P(X_1)P(X_2)$.\\

The main implication of this would be that $X_1,X_2$ should be statistically independent of each other even if they possess one or more common ancestors (not conditioned upon). This is the principle we would expect to hold in an experiment in which we have contrived to be `ignorant' about the starting conditions. To make this more rigorous, consider again the simple case of a single common cause, $X_1 \leftarrow C \rightarrow X_2$. Since now we are not conditioning on $C$, its values in \eqref{eqn:antiberk} must be summed over, leading to the probabilities:
\eqn{ \label{eqn:antiRP}
P(X_1,X_2) &=& \zum{c}{} \, P(X_1,X_2|C=c)P(C=c)  \nonumber \\
&=& \zum{c}{} \frac{1}{d_1d_2} \, \tr{ \Pi_{x_1} \otimes \Pi_{x_2} \cdot \rho_c} \, \frac{1}{d_C} \, \tr{\Pi_c \, \rho_{\trm{prep}}}  \, ,
}
where $P(C=c) = \tr{\Pi_c \, \rho_{\trm{prep}}}$ is the probability of obtaining $C=c$ when doing a SIC-instrument on the initial state $\rho_{\trm{prep}}$, and where $\rho_c := \mathcal{T}(\Pi_c)$ is the input state to the measurements $X_1,X_2$ conditioned on $C=c$, obtained by passing the post-measurement state of $C$ through some channel $\mathcal{T}$. We now introduce the notion of an \tit{unbiased} quantum channel (borrowing the terminology of Ref. \cite{COSTA17}): \\

\tbf{UB. Unbiased quantum channel:} A quantum channel $\mathcal{T}$ is \tit{unbiased} (or `maximally-mixed-state-preserving') iff it preserves the maximally mixed state, i.e. 
\eqn{
\mathcal{T}(\frac{1}{d_{\trm{in}}} \mathbb{I}_{\trm{in}})=\frac{1}{d_{\trm{out}}} \mathbb{I}_{\trm{out}} \, .
}
Note that when $d_{\trm{in}}=d_{\trm{out}}$ this reduces to the definition of a \tit{unital} (identity-preserving) channel.\\

We can now make the following observation: if $\mathcal{T}$ is unbiased and we restrict attention to quantum systems initially prepared in the maximally mixed state $\rho_{\trm{prep}} = \frac{1}{d_C}\mathbb{I}$, then
\eqn{ \label{eqn:antiRP2}
P(X_1,X_2) &=& \zum{c}{} \frac{1}{d_1d_2} \, \tr{ \Pi_{x_1} \otimes \Pi_{x_2} \cdot \mathcal{T}(\Pi_c)} \, \frac{1}{d^2_C} \, \nonumber \\
&=&  \frac{1}{d_1d_2} \, \tr{ \Pi_{x_1} \otimes \Pi_{x_2} \cdot \mathcal{T}(\frac{1}{d_C} \mathbb{I})} \, \nonumber \\
&=&  \frac{1}{d_1d_2} \, \tr{ \Pi_{x_1} \otimes \Pi_{x_2} \cdot \frac{1}{d_1d_2} \mathbb{I} } \,  \, \nonumber \\
&=& \frac{1}{d^2_1d^2_2} = P(X_1)P(X_2)\, ,
}
and hence \tbf{RP*} can be satisfied. We therefore see that there exists a special sub-class of experiments in which (i) systems are prepared in the maximally mixed state and (ii) evolve only through \tit{unbiased} quantum channels, for which quantum systems satisfy the condition \tbf{RP*}. Within this sub-class, we can combine \tbf{RP} and \tbf{RP*} into the following simple condition:\\

\noindent \tbf{SRP.} Symmetric Reichenbach Principle:\\
If neither of $X_1,X_2$ is a cause of the other, then they are statistically independent: $P(X_1 X_2)=P(X_1)P(X_2)$. \\ 

It is can be checked by inspection of the definitions that the set of physical Markov conditions for this sub-class, $\{ \trm{\tbf{SSO}, \tbf{BK}, \tbf{BK*}, \tbf{SRP}} \}$ is invariant under switching of the direction of causal arrows. We will refer to quantum systems observed under these special conditions as the class of \tit{causally reversible quantum systems}.

The restriction to maximally mixed exogenous inputs and unbiased processes has an important consequence, which is that if one does not condition on the \tit{ancestors} of a variable, then the other variables do not depend on whether it is un-measured. Formally this can be expressed as:\\

\tbf{IUM. Indifference to un-measurements:} If one doesn't condition on the ancestors of $Z$, then measuring or un-measuring $Z$ cannot affect the non-ancestors of $Z$. Formally, let $C^Z=\{ \oslash, \brm{undo} \}$ toggle between measuring and un-measuring $Z$ in a system whose causal relations are described by a DAG denoted $G(\tbf{A} \tbf{D} \tbf{R} Z)$, where $\tbf{A}$ are the causal ancestors of $Z$, $\tbf{D}$ are the descendants of $Z$, and $\tbf{R}$ are the remainder. Then:
\eqn{ \label{eqn:ium}
P(\tbf{D} \, \tbf{R}|\,C^Z=\brm{undo})=P(\tbf{D} \, \tbf{R}|\,C^Z=\oslash) \, .
}

The justification is intuitive: when $Z$'s ancestors are not conditioned on, the input to $Z$ is a maximally mixed state uncorrelated with any of its non-descendants $\tbf{R}$. Since measuring $Z$ and ignoring its outcome is the same as applying an unbiased channel from its input to its output, it is equivalent to an unbiased channel from its input to the inputs of its children. Hence, regardless of whether $Z$ is performed and its outcome ignored or not measured at all, the inputs to the children of $Z$ are the same: a maximally mixed state uncorrelated to the inputs to $\tbf{R}$. The iteration of this argument to each child of $Z$ then shows that the inputs to $Z$'s grand-children, and great-grand-children, etc, are similarly unaffected, hence all of $\tbf{D}$. This establishes \eqref{eqn:ium}.\\

It is interesting to note that the classical stochastic systems cannot so easily be made symmetric by the rejection of \tbf{PE} because they still suffer from the asymmetry between \tbf{FCC} and \tbf{BK}. To break this asymmetry, one must reject one or the other principle. It would be interesting to investigate whether this route would lead to new interesting classes of causally reversible classical systems besides the most obvious case of the deterministic classical systems -- we comment on this further in Sec. \ref{sec:discuss}.

\tit{Remark:} In principle, it is always possible to simulate any quantum phenomenon using a system that is causally reversible (given sufficient extra resources). For instance, preparation of an arbitrary pure state can be simulated by post-selecting on the outcome of a suitable SIC-instrument that has the desired pure state as one of its elements. An arbitrary channel can then be simulated by coupling a system to a suitably prepared ancilla and post-selecting on the outcomes of a SIC-instrument on the ancilla. Since we lose no fundamental generality in insisting upon the property of causal reversibility, we will continue restrict our attention to this class of systems from here onwards. 

We are now ready to obtain a general Quantum Markov Condition starting from a graphical interpretation of the physical Markov conditions \tbf{SSO}, \tbf{BK}, \tbf{BK*}, \tbf{SRP}. To this end, we extrapolate that the physical Markov conditions for quantum systems with \tit{arbitrary} causal structure are given by the following graphical condition (see Appendix \ref{app:graphrules} ):\\

\noindent \tbf{QMC.} Quantum Markov Condition (graphical version):\\
Let $\tbf{U}$,$\tbf{V}$,$\tbf{W}$ be disjoint subsets of variables in a DAG $G(\tbf{X})$. A distribution $P(\tbf{X})$ is said to satisfy the Quantum Markov Condition relative to $G(\tbf{X})$ iff $P(\tbf{U}\tbf{V}|\tbf{W})=P(\tbf{U}|\tbf{W})P(\tbf{V}|\tbf{W})$ holds whenever every path between $\tbf{U}$ and $\tbf{V}$ is blocked by $\tbf{W}$. A path between two variables is said to be `blocked' by the set $\tbf{W}$ iff at least one of the following conditions holds:\\
\tbf{g-SSO:} There is a chain $A \rightarrow C \rightarrow B$ along the path whose middle member $C$ is in $\tbf{W}$;\\
\tbf{g-BK:} There is a collider $A \rightarrow C \leftarrow B$ on the path where $C$ is \tit{not} in $\tbf{W}$ and has no descendants in $\tbf{W}$.\\
\tbf{g-BK*:} There is a fork $A \leftarrow C \rightarrow B$ on the path where $C$ is \tit{not} in $\tbf{W}$ and has no ancestors in $\tbf{W}$.\\

\tit{Remark:} The principle \tbf{SRP} is implicit in both graphical conditions \tbf{g-BK*} and \tbf{g-BK}. To see this, note that if \tbf{SRP} were false, then the graphical rules \tbf{g-BK*} and \tbf{g-BK} would be insufficient to indicate statistical independence; for then it would be possible to have variables $A,B$ such that neither is a cause of the other and with all paths between them blocked via \tbf{g-BK*} and \tbf{g-BK}, yet where they are still correlated. What prevents correlations in this case is precisely \tbf{SRP}. In more rigorous langage, the conditions \tbf{BK}, \tbf{BK*} only suggest that the graphical rules \tbf{g-BK*} and \tbf{g-BK} are \tit{neccessary} criteria for the path to be blocked, whereas \tbf{SRP} elevates them to \tit{sufficient} criteria.

Our methodology of postulating \tbf{QMC} first as a graphical criterion has the advantage that it immediately supplies a graphical algorithm, called a \tit{graph-separation criterion}, for efficiently determining by inspection of a DAG whether two subsets of variables are independent conditional on a third subset. It would be interesting to compare the present criterion to others that have been proposed in the literature, particularly those in Refs. \cite{HLP,PIEBRUK}. It is unclear whether the \tbf{QMC} can be expressed as a single factorization condition similar to Eq. \eqref{eqn:cmc}; this is left to future work.

For classes of quantum systems in which latent variables are suspected, i.e. where the observed behaviour $P(\tbf{X})$ is suspected to be only a marginal of an extended system with causal structure $G(\tbf{X},\tbf{L})$, it is natural to posit an extension of the \tbf{QMC} to this broader class in an analogous way to how we obtained the classical \tbf{CMC2}:\\

\tbf{QMC2.} Quantum Markov Condition (with latent variables):\\
There exists an extended distribution $P(\tbf{X},\tbf{L})$, such that $P(\tbf{X},\tbf{L})$ satisfies the \tbf{QMC} for the causal structure $G(\tbf{X},\tbf{L})$, and $P(\tbf{X})$ is obtained from $P(\tbf{X},\tbf{L})$ by marginalizing over the latent variables. \\

We can now finally define a \tit{quantum causal model}:\\

\tbf{Quantum Causal Model:}\\
A Quantum Causal Model consists of a pair $\{P(\tbf{X}), G(\tbf{X}) \}$ where $P(\tbf{X})$ satisfies the \tbf{QMC} and \tit{no fine-tuning} for the DAG $G(\tbf{X})$.\\

\subsection{Inference rules for quantum causal models \label{sec:qinterf}}

In this section we take up the third key question regarding the counterfactual inference rules for quantum causal models. We focus first on the case of interventions, and then deal with un-measurements. We will see that the very possibility of an inference rule for interventions is not guaranteed, and that the fundamental postulate of causal sufficiency \tbf{CS} cannot be upheld for arbitrary causal structures. To accomodate this, we impose the restriction that the causal structure must be `layered', which ultimately enables us to derive the necessary inference rules and uphold \tbf{CS}. 

\subsubsection{Interventions on quantum systems \label{sec:qinterv}}

An intervention on a variable $W$ in a quantum system may be usefully represented by associating a quantum instrument to $W$ that measures the local subsystem at the input, obtains an outcome $U=u$, and then re-prepares an arbitrary new state $\sigma_w$ with probability $P'(W=w)$ at the output, independently of the value of $U$ (thus breaking the causal connection between $W$ and its parents as required). Formally, we define:\\

\tbf{Quantum intervention:} An intervention on $W$ in a quantum system is associated with an instrument $\{ \mathcal{M}_{u w} \}$ whose elements have the form:
\eqn{ \label{eqn:simpleinterv}
\mathcal{M}_{u w}(\rho_{\trm{in}}) := \tr{\rho_{\trm{in}} F_u} P'(w) \, \sigma_w \, ,
} 

where $\{  F_u : u \in \trm{dom}(U) \}$ is an arbitrary POVM, $P'(w)$ an arbitrary probability, and $\sigma_w$ an arbitrary state, which together define the intervention. For simplicity, we may sometimes consider the case where $F_u = \frac{1}{d_W} \Pi_u$, $P'(w)=\frac{1}{d^2_W}$, and $\sigma_w=\Pi_w$, which we call a \tit{SIC-intervention}, because it represents doing a SIC-POVM on the input and then re-preparing a SIC state uniformly at random at the output. 

\tit{Remark:} In this most general definition, an intervention is associated with \tit{two} variables: the intervened-upon variable $W$ and a \tit{new} variable $U$ having the same domain as $W$ but treated as an independent variable. The reason for this is not due to any special feature of quantum theory. It arises naturally in the quantum setting due the usage of quantum instruments to formally represent measurements, because these make it explicit that measurements have both an `input' and an `output'. Classically we could do the same thing by formally including, as part of the intervention, an extra variable $U$ that represents the outcome of a measurement on the former parents of $W$. One then recovers the usual classical formalism by summing over the values of $U$. (One such example is the ``split-node" classical causal models defined in Ref.\cite{BARRETTQCM}).

This remark suggests a special case of quantum interventions that look more similar to the usual classical treatment of interventions, in which the `measurement of the parents' $U$ is simply discarded. We will call these \tit{simple interventions}:\\

\tbf{Quantum simple intervention:} A \tit{simple intervention} on $W$ in a quantum system is associated with an instrument $\{ \mathcal{M}_W \}$ whose elements have the form:
\eqn{ \label{eqn:discardinterv}
\mathcal{M}_w(\rho_{\trm{in}}) := \tr{\rho_{\trm{in}}} \, P'(w) \, \sigma_w \, .
} 
Similarly, we can define the special class of \tit{simple SIC interventions} by setting $P'(w)=\frac{1}{d^2_W}$, and $\sigma_w=\Pi_w$. In what follows, unless stated otherwise, we will restrict attention to quantum simple interventions and continue to neglect the variable $U$. 

To model an intervention as a counterfactual we introduce an associated control variable $C^{W}$ with possible values $\{ \oslash,\brm{do} \}$, such that $P(\tbf{X},W|C^W=\oslash)=P(\tbf{X},W)$ is the behaviour in the reference experiment and $P(\tbf{X},W|C^W=\brm{do})$ represents the probabilities when an intervention is performed on $W$. In the special case of interventions that specify a particular value of $W$, we can choose $P'(w)=\delta(w,w')$ and write $C^W=\brm{do}(W=w')$ (cf Sec. \ref{sec:ccms} ).\\

As in the classical case, quantum interventions are manipulations, so we adopt the same rule for updating the causal structure when a quantum intervention is performed on $W$, namely, that the incoming arrows to $W$ are deleted (For general interventions, we may include $U$ as a terminal node that is a child of all the former parents of $W$.) Since quantum causal models satisfy the graphical rules \tbf{g-SSO} and \tbf{g-BK}, it is reasonable to also assume that quantum interventions satisfy \tbf{NPE2} (cf the justification for \tbf{NPE2} in Sec. \ref{sec:passactclass} ). So far, there is essentially no difference between the quantum and classical definitions of an intervention. 

The difference arises in how we obtain the inference rule for interventions. Classically we obtained the rule \eqref{eqn:intervaxiom} by demanding that the new probabilities $P(\tbf{X}|C^W=\brm{do})$ should satisfy the \tbf{CMC} for the new causal graph. In the quantum case, evidently, we must replace the \tbf{CMC} with the \tbf{QMC}. However, it is then not clear whether this constraint is sufficient to guarantee the existence of an inference rule. In fact, we show in the next section that an inference rule for interventions on quantum systems does not exist for arbitrary causal structures.

\subsubsection{Impossibility of a general inference rule for quantum interventions  \label{sec:qintervimposs}}

\begin{figure}[!htb]
\centering\includegraphics[width=0.6\linewidth]{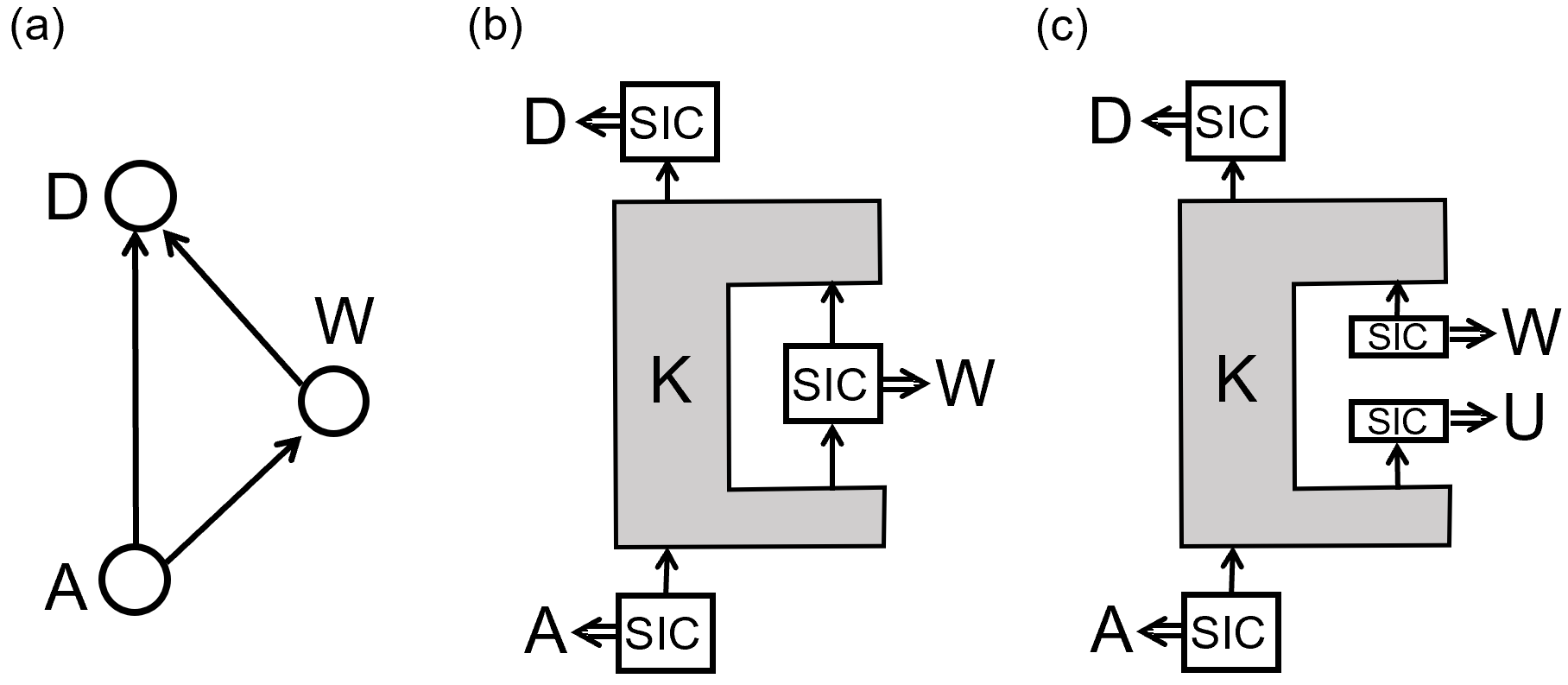}
\caption{(a) a causal diagram and (b) its most general possible circuit realization. Single arrows represent quantum systems and double arrows represent classical data. The boxes SIC-instruments. (c) is the circuit under an intervention of $W$. As explained in the text, the probabilities in this case are not uniquely specified by those in the pre-intervention circuit.}
\label{fig:qintervprob}
\end{figure}

Consider a system with the causal structure shown in Fig. \ref{fig:qintervprob} (a). The circuit diagram shown in Fig. \ref{fig:qintervprob} (b) represents the most general possible realization of this system. It is convenient to represent this circuit as a \tit{quantum comb} \cite{COMB1,COMB2,COSHRAP,POLLOCKPRA}. To do so, we introduce the Choi-Jamio\l kowski (CJ) matrix representation $M \in \mathcal{H}^{X_I} \otimes \mathcal{H}^{X_O}$ of a completely positive linear map $\mathcal{M}:\mathcal{H}^{X_I} \mapsto \mathcal{H}^{X_O}$ as:
\eqn{ \label{eqn:genbornce}
M := \zum{i=1,j=1}{d_{X_I}} \, \ket{i}\bra{j} \otimes \, \left( \mathcal{M}(\ket{j}\bra{i}) \right)^T
}
for some conventionally chosen orthonormal basis $\{ \ket{i} : i=1,2,\dots,d_{X_I} \}$ of $\mathcal{H}^{X_I}$, where $T$ denotes the transpose in that basis. Then the probabilities in the reference experiment for this circuit may be expressed as:

\eqn{  \label{eqn:genbornce2}
P(D,W,A) = \tr{ \Pi_{A} \otimes M^{W_I W_O}_{W} \otimes \frac{1}{d_D} \Pi_{D} \cdot K^{A W_I W_O D} } 
}
where $\Pi_{A},\Pi_{D}$ are the SIC-projectors on $\mathcal{H}^{A},\mathcal{H}^{D}$ corresponding to the outcomes of $A,D$ respectively, $M^{W_I W_O}_W$ is the CJ matrix representation of the SIC-instrument $\mathcal{M}_W:\mathcal{H}^{W_I} \mapsto \mathcal{H}^{W_O}$, and $K^{A W_I W_O D} \in \mathcal{H}^{A} \otimes \mathcal{H}^{W_I} \otimes \mathcal{H}^{W_O} \otimes \mathcal{H}^{D}$ is a positive linear matrix (called a `quantum comb') that represents the circuit fragment contained in the dashed lines in Fig. \ref{fig:qintervprob} (b). Since $M^{W_I W_O}_{W}$ is a SIC-instrument, its CJ matrix has the simple form $M^{W_I W_O}_W = \Pi_W \otimes ( \frac{1}{d_W}\Pi_W)$, which allows us to simplify \eqref{eqn:genbornce2} to:
\eqn{  \label{eqn:genbornce3}
P(D,W,A) = \frac{1}{d_W d_D} \tr{ \Pi_{A} \otimes \Pi_W \otimes \Pi_W \otimes \Pi_{D} \cdot K^{A W_I W_O D} } \, .
}

When $W$ is intervened upon, the circuit reduces to that of Fig. \ref{fig:qintervprob} (c), and the probabilities are obtained by replacing $M^{W_I W_O}_W$ in \eqref{eqn:genbornce} with the CJ matrix for a quantum intervention $\mathcal{M}_{UW}$ as defined in \eqref{eqn:simpleinterv}. Choosing the intervention on $W$ to be a SIC-intervention, the CJ matrix of $\mathcal{M}_{UW}$ is given by $M^{W_I W_O}_{U W} = \Pi_U \otimes ( \frac{1}{d_W}\Pi_W)$. Hence the post-intervention probabilities are:
\eqn{
P(D,U,W,A|C^W=\brm{do}) = \frac{1}{d_Wd_D} \tr{ \Pi_{A} \otimes \Pi_{U} \otimes \Pi_{W} \otimes \Pi_{D} \cdot K^{A W_I W_O D} } \, .
}
Note that $\{ \Pi_{A} \otimes \Pi_{U} \otimes \Pi_{W} \otimes \Pi_{D} \}$ is a set of $d^2_{A}d^4_{W}d^2_{D}$ linearly independent operators that span the space of linear operators on $\mathcal{H}^{A} \otimes \mathcal{H}^{W_I} \otimes \mathcal{H}^{W_O} \otimes \mathcal{H}^{D}$. Hence a specification of $P(D,W,A|C^W=\brm{do})$ is equivalent to a full specification of the comb representing the circuit fragment, $K^{A W_I W_O D}$. In order for the reference probabilities $P(D,W,A)$ to allow inference of $P(D,W,A|C^W=\brm{do})$, they must therefore be able to re-construct an arbitrary $K^{A W_I W_O D}$. However, from Eq. \ref{eqn:genbornce3} we see that $\{ \Pi_{A} \otimes \Pi_W \otimes \Pi_W \otimes \Pi_{D} \}$ are a set of only $d^2_{A}d^2_{W}d^2_{D}$ linearly independent operators, which is insufficient to span the full operator space, thus making it impossible in general to reconstruct an arbitrary $K^{A W_I W_O D}$ from $P(D,W,A)$.

We conclude that the causal model $\{P(A,D,W),G(A,D,W) \}$ is not sufficient to deduce what would happen under an arbitrary intervention; more information is needed. This violates the postulate of causal sufficiency \tbf{CS}, which is a core axiom of our framework. 

\tit{Remark:} It can be shown that this problem is not alleviated if we restrict ourselves to quantum simple interventions, and we conjecture that this remains true if one additionally restricts all processes to be unbiased. Instead, we argue that \tbf{CS} can be upheld by imposing constraints on the set of causal structures that are considered to be valid hypotheses for the reference experiment. In the next section, we motivate an ansatz that restricts the allowed causal structures and show that it restores causal sufficiency. 

\subsubsection{Inference rules for quantum interventions in layered DAGs \label{sec:qintervsoln}}

\begin{figure}[!htb]
\centering\includegraphics[width=0.6\linewidth]{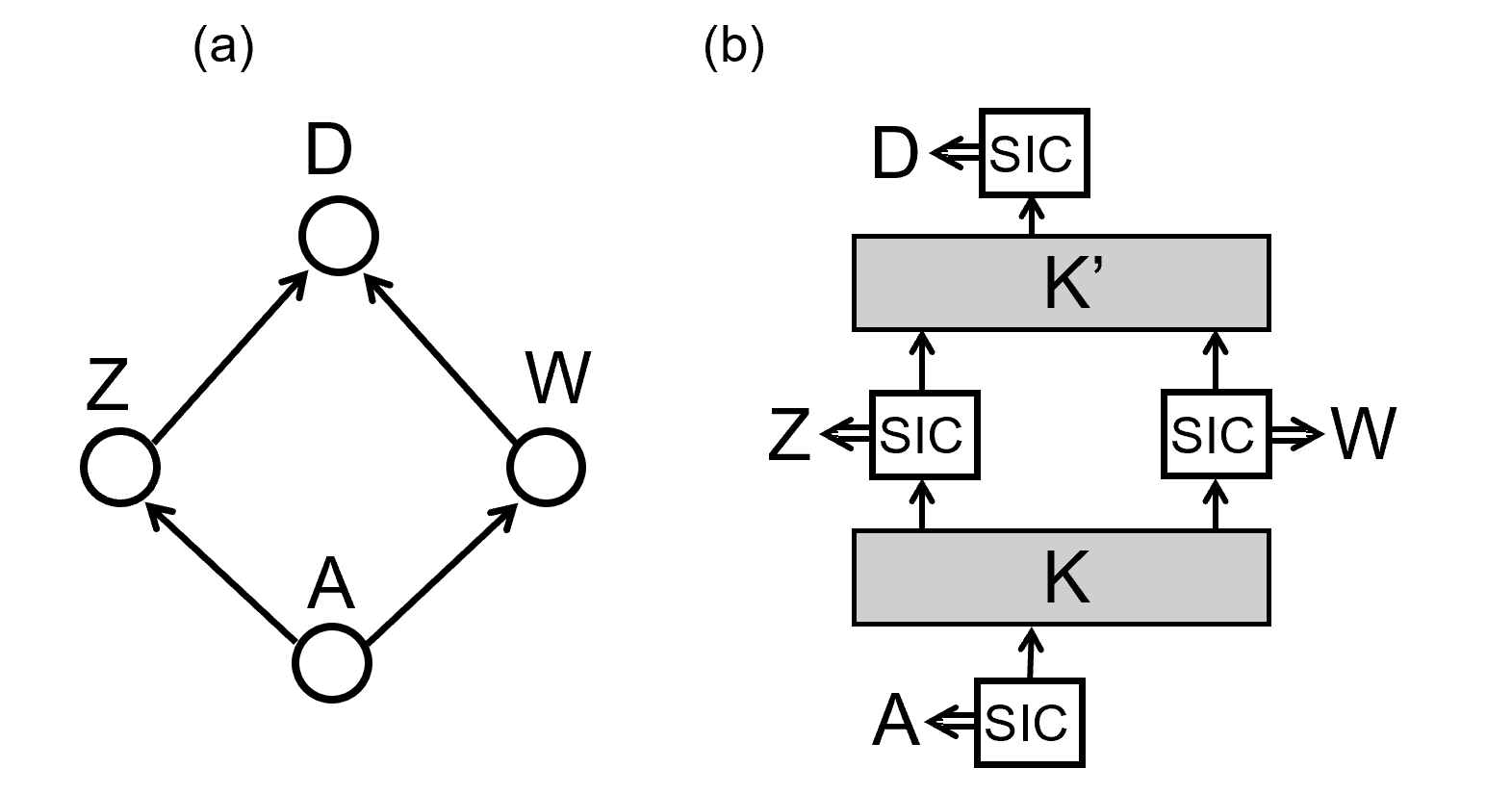}
\caption{(a) a causal graph and (b) a corresponding quantum circuit, similar to that shown in Fig. \ref{fig:qintervprob}, except with an additional SIC-instrument $Z$. In contrast to that case, the result of an intervention on $W$ (or on $Z$) in this circuit can be deduced directly from the reference probabilities $P(A,D,W,Z)$ for an arbitrary circuit. This enables us to derive rules for counterfactual inference.}
\label{fig:qintervsoln}
\end{figure}

Consider the causal graph shown in  Fig. \ref{fig:qintervsoln} (a) and corresponding circuit shown in Fig. \ref{fig:qintervsoln} (b). It is exactly the same circuit as that shown in Fig. \ref{fig:qintervprob}, except that an extra SIC-instrument $Z$ has now been introduced.   

In this example, the conflict with \tbf{CS} does not arise: since the tensor product of two SIC-POVMs is an informationally-complete POVM (though not itself symmetric), the statistics $P(A,D,W,Z)$ now are sufficient to reconstruct the two CPTP maps that comprise the circuit, implying that it ought to be possible to define an inference rule to compute $P(A,D,W,Z|C^W=\brm{do})$. From this example, we extrapolate the following ansatz describing a class of DAGs for which this solution is expected to work:\\

\tbf{LDAG. Layered DAG.} A DAG for a set of variables $\tbf{X}:=\{ X_i \}$ is said to be a \tit{layered} DAG or `LDAG' iff $\tbf{X}$ decomposes into $M$ disjoint subsets $\tbf{X} = \tbf{L}_1 \cup \tbf{L}_2 \cup \dots \cup \tbf{L}_M$ called `layers' such that no member of a layer is a cause of any other member of the same layer, and such that for any triplet $\tbf{L}_i,\tbf{L}_j,\tbf{L}_k$ with $i<j<k$, each path connecting $\tbf{L}_i$ to $\tbf{L}_k$ is intercepted by $\tbf{L}_j$, i.e. contains a causal chain $A \rightarrow B \rightarrow C$ whose middle member $B$ is in $\tbf{L}_j$.\\

We will now prove that, for any LDAG, there is an inference rule that can be used to obtain the probabilities under intervention, $P(A,D,W|C^W=\brm{do})$, using only those of the reference experiment, $P(A,D,W)$ plus the causal structure. 

First, consider the quantum causal model $\{ P(\tbf{X}), G(\tbf{X}) \}$ where $G(\tbf{X})$ is now assumed to be an LDAG. Let $\tbf{L}$ be the members (excluding $W$) of the layer containing the to-be-intervened-upon node $W$. The nodes $\tbf{R}$ can then be subdivided according to whether they causally precede or follow the layer: let $\tbf{R}_{D}$ be the subset of $\tbf{R}$ that are descendants of $\tbf{L}$ and let $\tbf{R}_A$ be the subset that are ancestors of $\tbf{L}$.

As already discussed, the post-intervention DAG $G(\tbf{X}|C^W=\brm{do})$ is obtained from $G(\tbf{X})$ by deleting the incoming arrows to $W$, which implies that the resulting graph is automatically also an LDAG. As we did for the classical case, we will derive the inference rule by postulating that the post-intervention probabilities satisfy the \tbf{QMC} relative to the new graph. This implies that the following conditional independence must hold:
\eqn{ \label{eqn:drsy}
P(\tbf{D} \tbf{R}_D|W \tbf{L} \tbf{A} \tbf{R}_A, \, C^W=\brm{do}) = P(\tbf{D} \tbf{R}_D|W \tbf{L}, \, C^{W}=\brm{do}) \, ,
}
which is simply a consequence of applying $\tbf{g-SSO}$ to the new LDAG. Another (less obvious) conditional independence implied by the \tbf{QMC} is:
\eqn{ \label{eqn:saray}
P(W|\tbf{L}  \tbf{A} \tbf{R}_A  , \, C^W=\brm{do}) &=& P(W| C^W=\brm{do}) \, \nonumber \\
 &:=&  P'(W) \, ,
}
which follows from the fact that, since $W$ has no ancestors in the intervened graph, every path connecting $W$ to any member of $\tbf{L} \tbf{A} \tbf{R}_A$ must contain a collider $A \rightarrow C \leftarrow B$ whose middle member $C$ is in $\tbf{D}$. Hence as long as we do not condition on $D$, the rule $\tbf{g-BK}$ implies that $\tbf{L} \tbf{A} \tbf{R}_A$ must be independent of $W$, which is just what \eqref{eqn:saray} says. Using these results, we obtain:

\eqn{ \label{eqn:qintervrule}
P(\tbf{X}|C^W=\brm{do}) &=& P(\tbf{D} \tbf{R}_D W \tbf{L}  \tbf{A} \tbf{R}_A |C^W=\brm{do}) \nonumber \\
&=& P(\tbf{D} \tbf{R}_D|W \tbf{L}  \tbf{A} \tbf{R}_A C^W=\brm{do}) \, P(W|\tbf{L}  \tbf{A} \tbf{R}_A  C^W=\brm{do}) \, P(\tbf{L}  \tbf{A} \tbf{R}_A|C^W=\brm{do}) \nonumber \\
&=& P(\tbf{D} \tbf{R}_D|W \tbf{L} , \, C^W=\brm{do})  \, P'(W) \, P(\tbf{L} \tbf{A} \tbf{R}_A|C^W=\brm{do}) \nonumber \\
&=& P(\tbf{D} \tbf{R}_D|W \tbf{L} , \, C^W=\oslash)  \, P'(W) \, P(\tbf{L} \tbf{A} \tbf{R}_A|C^W=\oslash) \, .
}
To obtain the third line in the above calculation we made use of \eqref{eqn:drsy} and \eqref{eqn:saray}. To obtain the last line we made use of \tbf{CNS} and \tbf{NPE2}. In the final line, all the terms on the RHS (except $P'(W)$, which is specified by the intervention) refer to probabilities in the reference experiment. Thus, Eq. \eqref{eqn:qintervrule} defines the inference rule for interventions on quantum systems whose causal structure is given by an LDAG. We summarize it as:\\

\tbf{QIR.} For quantum systems, the behaviour under an intervention on $W$ is related to the reference behaviour by (suppressing the $C^W=\oslash$ on the RHS):
\eqn{  \label{eqn:qintervaxiom}
P(\tbf{X}|C^W=\tbf{do}) = P'(W) \,P(\tbf{D} \tbf{R}_D|W \tbf{L}) \,  P(\tbf{L} \tbf{A} \tbf{R}_A) \, .
}
For a fine-grained intervention (that sets $W$ to a particular value) this becomes:
\eqn{  \label{eqn:qintervaxiomfine}
P(\tbf{X}|C^W=\tbf{do}(W=w')) = \delta(w,w') \,P(\tbf{D} \tbf{R}_D|W \tbf{L}) \,  P(\tbf{L} \tbf{A} \tbf{R}_A) \, .
}

For multiple interventions, this rule can simply be iterated. Thus, provided the causal structure of the system in the reference experiment is an LDAG, we can uphold \tbf{CS}. 

\tit{Remark:} Strictly speaking, we have only shown that the restriction to LDAGs is \tit{sufficient} for causal inference to be possible. To establish it as a necessary condition would require us to generalize the counter-example we gave earlier to arbitrary causal structures, that is, to show that for any DAG that is not an LDAG, full process tomography cannot be achieved. In the companion work Ref. \cite{JACQUES2}, we prove this using the \tit{process matrix} formalism for quantum causal models introduced in Ref. \cite{COSHRAP}. 

Under the joint assumptions that the causal structure is an LDAG, the class of causally reversible quantum systems has the curious feature that the state at any moment -- if we \tit{do not condition on the outcomes of any measurements} -- is always maximally mixed. In this sense, the dynamics is trivial, or, as some have put it, `eternal noise' \cite{COECKE}. Formally we may state it as follows:\\

\tbf{EN. Eternal noise:} In a causally symmetric quantum causal model on an LDAG, the unconditioned marginal distribution $P(\tbf{L})$ for any layer is the uniform random distribution over the outcomes of its variables. That is, \\
\eqn{
P(\tbf{L}) = \prod_{X_i \in \tbf{L}} \, P(X_i) = \prod_{X_i \in \tbf{L}} \, \frac{1}{d^2_{X_i}} \, .
}
This follows because if we don't condition on any other layers, the input state to any layer is equivalent to the maximally mixed state after propagation through some unbiased channel, hence is maximally mixed on the input Hilbert space of the given layer. (\tbf{EN} will be useful for proving some results in the Appendices).\\

\subsubsection{The generalized Urgleichung for un-measurements \label{sec:qunmeas}}

In Sec. \ref{sec:born} we discussed the inference rule for a quantum un-measurement in the simple case of a causal chain. For that case we found that the inference rule was given by the QBist Urgleichung, Eq. \eqref{eqn:causalurg}. In this section we will generalize this rule to arbitrary LDAGs. (For clarity in the equations, we will replace $C^{Z} = \brm{undo}$ with the shorter notation $\tbf{un}(Z)$, and will drop $C^{Z} = \oslash$ altogether, leaving it implicit whenever no value of $C^{Z}$ is specified).

\begin{figure}[!htb]
\centering\includegraphics[width=0.5\linewidth]{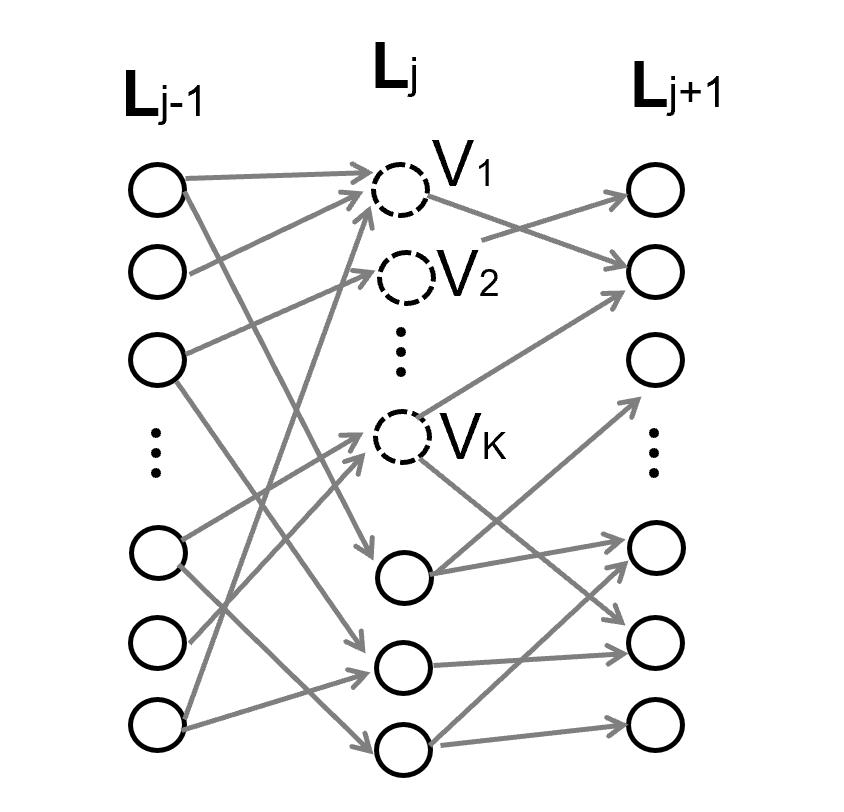}
\caption{A causal graph of three layers within a layered DAG. The main text explains how to infer the probabilities for an un-measurement of $K$ variables in the middle layer.}
\label{fig:genUrg}
\end{figure}

Let us consider the reference behaviour $P(\tbf{X} Z|C^Z=\oslash)=P(\tbf{X} Z)$ in a system with causal structure given by an LDAG $G(\tbf{X} Z)$, where $Z$ is the variable to be un-measured. Assuming $Z$ is contained in the $j_{\trm{th}}$ layer, let $\tbf{L}^{(-Z)}_j := \tbf{L}_j \setminus Z$ denote the elements of $\tbf{L}_j$ other than $Z$. From the \tbf{QMC} (more specifically, \tbf{SSO}) we can decompose the probabilities as:
\eqn{ \label{eqn:lsplit}
P(\tbf{X} Z) &=& \prod_{i} \, P(\tbf{L}_i|\tbf{L}_{i-1}) \nonumber \\
&=& P(\tbf{L}_M \tbf{L}_{M-1} \cdots \tbf{L}_{j+2}|\tbf{L}_{j+1}) \, P(\tbf{L}_{j+1}|\tbf{L}_{j} \tbf{L}_{j-1}) \,  P(\tbf{L}^{(-Z)}_j \, Z \, \tbf{L}_{j-1} \cdots \tbf{L}_{1}) \, .
}
Assuming the probabilities have a similar decomposition after the un-measurement leads us to posit the following ansatz:
\eqn{ \label{eqn:preansatz}
&& P(\tbf{X}|\brm{un}(Z)) := \nonumber \\
&& P(\tbf{L}_M \tbf{L}_{M-1} \cdots \tbf{L}_{j+2}|\tbf{L}_{j+1} , \, \brm{un}(Z)) \, 
P(\tbf{L}_{j+1}|\tbf{L}^{(-Z)}_{j} \tbf{L}_{j-1} , \, \brm{un}(Z)) \,
P(\tbf{L}^{(-Z)}_j \,\tbf{L}_{j-1} \cdots \tbf{L}_{1} | \, \brm{un}(Z)) \, .
}
Now we may invoke the principle \tbf{CSO} (recall Sec. \ref{sec:activemanips}) to deduce that:
\eqn{
P(\tbf{L}_M \tbf{L}_{M-1} \cdots \tbf{L}_{j+2}|\tbf{L}_{j+1} , \, \brm{un}(Z)) \, &=&
P(\tbf{L}_M \tbf{L}_{M-1} \cdots \tbf{L}_{j+2}|\tbf{L}_{j+1}), 
}
and
\eqn{
P(\tbf{L}^{(-Z)}_j \,\tbf{L}_{j-1} \cdots \tbf{L}_{1} | \brm{un}(Z)) &=&
P(\tbf{L}^{(-Z)}_j \,\tbf{L}_{j-1} \cdots \tbf{L}_{1})  \, .
}
Substituting these into \eqref{eqn:preansatz} we obtain:
\eqn{ \label{eqn:unmeasansatz}
P(\tbf{X}| \brm{un}(Z)) &:=& 
P(\tbf{L}_M \cdots \tbf{L}_{j+2}|\tbf{L}_{j+1}) \, P(\tbf{L}_{j+1}|\tbf{L}^{(-Z)}_{j} \tbf{L}_{j-1} , \, \brm{un}(Z)) \, P(\tbf{L}^{(-Z)}_j \,\tbf{L}_{j-1} \cdots \tbf{L}_{1}) \, ,
}
and the only term that is not the same as in the reference behaviour is the middle factor, $P(\tbf{L}_{j+1}|\tbf{L}^{(-Z)}_{j} \tbf{L}_{j-1} , \,\brm{un}(Z))$. This term represents the probabilities of the measurements in the layer $\tbf{L}_{j+1}$, conditional that all other measurements in $\tbf{L}^{(-Z)}_j$ were performed, and conditional on the values of the preceding layer $\tbf{L}_{j-1}$. More generally, one might consider un-measuring $K$ variables in the $j_{\trm{th}}$ layer, as shown in Fig. \ref{fig:genUrg}. 

By inspection of the figure, one sees a close analogy with the scenario in which the original Urgleichung was derived, except that now the relevant `causal chain' involves entire layers, $\tbf{L}_{j-1} \rightarrow \tbf{L}_{j} \rightarrow \tbf{L}_{j+1}$. This suggests that it might be more straightforward to first derive the inference rule for the un-measurement of the \tit{entire} middle layer $\tbf{L}_{j}$, and then specialize this to the case where only a subset, or the single member $Z$, is un-measured. We may therefore pose the problem as follows. Let $\rho_{\tbf{L}_{j-1}}$ be an operator on $\mathcal{H}_{\tbf{L}_j}$ representing the quantum state input to all measurements in the layer $\tbf{L}_j$, conditional on the outcomes $\tbf{L}_{j-1}$ measured at the previous layer. Then the desired probabilities can be expressed in the form: 
\eqn{ \label{eqn:genurgstart}
P(\tbf{L}_{j+1}|\tbf{L}_{j-1} , \, \brm{un}(\tbf{L}_j)) &=& \tr{\rho_{\tbf{L}_{j-1}} D_{\tbf{L}_{j+1}} } \, , 
}
where $D_{\tbf{L}_{j+1}}$ is some POVM comprised of operators on $\mathcal{H}_{\tbf{L}_j}$, whose outcomes correspond to the variables in $\tbf{L}_{j+1}$. Let us label the variables in $j_{\trm{th}}$ layer as $\tbf{L}_j := V_1 \cup \dots \cup V_N$, and associate the layer $\tbf{L}_j$ to the joint Hilbert space $\mathcal{H}_{\tbf{L}_j}:=\mathcal{H}_{V_1} \otimes \dots \otimes \mathcal{H}_{V_N}$. Since each $V_i$ is the outcome of a SIC-instrument, the whole layer $\tbf{L}_j$ is associated with a tensor product of SIC-instruments, which is itself an IC-instrument on $\mathcal{H}_{\tbf{L}_j}$. Hence the reference probabilities $P(\tbf{L}_{j+1}|\tbf{L}_{j}),P(\tbf{L}_{j}|\tbf{L}_{j-1})$ contain sufficient information to fully reconstruct the operators $\rho_{\tbf{L}_{j-1}}, D_{\tbf{L}_{j+1}}$. This means we can write these operators as functions of the probabilities and insert the resulting expressions into the RHS of Eq. \eqref{eqn:genurgstart} to obtain the desired inference rule for un-measuring $\tbf{L}_{j}$. The main obstacle is that a tensor product of SIC-instruments is not itself a SIC-instrument, and so the final equation will have a different form than the Urgleichung. Indeed, one might expect it be rather complicated and ugly. Remarkably, following the procedure just outlined, we arrive at the unexpectedly simple result (derived in Appendix \ref{app:genurg1} ): 

\eqn{ \label{eqn:genurgall}
P(\tbf{L}_{j+1}|\tbf{L}_{j-1} , \, \brm{un}(\tbf{L}_j)) &=&  \zum{\tbf{v} \tbf{v}'}{} \, P(\tbf{L}_{j+1}|\tbf{v}') \left( \prod_{n=1}^{N} \, \left[ (d_n+1)\delta_{v_n v'_n}-\frac{1}{d_n}  \right] \right) \, P(\tbf{v}|\tbf{L}_{j-1}) ,
}
where $d_n$ is the dimension of $\mathcal{H}_{V_n}$, and the summation over each $v_n$ runs up to $d^2_n$ (i.e. because that is the number of outcomes of the $n_{\trm{th}}$ SIC-instrument). Notice that this equation gives us $P(\tbf{L}_{j+1}|\tbf{L}_{j-1} , \, \brm{un}(\tbf{L}_j))$ purely as a function of the reference probabilities, i.e. the terms on the RHS are understood to be conditioned on $C^{\tbf{L}_j}=\oslash$, so it gives us the desired inference rule for an un-measurement of $\tbf{L}_j$.

We next tackle the question of what form this rule should take when we only wish to un-measure a subset of the $\tbf{L}_j$. Without loss of generality, we can partition $\tbf{L}_j := \{U_i : i=1,2,\dots ,K \} \cup \{W_i : i=1,2,\dots,N-K \} := \tbf{U} \cup \tbf{W}$ for some $K < N$, and contemplate an un-measurement of all $\tbf{U}$, while keeping the measurements $\tbf{W}$ in place. Our goal is then to calculate the probabilities for $\tbf{L}_{j+1}$ conditional that $\tbf{U}$ are un-measured, and conditional that $\tbf{W}$ are measured and attain specific values $\tbf{W} =\{w_1 \dots w_{N-K} \}$. Since a SIC-instrument $W_i$ with outcome $W_i=w_i$ projects the measured sub-system into the post-measurement state $\Pi_{w_i}$, we can express the desired probabilities as:

\eqn{ \label{eqn:genurgstart2}
P(\tbf{L}_{j+1}|\tbf{L}_{j-1} \tbf{w}, \, \brm{un}(\tbf{U})) &=& \tr{\left( \rho^{\tbf{U}}_{\tbf{L}_{j-1}}\otimes \Pi_{w_{1}} \otimes \cdots \otimes \Pi_{w_{N-K}} \right) D_{\tbf{L}_{j+1}} } \, , 
}
where $\rho^{\tbf{U}}_{\tbf{L}_{j-1}} := \trc{W}{\rho_{\tbf{L}_{j-1}} }$ is the reduced state of the input to the first $K$ measurements, defined on the Hilbert space $\mathcal{H}_{U_1} \dots \mathcal{H}_{U_K}$, and conditioned on the outcomes $\tbf{L}_{j-1}$ from the previous layer. We can then carry out the calculation in exactly the same way as before. To do so, we partition the layer as $\tbf{L}_j := \{V_i : i=1,2,\dots ,K \} \cup \{V_i : i=1,2,\dots,N-K \} := \tbf{U} \cup \tbf{W}$, where $\tbf{U}$ are the variables to be un-measured. Hence $\tbf{U}$ takes values that are $K$-tuples $\tbf{u}:=(v_1,\dots,v_K)$ and $\tbf{W}$ takes values that are $(N-K)$-tuples $\tbf{w}:=(v_{K+1},\dots,v_{N})$. It will also be useful to define the variable $\tbf{V}$ covering the whole layer, having the $N$-tuple values $\tbf{v}:=(v_{1},\dots,v_{N})$. We then obtain (details in Appendix \ref{app:genurg2}):

\eqn{ \label{eqn:genurgsubset}
P(\tbf{L}_{j+1}|\tbf{L}_{j-1} \tbf{w} , \, \brm{un}(\tbf{U})) &=& \zum{\tbf{u}' \tbf{u}}{}  \, P(\tbf{L}_{j+1}|\tbf{u}\tbf{w}) \, \left( \prod^{K}_{n=1} \, \left[ (d_n+1)\delta_{v'_n v_n}-\frac{1}{d_n} \right] \right) \, P(\tbf{u}'|\tbf{L}_{j-1}) \, ,
}
This equation tells us the probabilities conditional on the outcomes $\tbf{W}=\tbf{w}$. It can be combined with $P(\tbf{W} |\tbf{L}_{j-1} , \, \brm{un}(\tbf{U}))$ to find the result after marginalizing over $\tbf{W}$, namely:

\eqn{ \label{eqn:qLTP}
P(\tbf{L}_{j+1}|\tbf{L}_{j-1} , \, \brm{un}(\tbf{U})) &=& \zum{\tbf{w}}{} \, P(\tbf{L}_{j+1}|\tbf{L}_{j-1} \tbf{w} , \, \brm{un}(\tbf{U})) \, P(\tbf{w} |\tbf{L}_{j-1} \brm{un}(\tbf{U})) \, \nonumber \\
&=& \zum{\tbf{w}}{} \, P(\tbf{L}_{j+1}|\tbf{L}_{j-1} \tbf{w}  , \, \brm{un}(\tbf{U})) \, P(\tbf{w} |\tbf{L}_{j-1}) \, ,
}
where in the last line we used the fact that, as a consequence of \tbf{CNS},
\eqn{
P(\tbf{w} |\tbf{L}_{j-1} , \, \brm{un}(\tbf{U}) ) = P(\tbf{w} |\tbf{L}_{j-1}) \nonumber \, ,
}
i.e. the probabilities for the measurements of $\tbf{W}$ are the same whether or not the measurements of $\tbf{U}$ are performed or not. 

It is illuminating to consider some special cases of this inference rule. First, consider the case that $N=1$, that is, the layer $\tbf{L}_j$ consists of a single measurement $V_1 := Z$ which is to be un-measured. In that case Eq. \eqref{eqn:genurgsubset} reduces to:
\eqn{ \label{eqn:oneZ}
P(\tbf{L}_{j+1}|\tbf{L}_{j-1} \, , \brm{un}(Z))
&=& \zum{z' z}{}  \, P(\tbf{L}_{j+1} |z ) \, \left[ (d_Z+1)\delta_{z' z}-\frac{1}{d_Z} \right] \, P(z'|\tbf{L}_{j-1}) \, \nonumber \\
&=& \zum{z}{}  \, P(\tbf{L}_{j+1} |z ) \, \left[ (d_Z+1) P(z|\tbf{L}_{j-1}) -\frac{1}{d_Z} \right] \, ,
}
which recovers the usual Urgleichung, as expected. Next, consider the case where the probabilities at the layer $\tbf{L}_{j+1}$ only depend on the subsystems \tbf{W} of the previous layer $\tbf{L}_{j}$ (this could occur if, for instance, the un-measured \tbf{U} variables are all terminal nodes in the DAG, meaning that the sub-systems are discarded after measurement). In this case it should not make any difference to our considerations whether the \tbf{U} are un-measured, or simply measured and then discarded. In other words, we would expect to recover the usual rule for non-disturbing measurements in this case. And indeed, substituting our assumption $P(\tbf{L}_{j+1}|\tbf{U}\tbf{W}) \mapsto P(\tbf{L}_{j+1}|\tbf{W})$ into the RHS of \eqref{eqn:genurgsubset} we obtain:
\eqn{
P(\tbf{L}_{j+1}|\tbf{L}_{j-1} \tbf{w} , \, \brm{un}(\tbf{U})) &=& \zum{\tbf{u}' \tbf{u}}{}  \, P(\tbf{L}_{j+1}|\tbf{w}) \, \left( \prod^{K}_{n=1} \, \left[ (d_n+1)\delta_{v'_n v_n}-\frac{1}{d_n} \right] \right) \, P(\tbf{u}'|\tbf{L}_{j-1}) \, \nonumber \\
&=& \zum{\tbf{u}'}{}  \, P(\tbf{L}_{j+1}|\tbf{w}) \, \left( \prod^{K}_{n=1} \, \zum{v_n}{} \, \left[ (d_n+1) \delta_{v'_n v_n}-\frac{1}{d_n} \right] \right)P(\tbf{u}'|\tbf{L}_{j-1}) \, \nonumber \\
&=& \zum{\tbf{u}'}{}  \, P(\tbf{L}_{j+1}|\tbf{w}) \, \left( \prod^{K}_{n=1} \, \left[ (d_n+1)-(d^2_n)\frac{1}{d_n} \right] \right)P(\tbf{u}'|\tbf{L}_{j-1}) \, \nonumber \\
&=& \zum{\tbf{u}'}{}  \, P(\tbf{L}_{j+1}|\tbf{w}) P(\tbf{u}'|\tbf{L}_{j-1}) \, \nonumber \\
&=& P(\tbf{L}_{j+1}|\tbf{w}) \, ,
}
and inserting this into the RHS of \eqref{eqn:qLTP} gives us:
\eqn{
P(\tbf{L}_{j+1}|\tbf{L}_{j-1} , \, \brm{un}(\tbf{U})) &=& \zum{\tbf{w}}{} \, P(\tbf{L}_{j+1}|\tbf{w}) \, P(\tbf{w} |\tbf{L}_{j-1}) \, ,
}
which recovers the rule for non-disturbing measurements, Eq. \eqref{eqn:nondisturb}.

The question still stands as to what should be the causal graph $G(\tbf{X})$ that holds for the counterfactual behaviour $P(\tbf{X}|\,\brm{un}(Z))$. Given that un-measurements do not break causal chains (property (ii) of un-measurements, Sec. \ref{sec:activemanips} ), the following is a natural postulate:\\

\tbf{UNDAG. The DAG after an un-measurement:}\\ 
Let $G(\tbf{X},Z)$ represent the causal structure of a quantum system in the reference experiment, where $Z$ is not an exogenous node. Then the DAG conditional on un-measuring $Z$, denoted $G(\tbf{X}|C^Z=\brm{undo})$, is obtained from $G(\tbf{X},Z)$ by the following procedure:\\
(i) Directly connect every parent of $Z$ to every child of $Z$;\\
(ii) Delete $Z$ and its incoming and outgoing arrows.\\

It is natural to ask what conditional independences are implied by the \tbf{QMC} in the DAG after un-measuring $Z$. To obtain the answer, note that the removal of $Z$ in the manner described in \tbf{UNDAG} cannot un-block any path between variables that were previously blocked (unless they were blocked by $Z$) and it cannot block any path that was previously un-blocked. This means that the conditional independences implied by the DAG $G(\tbf{X}|C^Z=\brm{undo})$ after the un-measurement are simply the set of conditional independences implied by the original LDAG $G(\tbf{X} \, Z)$, excluding those that involve $Z$. 

The next important question is whether the probabilities after the un-measurement continue to satisfy these conditional independences, i.e. whether $P(\tbf{X}|\,C^Z=\brm{undo})$ satisfies the \tbf{QMC} relative to the new DAG $G(\tbf{X}|C^Z=\brm{undo})$. True enough, we effectively \tit{assumed} that this is the case in order to obtain the general decomposition into layers of Eq. \eqref{eqn:unmeasansatz}. However, this decomposition only relies on the rule \tbf{SSO}, which does not apply between layers. In particular, we have \tit{not} assumed that the probabilities $P(\tbf{L}_{j+1}|\tbf{L}_{j-1} \tbf{W} , \, \brm{un}(\tbf{U}))$ as defined by Eq. \eqref{eqn:genurgsubset} obey the constraints of the \tbf{QMC} relative to $G(\tbf{X}|C^Z=\brm{undo})$. In fact, establishing this turns out to be highly non-trivial and we were not able to do it in full generality. Instead, it is left as a conjecture. Evidence in its favour is presented in Appendix \ref{app:qunmeas}, where it is shown to be true for a significant class of conditional independence relations.

\tit{Remark:} It is important to note that the DAG $G(\tbf{X}|C^Z=\brm{undo})$ after un-measuring $Z$ is not necessarily an LDAG. When this occurs, it may not be possible to perform further un-measurements or interventions (i.e. because by not being an LDAG the new model does not meet the requirements of a reference experiment, and the desired inference rules may not exist). However, the rule \eqref{eqn:genurgsubset} is sufficiently general to allow unmeasurements of any sub-set of variables within a single layer. Moreover, variables can be un-measured in multiple layers, provided each affected layer is sandwiched between a pair of layers that are not themselves subject to un-measurements. Hence the fact that un-measurements do not preserve the LDAG structure does not entirely prevent us from making inferences about multiple un-measurements, so long as these meet the above requirements.

\begin{figure}[!htb]
\centering\includegraphics[width=0.5\linewidth]{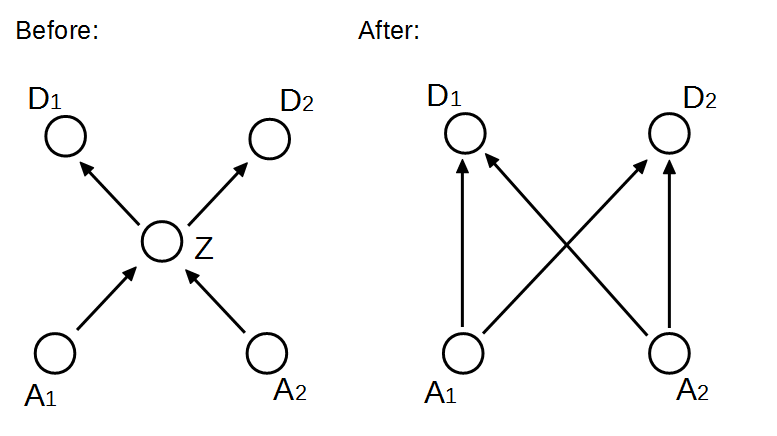}
\caption{The causal graph before and after un-performing the measurement $Z$; the causal pathways from its parents to its children are preserved.}
\label{fig:qunmeas}
\end{figure}

\subsection{Discussion \label{sec:discuss}}

In this work we have seen how a quantum causal model can be constructed within a general framework for causal modeling that defines causality in terms of probabilities for observations in counterfactual experiments. We proposed that quantum reference measurements should be informationally complete, and on grounds of simplicity we adopted the convention of using SIC-instruments (definition \tbf{QRM} in Sec. \ref{sec:born}). Using this as our guide, we identified a set of physical Markov conditions applicable to quantum systems, namely $\{ \tbf{RP}, \tbf{SSO}, \tbf{BK}, \tbf{BK*}, \tbf{PE} \}$. We drew attention to a quasi-symmetry in this set under causal inversion, and we proposed to make the symmetry complete by dropping \tbf{PE}, and pairing \tbf{RP} with its causal inverse \tbf{RP*} to form the `symmetric Reichenbach principle' \tbf{SRP}. We argued that the resulting set of causally symmetric Markov conditions, $\{ \tbf{SRP}, \tbf{SSO}, \tbf{BK}, \tbf{BK*} \}$, could be enforced by restricting to experiments involving unbiased quantum processes acting on maximally mixed input states, which we defined as the set of causally symmetric quantum systems. We then converted the causally symmetric Markov conditions into graphical rules that we used to obtain the corresponding rules for arbitrary causal structures, as given by the \tbf{QMC} (cf Sec. \ref{sec:qmc}). 

Using the \tbf{QMC}, we attempted to derive counterfactual inference rules for interventions and quantum un-measurements in Sec. \ref{sec:qinterf}. The two main challenges to this task were (i) upholding the principle of causal sufficiency \tbf{CS} (Sec. \ref{sec:experiments}) for interventions on quantum systems, and (ii) defining an inference rule for un-measurements on quantum systems, which in general are disturbing measurements. The first was overcome by positing that the causal structure in the reference experiment should be layered (see definition \tbf{LDAG} in Sec. \ref{sec:qintervsoln}), while the second was overcome by adapting an existing result from the literature on QBism (the Urgleichung, Eq. \eqref{eqn:urg}) to our framework and generalizing it to more general causal structures, as described by the inference rule Eq. \eqref{eqn:genurgsubset}. We now anticipate and address some questions about this framework.\\

\tit{What is the physical significance of the restriction to LDAGs?} \\  
This restriction implies that it must always be possible -- in principle -- to perform measurements on a quantum system so as to have the causal structure conform to that of an LDAG. We note that this structure is strongly reminiscent of the network structure of a \tit{Markovian} quantum process as discussed elsewhere in Eg. \cite{POLLOCKPRL,POLLOCKPRA,COSHRAP}. However, to make this comparison precise is difficult because the aforementioned works are primarily concerned with Markovianity as a property of \tit{process tensors} \cite{POLLOCKPRL} or \tit{process matrices} \cite{COSHRAP}, whereas the present framwork formulates the \tbf{QMC} only at the level of probabilities. Nevertheless, our model was explicitly constructed with operator representations of quantum networks in mind, and so we expect a link to exist between our definitions and those pertaining to entities like process tensors and process matrices. In particular, we note that the decomposition of the first line in \eqref{eqn:lsplit} is essentially identical to the definition of a classical Markovian process except that the measurements $\tbf{L}_i$ do not necessarily occur simultaneously at a single time-step, but are ordered by the more general consideration of causal influence in the LDAG. This enables us to apply the \tit{Corollary} in Ref. \cite{POLLOCKPRL} to deduce that any representation of our quantum causal model in terms of process tensors will necessarily result in a process tensor that is Markovian by the definition of that work, at least with regards to measurements and interventions performed on whole layers. Thus, the restriction to LDAGs may be understood as imposing the convention that the reference experiment should be Markovian.\\

\tit{What is the significance of the causal symmetry?} \\ 
We can formalize the causal symmetry of the \tbf{QMC} in the following way: suppose that $P(\tbf{X})$ satisfies the \tbf{QMC} relative to a DAG $G(\tbf{X})$. Let $G^{*}(\tbf{X})$ be the DAG obtained from $G(\tbf{X})$ by reversing the directions of all arrows. Then $P(\tbf{X})$ also satisfies the \tbf{QMC} relative to $G^{*}(\tbf{X})$. Thus if one were ignorant of the overall direction of the causal structure in a system -- though we must suspend disbelief to imagine how one could ever be so ignorant -- then one would be powerless to determine this direction just from examining the behaviour of the system in the reference experiment (\tit{sans} other information about eg. the actual timing of the measurements). To obtain information about the causal direction one would have to either perform an intervention or an un-measurement and see how the system responds. This leaves open the intriguing possibility of interpreting the \tit{direction} of causal structure as a property that has no definite value prior to an act of \tit{manipulation}, in analogy to how the properties of a quantum system are regarded to have no definite values prior to their measurement. Similar ideas have been explored in related frameworks by other authors \cite{ORESH, GUERIN,OCB}, and we show in a companion work (Ref. \cite{JACQUES2}) that the causal symmetry we defined here at the level of probabilities can also be extended to the process matrix causal modeling framework of Ref. \cite{COSHRAP}. More work is needed to understand the relevance of such results to the quantum gravity setting, where space-time relations are treated as dynamical variables that may be quantized.

\tit{What is the classical limit of causally symmetric quantum causal models?} \\ 
We remarked in Sec. \ref{sec:soff} that a `classical limit' can be obtained from a quantum network by restricting all measurements to be performed in a single basis, and restricting all states to be diagonal in this same basis. In the present framework, doing so leads to the result that the reference measurements can be non-disturbing, just as in the case of classical stochastic systems. However, if applied to the class of causally reversible quantum systems, it appears unlikely that the classical limit would produce a model that satisfies \tbf{FCC}, as this would break the symmetry between $\tbf{BK}*$, $\tbf{BK}$. The possibility of a stochastic classical causal model violating \tbf{FCC} is intriguing, for it would indicate that the failure of \tbf{FCC} (and for that matter, causal reversibility) is not by itself a uniquely quantum phenomenon. It may be, for instance, that there is no fundamental difference between classical and quantum systems in terms of the physical Markov conditions that each can satisfy, but that the difference lies entirely in the form of the counterfactual inference rules. More work is needed to investigate this possibility.\\

\tit{Is there a quantum advantage to causal inference?} \\ 

We remarked in Sec. \ref{sec:soff} that in order to establish a basis for comparison of the different powers of quantum versus classical systems in causal inference, it is necessary to establish a quantum analog of the classical `passive observational scheme'. Our concept of a `reference observational scheme' is a natural candidate for the quantum analog of a passive observational scheme, and it may be fruitful to make comparisons between the quantum and classical causal models on this basis, such as comparing our present constraint of causal symmetry (characteristic of our reference observational scheme) with the different notion of \tit{informationally symmetric measurements} introduced in Refs. \cite{RIED, RIEDPHD}. Another interesting avenue to explore is the role of un-measurements in the task of inferring causal structure. As we have shown, un-measurements can reveal causal structure in a way that does not break causal chains, which raises questions about their limitations and advantages in comparison to interventions. Furthermore, the concept of an un-measurement might be carried over to classes of classical systems in which measurements are known to be disturbing. Situations abound in which the mere act of measuring an experimental subjet may change that subject's behaviour, and it would be interesting to see whether our framework can be used to model the response to such measurements using an appropriately tailored inference rule for un-measurements.

\section{Conclusions}

Science is primarily concerned with the acquisition of knowledge about the world, but in order to achieve this role it must also be concerned with understanding the limitations on how such knowledge can be acquired. On a manipulationist interpretation of causality, one is concerned with the information available to an observer prior to performing `manipulations' on the system, as encoded in the observer's probability assignments for measurements in a conventionally defined reference experiment. From this point of view, the reference measurements should be maximally informative about counterfactual experiments, hence should be informationally complete. For quantum systems, this leads naturally to the idea of SIC-instruments as the elementary causal relata, and this in turn leads to the quantum physical Markov Conditions and then to inference rules for interventions and un-measurements. 

Curiously, we have found that the reference experiment may be defined so that the physical Markov conditions of quantum systems are symmetric under causal inversion. If we reject the assumption that this symmetry represents ignorance of a true underlying causal orientation, then we may entertain the idea that the asymmetry of the direction of cause and effect is introduced at the moment that an observer interacts with a system by interacting with it, either to make an intervention or perform an un-measurement.

These results suggest two main conclusions. Firstly, it has been shown that the manifest asymmetry of the cause-effect relationship on a manipulationist and probabilistic account of causality may yet be compatible with the symmetry of physics at the fundamental level. That is to say, if one regards the reference experiment as representing the state of maximal information that is in principle available about a system for making inferences, and if one defines this experiment in the manner that makes it causally symmetric as we have done here, then it can be maintained that the very direction of causal structure itself need not be treated as a feature intrinsic to the observer-independent world, but may emerge through the very process of the observer's interaction with the system. This idea is explored further in the companion work \cite{JACQUES2}.

Secondly, our introduction of the concept of `un-measurement' (i.e. the enforcement of a lack of some measuring device at some location) as an essential part of causal inference on quantum systems is quite novel. Un-measurements are interesting in that they preserve the causal paths, and yet appear to depend upon (and thus have the potential to reveal) causal structure beyond what can be ascertained from the reference experiment. Future work is needed to understand whether un-measurements have a special role to play in the inference of causal structure in both quantum and classical settings.   

\acknowledgments
I thank G. Barreto Lemos and C. Duarte for detailed feedback on early versions; C. Fuchs, B. Stacey and J. DeBrota for help in understanding the mathematics of QBism; P. Kauark Leite, Romeu Rossi Jr. and R. Chaves for stimulating discussions that influenced the course of the paper; and an anonymous referee for pointing out a serious error that has now been fixed. This work was supported in part by the John E. Fetzer Memorial Trust.

\appendix

\section{Graphical rules from principles \label{app:graphrules}}
We present here a general procedure for taking a set of principles that apply to the three special cases (common cause, chain, and common effect) and using them to obtain a graph-separation rule. So far as we know, no such procedure has been proposed before in the literature. The standard route is to postulate the physical Markov conditions in general causal structures, in the form of a factorization rule like that of Eq. \ref{eqn:cmc}, and derive a graph separation rule from that. Our approach is potentially more powerful, as it allows us to formulate physical Markov conditions for general causal structures that might not be possible to express as a single factorization condition. (Indeed, as we pointed out in Sec. \ref{sec:qmc}, we have not been able to find a factorization condition that expresses the \tbf{QMC}).

To begin with, we assume that any graphical rule examines the undirected paths between two variables and determines whether the path is blocked or unblocked according to whether it contains one of the following triplets:\\
(i) a chain $A \rightarrow C \rightarrow B$;\\
(ii) a fork $A \leftarrow C \rightarrow B$;\\
(iii) a collider $A \rightarrow C \leftarrow B$.\\
We further assume that whether the path is blocked depends in each case on whether $C$ is conditioned or unconditioned. In cases (ii) (resp. (iii)), if $C$ is not conditioned upon, then we also need to consider whether the ancestors (resp. descendants) of $C$ are conditioned upon. Therefore there are a total of eight cases to consider, as shown in Table \ref{table:graphrules}. 

Since a path can be blocked or unblocked in each of the eight cases, there are $2^8=256$ logically possible graph separation rules under these restrictions. To find the correct rule for a class of systems, we look at each of the 8 cases in turn and apply the principles derived for that case. The results are summarized for the \tbf{CMC} and \tbf{QMC} in Table \ref{table:graphrules}.

\tit{Remark:} A strictly conservative reading of $\tbf{FCC}$ would not exclude the possibility that a path $A \leftarrow C \rightarrow B$ might be blocked when only the ancestors of $C$ are conditioned upon, as we have assumed in the table. Allowing such a possibility would open the door to a variety of additional possible rules according to which conditioning on some subset of $C$'s ancestors would be sufficient to block the path. Since there seems to be little physical motivation for such cases, we have opted here to exclude these possibilities by fiat, by taking an "all or nothing" reading of \tbf{FCC} according to which the path is assumed to be unblocked unless $C$ is conditioned upon.

\begin{table*}[ht] 
\caption{Key properties of the Quantum Markov Condition \label{table:graphrules}}
\centering
\begin{tabular}{p{0.1\linewidth}p{0.15\linewidth}|p{0.15\linewidth}p{0.15\linewidth}|p{0.15\linewidth}p{0.15\linewidth}|}
\cline{3-6}
 & & \multicolumn{2}{ c| }{\tbf{Classical}}  & \multicolumn{2}{ c| }{\tbf{Quantum}} \\ 
 \cline{2-6}
 & \multicolumn{1}{ |c| }{$C$ conditioned?} & \multicolumn{1}{ c| }{Blocked?} & \multicolumn{1}{ c| }{Reason} & \multicolumn{1}{ c| }{Blocked?} & \multicolumn{1}{ c| }{Reason} \\ 
 \cline{1-6}
\multicolumn{1}{ |c| }{$A\rightarrow C \rightarrow B$} &  \multicolumn{1}{ c| }{No} & \multicolumn{1}{ c| }{$\times$} & \multicolumn{1}{ c| }{\tbf{CIC}} & \multicolumn{1}{ c| }{$\times$} & \multicolumn{1}{ c| }{\tbf{CIC}}      \\ 
\cline{1-6}
\multicolumn{1}{ |c| }{$A\leftarrow C \rightarrow B$} &  \multicolumn{1}{ c| }{No, nor ancestors} & \multicolumn{1}{ c| }{$\times$} & \multicolumn{1}{ c| }{\tbf{PE}} & \multicolumn{1}{ c| }{\checkmark} & \multicolumn{1}{ c| }{\tbf{RP*}}      \\ 
\cline{1-6}
\multicolumn{1}{ |c| }{$A\leftarrow C \rightarrow B$} &  \multicolumn{1}{ c| }{No, ancestors only} & \multicolumn{1}{ c| }{$\times$} & \multicolumn{1}{ c| }{\tbf{FCC}} & \multicolumn{1}{ c| }{$\times$} & \multicolumn{1}{ c| }{\tbf{BK*}}      \\ 
\cline{1-6}
\multicolumn{1}{ |c| }{$A\rightarrow C \leftarrow B$} & \multicolumn{1}{ c| }{No, nor descendants} & \multicolumn{1}{ c| }{\checkmark} & \multicolumn{1}{ c| }{\tbf{RP}} & \multicolumn{1}{ c| }{\checkmark} & \multicolumn{1}{ c| }{\tbf{RP}}      \\ 
\cline{1-6}
\multicolumn{1}{ |c| }{$A\rightarrow C \leftarrow B$} & \multicolumn{1}{ c| }{No, descendants only} & \multicolumn{1}{ c| }{$\times$} & \multicolumn{1}{ c| }{\tbf{BK}} & \multicolumn{1}{ c| }{$\times$} & \multicolumn{1}{ c| }{\tbf{BK}}      \\ 
\cline{1-6}
\multicolumn{1}{ |c| }{$A\rightarrow C \rightarrow B$} &  \multicolumn{1}{ c| }{Yes} & \multicolumn{1}{ c| }{\checkmark} & \multicolumn{1}{ c| }{\tbf{SSO}} & \multicolumn{1}{ c| }{\checkmark} & \multicolumn{1}{ c| }{\tbf{SSO}}      \\ 
\cline{1-6}
\multicolumn{1}{ |c| }{$A\leftarrow C \rightarrow B$} &  \multicolumn{1}{ c| }{Yes} & \multicolumn{1}{ c| }{\checkmark} & \multicolumn{1}{ c| }{\tbf{FCC}} & \multicolumn{1}{ c| }{$\times$} & \multicolumn{1}{ c| }{\tbf{BK*}}      \\ 
\cline{1-6}
\multicolumn{1}{ |c| }{$A\rightarrow C \leftarrow B$} &  \multicolumn{1}{ c| }{Yes} & \multicolumn{1}{ c| }{$\times$} & \multicolumn{1}{ c| }{\tbf{BK}} & \multicolumn{1}{ c| }{$\times$} & \multicolumn{1}{ c| }{\tbf{BK}}      \\ 
\cline{1-6}
\end{tabular}
\end{table*}

In the table, $\tbf{CIC}$ stands for `causation implies correlation', which we recall is implicit in our very definition of causation \tbf{MC}. It ensures that $A,B$ are correlated even when $A$ is only an indirect cause of $B$, hence that the chain is unblocked when $C$ is not conditioned upon. Taking this for granted reduces the set of `reasonable' possible graph separation rules to $128$; it would be interesting to investigate how many of these might correspond to familiar physical systems.

\section{Inference rule for un-measurement of a layer \label{app:genurg1} }

The aim is to derive $P(\tbf{L}_{j+1}|\tbf{L}_{j-1} ,\, \brm{un}(\tbf{L}_j))$ as a function of $P(\tbf{L}_{j+1}|\tbf{L}_{j}),P(\tbf{L}_{j}|\tbf{L}_{j-1})$. Recall \eqref{eqn:genurgstart}:
\eqn{ \label{eqn:qi}
P(\tbf{L}_{j+1}| \tbf{L}_{j-1} ,\, \brm{un}(\tbf{L}_j)) = \tr{\rho_{\tbf{L}_{j-1}} \, D_{\tbf{L}_{j+1}}}
}
where $\rho_{\tbf{L}_{j-1}} $ is a density operator and $\{D_{\tbf{L}_{j+1}}\}$ is some POVM, both defined on the Hilbert space $\mathcal{H}_{\tbf{L}_j} := \mathcal{H}_{V_1} \otimes \dots \otimes \mathcal{H}_{V_N}$. Let $d_n$ be the dimension of $\mathcal{H}_{V_n}$ and the total dimension of the layer $d_{\tbf{L}_j}:= d_{1} d_{2} \dots d_{N}$. We consider each variable in the layer to represent a SIC instrument with SIC-POVM $\{E_{v_n} : v_n = 1,2,\dots,d^2_n \}$ on $\mathcal{H}_{V_n}$, and we define the total IC on the whole layer $\{E_{(v_1,v_2,\dots, v_N)} \} := \{ E_{\tbf{v}}\}$ as the tensor product of these:

\eqn{
E_{\tbf{v}} &=& E_{(v_{1}, v_{2}, \,\dots \, ,v_{N})} \nonumber \\
&:=& E_{v_{1}} \otimes E_{v_{2}} \otimes \dots \otimes E_{v_{N}} \, .
}

Since a tensor product of SIC-POVMs is an IC-POVM (though evidently not a SIC), we have the expansions:

\eqn{ \label{eqn:rhoDall}
\rho_{\tbf{L}_{j-1}} &=& \zum{\tbf{r}}{} \, \alpha_{\tbf{r}}(\tbf{L}_{j-1}) \, E_{\tbf{r}} \, , \nonumber \\
D_{\tbf{L}_{j+1}} &=& \zum{\tbf{s}}{} \, \beta_{\tbf{s}}(\tbf{L}_{j+1}) \, E_{\tbf{s}} \, . \nonumber
}

To write the coefficients $\alpha_{\tbf{r}}(\tbf{L}_{j-1}), \beta_{\tbf{s}}$ in terms of probabilities, first note that:

\eqn{
P(\tbf{v}|\tbf{L}_{j-1}) &=& \zum{\tbf{r}}{} \, \alpha_{\tbf{r}}(\tbf{L}_{j-1}) \, \tr{ E_{\tbf{r}} E_{\tbf{v}} } \nonumber \\
&:=& \zum{\tbf{r}}{} \, \alpha_{\tbf{r}}(\tbf{L}_{j-1}) \, M_{\tbf{r} \tbf{v}} \\
P(\tbf{L}_{j+1}|\tbf{v}') &=& \zum{\tbf{s}}{} \, \beta_{\tbf{s}}(\tbf{L}_{j+1}) \, \tr{ d_{\tbf{L}_j} E_{\tbf{v}'} E_{\tbf{s}} } \, \nonumber \\
&:=&  \zum{\tbf{s}}{} \, \beta_{\tbf{s}}(\tbf{L}_{j+1}) \, d_{\tbf{L}_j} M_{\tbf{v}' \tbf{s}} \, .
}
To invert these equations, let $\tbf{M}$ be the $d^2_{\tbf{L}_j} \times d^2_{\tbf{L}_j}$ matrix with components $M_{\tbf{v} \tbf{v}'}$. Note that
\eqn{
M_{\tbf{v} \tbf{v}'} &:=& \tr{ E_{\tbf{v}} E_{\tbf{v}'} } \nonumber \\
&=& \tr{ \right(E_{v_{1}} \otimes\dots \otimes E_{v_{N}} \left) \right(E_{v'_{1}} \otimes\dots \otimes E_{v'_{N}} \left)  } \nonumber \\
&=& \tr{E_{v_{1}}E_{v'_{1}}}\tr{E_{v_{2}}E_{v'_{2}}} \dots \tr{E_{v_{N}}E_{v'_{N}}} \nonumber \\
&:=& \prod_{n=1}^{N} \, M_{v_{n} v'_{n}} 
}
which implies that
\eqn{
\tbf{M} = \tbf{M}_{1} \otimes \tbf{M}_{2} \otimes \dots \otimes \tbf{M}_{N} \, ,
}
where $\tbf{M}_{n}$ is the usual $d^2_n \times d^2_n$ matrix of components $M_{v_{n} v'_{n}} = \tr{E_{v_n}E_{v'_n}}$ constructed from the SIC-POVM $\{E_{v_n} \}$ on the $n_{\trm{th}}$ sub-space. Since the inverse of a tensor product of matrices is the tensor product of the inverses, we have:
\eqn{
\tbf{M}^{-1}= \tbf{M}^{-1}_{1} \otimes \tbf{M}^{-1}_{2} \otimes \dots \otimes \tbf{M}^{-1}_{N} \, ,
}
or in component form:
\eqn{
M^{-1}_{\tbf{v}\tbf{v}'} = \prod_{n=1}^{N} \, M^{-1}_{v_{n}v'_{n}} \, .
}
For the special case of SICs we have:
\eqn{ \label{eqn:SICMs}
M_{v_{n} v'_{n}} &=& \tr{E_{v_n}E_{v'_n}} = \left[ \frac{d_n \delta_{v_n v'_n}+1}{d^2_n(d_n+1)} \right] \, , \nonumber \\
M^{-1}_{v_{n} v'_{n}} &=& \left[ d_n(d_n+1)\delta_{v_n v'_n}-1 \right] \, .
}
Therefore we can now write:
\eqn{ \label{eqn:alphabeta}
\alpha_{\tbf{r}}(\tbf{L}_{j-1}) &=& \zum{\tbf{v}}{} \, M^{-1}_{\tbf{r}\tbf{v}}  \, P(\tbf{v}|\tbf{L}_{j-1}) \nonumber \\
\beta_{\tbf{s}}(\tbf{L}_{j+1}) &=& \frac{1}{d_{\tbf{L}_j}} \zum{\tbf{v}'}{} \, M^{-1}_{\tbf{v}'\tbf{s}} \, P(\tbf{L}_{j+1}|\tbf{v}') \, .
}

Substituting \eqref{eqn:alphabeta} and \eqref{eqn:rhoDall} into \eqref{eqn:qi} gives us:
\eqn{ \label{eqn:qi2}
P(\tbf{L}_{j+1}|\brm{un}(\tbf{L}_j)) &=&  \zum{\tbf{r} \tbf{s}}{} \, \alpha_{\tbf{r}}(\tbf{L}_{j-1}) \, \beta_{\tbf{s}}(\tbf{L}_{j+1}) \, M_{\tbf{r} \tbf{s}} \nonumber \\
&=&  \frac{1}{d_{\tbf{L}_j}} \, \zum{\tbf{r} \tbf{s} \tbf{v} \tbf{v}'}{} \, M^{-1}_{\tbf{r}\tbf{v}} \, M^{-1}_{\tbf{v}'\tbf{s}} \, M_{\tbf{r} \tbf{s}} \,\, P(\tbf{v}|\tbf{L}_{j-1}) P(\tbf{L}_{j+1}|\tbf{v}')\nonumber \\
&=&  \frac{1}{d_{\tbf{L}_j}} \, \zum{\tbf{v} \tbf{v}'}{} \, \left[ \zum{\tbf{r} \tbf{s}}{} \, M^{-1}_{\tbf{r}\tbf{v}} \, M^{-1}_{\tbf{v}' \tbf{s}} \, M_{\tbf{r} \tbf{s}} \right] \, P(\tbf{v}|\tbf{L}_{j-1}) P(\tbf{L}_{j+1}|\tbf{v}') \, .
}
We can simplify this expression by reducing the term in the square brackets. To do that, notice that the summation can be brought inside the product:
\eqn{ \label{eqn:sqbrac}
 && \zum{\tbf{r} \tbf{s}}{} \, M^{-1}_{\tbf{r}\tbf{v}} \, M^{-1}_{\tbf{v}' \tbf{s}} \, M_{\tbf{r} \tbf{s}} = \zum{\tbf{r} \tbf{s}}{} \, \prod_{n=1}^{N} M^{-1}_{r_{n} v_{n}} \, M^{-1}_{v'_{n} s_{n}} \, M_{r_{n} s_{n}} \nonumber \\
 &:=& \zum{\tbf{r} \tbf{s}}{} \, \prod_{n=1}^{N} \, A_{r_{n} s_{n} v_{n} v'_{n}} \nonumber \\
 &=& \zum{r_1 s_1}{} \zum{r_2 s_2}{} \dots \zum{r_N s_N}{} \, \prod_{n=1}^{N} \, A_{r_{n} s_{n} v_{n} v'_{n}} \nonumber \\
 &=& \zum{r_1 s_1}{}\, A_{r_{1} s_{1} v_{1} v'_{1}} \, \zum{r_2 s_2}{} \, A_{r_{2} s_{2} v_{2} v'_{2}} \, \dots \, \zum{r_N s_N}{}\, A_{r_{N} s_{N} v_{N} v'_{N}} \nonumber \\
 &=&  \prod_{n=1}^{N} \, \zum{r_n s_n}{}\, A_{r_{n} s_{n} v_{n} v'_{n}} \nonumber \\
 &=& \prod_{n=1}^{N} \, \left[ \zum{r_n s_n}{}\,  M^{-1}_{r_{n} v_{n}} \, M^{-1}_{v'_{n} s_{n}} \, M_{r_{n} s_{n}} \right] \, .
}
The expression in square brackets can now be evaluated because $M_{v_{n} v'_{n}}$ and its inverse are composed of SIC-POVM elements on the relevant sub-space, i.e. are given by (\ref{eqn:SICMs}). Substituting in these expressions, we obtain (suppressing the $n$ sub-index):
\eqn{ \label{eqn:eval3M}
&& \zum{r s}{}\,  M^{-1}_{r v} \, M^{-1}_{v' s} \, M_{r s} \nonumber \\ 
&=& \frac{1}{d^2(d+1)} \,  \zum{r s}{} \,  \left[ d(d+1)\delta_{r v}-1 \right] \left[ d(d+1)\delta_{v' s}-1 \right] \left[d \delta_{r s}+1 \right] \nonumber \\
&=& \frac{1}{d^2(d+1)} \,  \zum{r s}{} \,  \left[ d^3(d+1)^2\delta_{r v}\delta_{v' s}\delta_{r s}-d^2(d+1)\delta_{v' s}\delta_{r s}-d^2(d+1)\delta_{r v}\delta_{r s} \right. \nonumber \\
&& \left. + d \delta_{r s} + d^2(d+1)^2\delta_{r v}\delta_{v' s}-d(d+1)\delta_{v' s}-d(d+1)\delta_{r v}+1 \right] \nonumber \\ 
&=& \frac{1}{d^2(d+1)} \, \left[ d^3(d+1)^2 (\delta_{v v'})-d^2(d+1)(1)-d^2(d+1)(1)+ d(d^2) \right. \nonumber \\
&& \left. + d^2(d+1)^2(1)(1)-d(d+1)(d^2)-d(d+1)(d^2)+1(d^4) \right] \nonumber \\
&=& \left[ d(d+1)\delta_{v v'}-1-1+\frac{d}{(d+1)} + (d+1)-d-d+\frac{d^2}{(d+1)} \right] \nonumber \\
&=& \left[ d_n(d_n+1)\delta_{v_n v'_n}-1 \right] \, .
}
Substituting this into (\ref{eqn:sqbrac}) gives:
\eqn{ 
\prod_{n=1}^{N} \, \left[ \zum{r_n s_n}{}\,  M^{-1}_{r_{n} v_{n}} \, M^{-1}_{v'_{n} s_{n}} \, M_{r_{n} s_{n}} \right] = \prod_{n=1}^{N} \, \left[ d_n(d_n+1)\delta_{v_n v'_n}-1  \right] \, ,
}
which can finally be substituted into (\ref{eqn:qi2}) to obtain the result:
\eqn{
P(\tbf{L}_{j+1}|\tbf{L}_{j-1} , \, \brm{un}(\tbf{L}_j)) &=& \frac{1}{d_{\tbf{L}_j}} \, \zum{\tbf{v} \tbf{v}'}{} \, \left( \prod_{n=1}^{N} \, \left[ d_n(d_n+1)\delta_{v_n v'_n}-1  \right] \right) \, P(\tbf{v}|\tbf{L}_{j-1}) P(\tbf{L}_{j+1}|\tbf{v'}) \, \nonumber \\
&=& \, \zum{\tbf{v} \tbf{v}'}{} \, P(\tbf{L}_{j+1}|\tbf{v}') \left( \prod_{n=1}^{N} \, \left[ (d_n+1)\delta_{v_n v'_n}-\frac{1}{d_n}  \right] \right) \, P(\tbf{v}|\tbf{L}_{j-1}) \, ,
}
which is Eq. \eqref{eqn:genurgall} as promised.

\section{Inference rule for un-measurement of an arbitrary subset of a layer \label{app:genurg2} }
In this section we generalize the result of the preceding Appendix to the case where only a subset of variables in the layer $\tbf{L}_j$ are un-measured. We partition the layer as $\tbf{L}_j := \{V_i : i=1,2,\dots ,K \} \cup \{V_i : i=1,2,\dots,N-K \} := \tbf{U} \cup \tbf{W}$, where $\tbf{U}$ are the variables to be un-measured. Hence $\tbf{U}$ takes values that are $K$-tuples $\tbf{u}:=(v_1,\dots,v_K)$ and $\tbf{W}$ takes values that are $(N-K)$-tuples $\tbf{w}:=(v_{K+1},\dots,v_{N})$. It will also be useful to define the variable $\tbf{V}$ covering the whole layer, having the $N$-tuple values $\tbf{v}:=(v_{1},\dots,v_{N})$. Starting with \eqref{eqn:genurgstart2}, 

\eqn{ \label{eqn:subsetstart}
P(\tbf{L}_{j+1}|\tbf{L}_{j-1} \tbf{W}, \, \brm{un}(\tbf{U})) &=& \tr{\left( \rho^{\tbf{U}}_{\tbf{L}_{j-1}}\otimes \Pi_{v_{K+1}} \otimes \cdots \otimes \Pi_{v_{N}} \right) D_{\tbf{L}_{j+1}} } \, , 
}

we expand $\rho^{\tbf{U}}_{\tbf{L}_{j-1}}$, $D_{\tbf{L}_{j+1}}$ as: 

\eqn{ \label{eqn:rhoDsubset}
\rho^{\tbf{U}}_{\tbf{L}_{j-1}} &=& \zum{\tbf{u}'}{} \, \alpha_{\tbf{u}'}(\tbf{L}_{j-1}) \, E_{\tbf{u}'} \, , \nonumber \\
D_{\tbf{L}_{j+1}} &=& \zum{\tbf{v}'}{} \, \beta_{\tbf{v}'}(\tbf{L}_{j+1}) \, E_{\tbf{v}'} \, .
}

where primes are used to denote the dummy indices, eg. $\tbf{u}'$ is a dummy index for $\tbf{u}$. Inverting the above equations (following the same procedure as in Appendix \ref{app:genurg1}) we obtain:

\eqn{ \label{eqn:alphabeta2}
\alpha_{\tbf{u}'}(\tbf{L}_{j-1}) &=& \zum{\tbf{u}}{} \, M^{-1}_{\tbf{u}' \tbf{u}}  \, P(\tbf{u}|\tbf{L}_{j-1}) \nonumber \\
\beta_{\tbf{v}'}(\tbf{L}_{j+1}) &=& \frac{1}{d_{\tbf{L}_j}} \zum{\tbf{v}''}{} \, M^{-1}_{\tbf{v}'' \tbf{v}'} \, P(\tbf{L}_{j+1}|\tbf{v}'') \, .
}

Substituting \eqref{eqn:rhoDsubset} into \eqref{eqn:subsetstart} and expanding $\alpha_{\tbf{u}'}(\tbf{L}_{j-1})$, $\beta_{\tbf{v}'}(\tbf{L}_{j+1}) $ using \eqref{eqn:alphabeta2}, and identifying $\tr{(E_{\tbf{u}'}\otimes E_{\tbf{w}}) E_{\tbf{v}'}} := M_{\tbf{v} \tbf{v}'}$, we obtain:
\eqn{
P(\tbf{L}_{j+1}|\tbf{L}_{j-1} \tbf{W} , \, \brm{un}(\tbf{U}) ) &=& 
\zum{\tbf{u}'}{} \zum{\tbf{u}}{} \zum{\tbf{v}'}{} \zum{\tbf{v}''}{}  \, \frac{1}{d_{\tbf{u}}} M^{-1}_{\tbf{u}' \tbf{u}} \, M^{-1}_{\tbf{v}'' \tbf{v}'} \, M_{\tbf{v} \tbf{v}'}  P(\tbf{u}|\tbf{L}_{j-1}) P(\tbf{L}_{j+1}|\tbf{v}'') \, ,
}
and we can use the procedure \eqref{eqn:sqbrac} to move the summations of $\tbf{u}',\tbf{v}'$ inside the product:
\eqn{
&=& \zum{\tbf{u} \tbf{v}''}{}  \, \frac{1}{d_{\tbf{u}}} \, \prod^{K}_{n=1} \, \left[ \zum{u'_n v'_n}{} M^{-1}_{v'_{n} v_{n}} \, M^{-1}_{v''_{n} v'_{n} } \, M_{v_n v'_n} \right] \, \prod^{N}_{m=K+1} \, \left[ \zum{v'_m}{} M^{-1}_{v''_{m} v'_{m} } \, M_{v_m v'_m} \right] \, P(\tbf{u}|\tbf{L}_{j-1}) P(\tbf{L}_{j+1}|\tbf{v}'') \, .
}
The sums inside first product can be evaluated using \eqref{eqn:eval3M}, while the sum inside the second product are (suppressing the $m$ sub-index): 
\eqn{ \label{eqn:eval2M}
&& \zum{v'}{} \, M^{-1}_{v'' v'} \, M_{v v'} \nonumber \\ 
&=& \zum{v'}{} \, \left[ d \left(d+1 \right) \delta_{v'' v'}-1 \right]\left[ \frac{d \delta_{v v'} +1}{d^2(d+1)} \right] \, \nonumber \\ 
&=& \zum{v'}{} \, \left[ \delta_{v'' v'} \delta_{v v'} + \frac{\delta_{v'' v'}}{d}-\frac{\delta_{v v'}}{d(d+1)}-\frac{1}{d^2(d+1)}  \right] \, \nonumber \\
&=& \delta_{v'' v} \, .
}
Hence,
\eqn{
P(\tbf{L}_{j+1}|\tbf{L}_{j-1} \tbf{w}, \,  \brm{un}(\tbf{U}) ) 
&=& \zum{\tbf{u} }{}\zum{ \tbf{v}''}{}  \, \frac{1}{d_{\tbf{u}}} \,  \left( \prod^{K}_{n=1} \, \left[ d_n(d_n+1)\delta_{v_n v''_n}-1 \right] \, \prod^{N}_{m=K+1} \, \delta_{v''_m v_m}  \right) \, P(\tbf{u}|\tbf{L}_{j-1}) P(\tbf{L}_{j+1}|\tbf{v}'') \, \nonumber \\
&=& \zum{\tbf{u}}{} \zum{\tbf{u}'' \tbf{w}''}{}  \,  \left( \prod^{K}_{n=1} \, \left[ (d_n+1)\delta_{v_n v''_n}-\frac{1}{d_n} \right] \, \delta_{\tbf{w}'' \tbf{w}} \right) \, P(\tbf{u}|\tbf{L}_{j-1}) P(\tbf{L}_{j+1}|\tbf{u}'' \tbf{w}'') \, \nonumber \\
&=& \zum{\tbf{u} \tbf{u}''}{}  \,  \left( \prod^{K}_{n=1} \, \left[ (d_n+1)\delta_{v_n v''_n}-\frac{1}{d_n} \right] \right) \, P(\tbf{u}|\tbf{L}_{j-1}) P(\tbf{L}_{j+1}|\tbf{u}'' \tbf{w}) \, \nonumber \\
&=& \zum{(v_1 \dots v_K)}{} \zum{(v''_1 \dots v''_K)}{}  \,  \left( \prod^{K}_{n=1} \, \left[ (d_n+1)\delta_{v_n v''_n}-\frac{1}{d_n} \right] \right) \, P(v_1 \dots v_K|\tbf{L}_{j-1}) P(\tbf{L}_{j+1}|v''_1 \dots v''_K v_{K+1} \dots v_{N}) \, \nonumber \\
&=& \zum{\tbf{u}' \tbf{u}}{}  \, P(\tbf{L}_{j+1}|\tbf{u}\tbf{w}) \, \left( \prod^{K}_{n=1} \, \left[ (d_n+1)\delta_{v'_n v_n}-\frac{1}{d_n} \right] \right) \, P(\tbf{u}'|\tbf{L}_{j-1}) \, ,
}
where in the last line we used judicious re-labelling of the indices to obtain the same form as Eq. \eqref{eqn:genurgsubset}.

\section{Preservation of conditional independences by un-measurements \label{app:qunmeas}}

Here we provide evidence for the conjecture in Sec. \ref{sec:qunmeas} that if $P(\tbf{X},Z)$ satisfies the \tbf{QMC} relative to $G(\tbf{X},Z)$, then $P(\tbf{X}| \brm{un}(Z))$ satisfies the \tbf{QMC} relative to the new DAG $G(\tbf{X}|\brm{un}(Z))$ given by the definition \tbf{UNDAG}. Given the ansatz \eqref{eqn:unmeasansatz}, we may restrict attention to the layers immediately preceding and following $\tbf{L}_j$ (which contains the variables to be un-measured). Our strategy is to use the decomposition
\eqn{
P(\tbf{L}_{j+1} \, \tbf{W} \, \tbf{L}_{j-1} | \brm{un}(\tbf{U})) &=& P(\tbf{L}_{j+1} | \tbf{W} \, \tbf{L}_{j-1}, \, \brm{un}(\tbf{U})) \, P(\tbf{W} \, \tbf{L}_{j-1} | C^{\tbf{U}}=\oslash) \, 
}
(due to \tbf{CNS}), and then expand $P(\tbf{L}_{j+1} | \tbf{W} \, \tbf{L}_{j-1}, \, \brm{un}(\tbf{U}))$ in terms of reference probabilities using the inference rule \eqref{eqn:genurgsubset}. 

\tit{Notation:} We will use the standard notation $(\tbf{X} \perp \tbf{Y} | \tbf{Z} )$ to mean that $X$ and $Y$ are independent conditional on $\tbf{Z}$, i.e. that $P(\tbf{X}|\tbf{Y} \tbf{Z}) = P(\tbf{X}|\tbf{Z})$. Conditional independence relations of this form are subject to the \tit{semi-graphoid axioms}:\\

\noindent \tbf{SG1.} Symmetry: $(X \perp Y |Z) \Leftrightarrow (Y \perp X |Z)$  \\
\tbf{SG2.} Decomposition:  $(X \perp YW |Z) \Rightarrow (X \perp Y |Z)$   \\
\tbf{SG3.} Weak union:  $(X \perp YW |Z) \Rightarrow (X \perp Y |ZW)$   \\
\tbf{SG4.} Contraction: $(X \perp Y |ZW) \& (X \perp W |Z) \Rightarrow (X \perp YW |Z)$  .\\

For convenience, we will re-label $\tbf{L}_{j+1}:=\tbf{L}$, $\tbf{L}_{j-1}:=\tbf{M}$, and let $\tbf{L}_1,\tbf{L}_2,\tbf{L}_3$ be arbitrary disjoint subsets of $\tbf{L}$, with similar definitions for $\tbf{W},\tbf{M}$. As usual, we have $\tbf{L}_j := \tbf{V} := \tbf{U}\cup \tbf{W}$, and we will omit the conditional $C^{\tbf{U}}=\oslash$ from the reference probabilities. Finally, to simplify the expressions, it is useful to define:
\eqn{
\Delta_{\tbf{u} \tbf{u}'} :=  \prod^{K}_{n=1} \, \left[ (d_n+1)\delta_{v'_n v_n}-\frac{1}{d_n} \right] \, .
}
We now prove the following theorem:\\

\tbf{Theorem:} A conditional independence relation $R \in P(\tbf{L} \, \tbf{W} \, \tbf{M})$ also holds in $P(\tbf{L} \, \tbf{W} \, \tbf{M} | \brm{un}(\tbf{U}))$ whenever either (i) $R$ involves only two out of the three sets $\tbf{L} , \tbf{W} , \tbf{M}$; or (ii) $R$ consists of the three sets in some permutation. That is, whenever one of the following cases applies:\\

\noindent \tbf{Case 1.} $R = (\tbf{L}_1 \, \tbf{W}_1 \, \perp \tbf{L}_2 \, \tbf{W}_2| \tbf{L}_3 \, \tbf{W}_3)$ ; \\
\tbf{Case 2.} $R = (\tbf{W}_1 \, \tbf{M}_1 \, \perp \tbf{W}_2 \, \tbf{M}_2| \tbf{W}_3 \, \tbf{M}_3)$; \\
\tbf{Case 3.} $R = (\tbf{L}_1 \, \tbf{M}_1 \, \perp \tbf{L}_2 \, \tbf{M}_2| \tbf{L}_3 \, \tbf{M}_3)$; \\
\tbf{Case 4.} $R = (\tbf{L}_1 \perp \tbf{W}_2 | \tbf{M}_3)$ ; \\
\tbf{Case 5.} $R = (\tbf{M}_3 \perp \tbf{L}_1 | \tbf{W}_2)$ ; \\
\tbf{Case 6.} $R = (\tbf{W}_2 \perp \tbf{M}_3 | \tbf{L}_1)$ . \\

The following Lemmas will be useful.

\tbf{Lemma 1.} Using the property of `eternal noise' \tbf{EN} (Sec. \ref{sec:discuss}) we have $\zum{\tbf{u}'}{} \, \Delta_{\tbf{u} \tbf{u}'} P(\tbf{u}') = P(\tbf{u}).$\\
\tit{Proof:} 
\eqn{
\zum{\tbf{u}'}{} \, \Delta_{\tbf{u} \tbf{u}'} P(\tbf{u}') &=& \zum{\tbf{u}'}{} \, \left( \prod^{K}_{n=1} \, \left[ (d_n+1)\delta_{v'_n v_n}-\frac{1}{d_n} \right] \right) \, P(\tbf{u}') \nonumber \\
&=& \left( \prod^{K}_{n=1} \, \zum{v'_n}{} \,  \left[ (d_n+1)\delta_{v'_n v_n}-\frac{1}{d_n} \right] \right) \, \prod_n \, P(v'_n)  \nonumber \\
&=& \prod^{K}_{n=1} \, \left[ (d_n+1)P(v_n)-\frac{1}{d_n} \right]  \nonumber \\
&=& \prod^{K}_{n=1} \, \left[ (d_n+1)\frac{1}{d^2_n}-\frac{1}{d_n} \right] \nonumber \\
&=& \prod^{K}_{n=1} \, \frac{1}{d^2_n} = P(\tbf{u}) \, . \qquad \Box 
}

\tbf{Lemma 2.} For any $\tbf{L}_1 \subseteq \tbf{L}$, we have $P(\tbf{L}_1|\tbf{U} \tbf{W}) = P(\tbf{L}_1|\pa{\tbf{L}_1})$.\\
\tit{Proof:} Since these are reference probabilities ($C^{\tbf{U}}=\oslash$) they satisfy the \tbf{QMC} relative to the given LDAG. We see that any path connecting $\tbf{L}_1$ to one of its non-parents $V \in \tbf{V}$ in the preceding layer must either contain an unconditioned collider in $\tbf{L}$ or an unconditioned fork in $\tbf{M}$ and hence be blocked. Hence $(\tbf{L}_1 \perp V )$ for all non-parents $V \in \tbf{V}$. $\Box$

\tbf{Lemma 3.} For any $\tbf{M}_3 \subseteq \tbf{M}$ we have $P(\tbf{U}|\tbf{M}_{3}) = P(\tbf{U}_{3}|\tbf{M}_3) P(\tbf{U}_R)$ where $\tbf{U}_3 := \ch{\tbf{M}_3} \cap \tbf{U}$ are the children of $M_3$ in $\tbf{U}$, and $\tbf{U}_R$ its complement in $\tbf{U}$. \\ 
\tit{Proof:} Note that $P(\tbf{U}|\tbf{M}_{3})=P(\tbf{U}_R|\tbf{U}_3 \tbf{M}_{3})P(\tbf{U}_3 |\tbf{M}_{3})$. Then note that every path from $\tbf{U}_R$ to $\tbf{U}_3$ and hence also to $\tbf{M}_3$ must contain an unconditioned collider in $\tbf{L}$ or an unconditioned fork in $\tbf{M}$ and hence be blocked. Hence $(\tbf{U}_R \perp \tbf{U}_3 \tbf{M}_{3} )$, so $P(\tbf{U}|\tbf{M}_{3})=P(\tbf{U}_R)P(\tbf{U}_3 |\tbf{M}_{3})$. $\Box$\\

\tbf{Lemma 4.} For any disjoint subsets $\tbf{V}_1,\tbf{V}_2,\tbf{M}_1,\tbf{M}_2$ of $\tbf{V},\tbf{M}$ respectively, we have: 
\eqn{
\zum{\tbf{M}_2}{} \, P(\tbf{V}_1 |\tbf{M}_1 \, \tbf{M}_2) P(\tbf{V}_2 | \tbf{M}_2) P( \tbf{M}_2) 
&=& P(\tbf{V}_1 |\tbf{M}_1 ) P(\tbf{V}_2) 
}
\tit{Proof:} By definition we can write:
\eqn{ \label{eqn:lem4whole}
\zum{\tbf{M}_2}{} \, P(\tbf{V}_1 |\tbf{M}_1 \, \tbf{M}_2) P(\tbf{V}_2 | \tbf{M}_2) P(\tbf{M}_2) &:=& \zum{\tbf{M}_2}{} \, \tr{ \rho_{\tbf{M}_1,\tbf{M}_2} \cdot E_{\tbf{V}_1} } \tr{ \rho_{\tbf{M}_2} \cdot E_{\tbf{V}_2} }  P(\tbf{M}_2) \nonumber \\
&:=& \zum{\tbf{M}_2}{} \, \tr{ \rho'_{\tbf{M}_1,\tbf{M}_2} \cdot \left( E_{\tbf{V}_1} \otimes E_{\tbf{V}_2} \right) } P(\tbf{M}_2) \, ,
}
where
\eqn{ \label{eqn:lem4rhopart}
\rho'_{\tbf{M}_1,\tbf{M}_2} := \mathcal{T}( \Pi_{\tbf{M}_1} \otimes \Pi_{\tbf{M}_2})
}
for some unbiased channel $\mathcal{T}:\mathcal{H}_{\tbf{M}_1 \tbf{M}_2 } \mapsto \mathcal{H}_{\tbf{V}_1 \tbf{V}_2 }$, and where 
\eqn{
\Pi_{\tbf{V}} := \otimes_{V_i \in \tbf{V}} \, \Pi_{V_i} \, , \nonumber \\
E_{\tbf{V}} := \otimes_{V_i \in \tbf{V}} \, \frac{1}{d_{V_i}} \Pi_{V_i} \, , \nonumber
}
with $\Pi_{V_i}$ the SIC-projectors on the subspace $\mathcal{H}_{V_i}$. Substituting \eqref{eqn:lem4rhopart} into \eqref{eqn:lem4whole} and using \tbf{EN} to decompose $P(\tbf{M}_2)$, we obtain:
\eqn{
\zum{\tbf{M}_2}{} \, P(\tbf{V}_1 |\tbf{M}_1 \, \tbf{M}_2) P(\tbf{V}_2 | \tbf{M}_2) P(\tbf{M}_2)  
&=& \tr{  \mathcal{T}( \Pi_{\tbf{M}_1} \otimes \frac{1}{d_{\tbf{M}_2}} \mathbb{I}) \cdot \left( E_{\tbf{V}_1} \otimes E_{\tbf{V}_2} \right) } \nonumber \\
&=& \tr{  \mathcal{T}( \Pi_{\tbf{M}_1}) \cdot E_{\tbf{V}_1} }\, \tr{\frac{1}{d_{\tbf{V}_2} }\mathbb{I} \cdot  E_{\tbf{V}_2} } \nonumber \\
&=& P(\tbf{V}_1 |\tbf{M}_1) P(\tbf{V}_2) \,. \qquad \Box \, 
}

\tbf{Lemma 5.} Let $\tbf{L}_1$, $\tbf{W}_2$, $\tbf{M}_3$ denote arbitrary subsets and $\tbf{L}_{R1}$, $\tbf{W}_{R2}$, $\tbf{M}_{R3}$ their complements in $\tbf{L},\, \tbf{W}, \, \tbf{M}$ respectively. Then:
\eqn{
P(\tbf{L}_1 | \tbf{W}_2 \, \tbf{M}_3 , \, \brm{un}(\tbf{U})) &=& \zum{\tbf{L}_{R1} \tbf{W}_{R2} \tbf{M}_{R3}}{} \, P(\tbf{L} | \tbf{W} \, \tbf{M} , \, \brm{un}(\tbf{U})) P(\tbf{W}_{R2} \, \tbf{M}_{R3}) \nonumber \\
&=& \zum{\tbf{W}_{R2}}{} \zum{\tbf{u}' \tbf{u}}{}  \, P(\tbf{L}_1|\tbf{u} \, \tbf{W}) \, \Delta_{\tbf{u} \tbf{u}'} \, \zum{\tbf{M}_{R3}}{} \, P(\tbf{u}'|\tbf{M}_{R3} \, \tbf{M}_3) P(\tbf{W}_{R2}| \tbf{M}_{R3}) P(\tbf{M}_{R3}) \nonumber \\
&=& \zum{\tbf{u}' \tbf{u}}{}   \zum{\tbf{W}_{R2}}{} \, P(\tbf{L}_1|\tbf{u} \, \tbf{W}_2 \, \tbf{W}_{R2}) P(\tbf{W}_{R2}) \, \Delta_{\tbf{u} \tbf{u}'} \, \zum{\tbf{M}_{R3}}{} \, P(\tbf{u}'|\tbf{M}_3)  P(\tbf{M}_{R3})\nonumber \\
&=& \zum{\tbf{u}' \tbf{u}}{}  \, P(\tbf{L}_1|\tbf{u} \, \tbf{W}_2 ) \, \Delta_{\tbf{u} \tbf{u}'} \, P(\tbf{u}'|\tbf{M}_3)  \, . \qquad \Box
}
Note: in the first line we used $\tbf{IUM}$ (Sec. \ref{sec:qmc}) to identify $P(\tbf{W}_{R2} \tbf{M}_{R3}|\brm{un}(\tbf{U})):=P(\tbf{W}_{R2} \tbf{M}_{R3})$; in the third line we used \tbf{Lemma 4}; and in the last line we used the fact that $P(\tbf{W}_{R2})=P(\tbf{W}_{R2}|\tbf{U} \tbf{W}_2)$.\\

\tbf{Lemma 6.} For any disjoint subsets $\tbf{U}_1,\tbf{U}_2$ of $\tbf{U}$ and $\tbf{M}_2$ of $\tbf{M}$, we have $\zum{\tbf{u}_1 \tbf{u}'_1}{} \, \Delta_{\tbf{u}_1 \tbf{u}'_1} P(\tbf{u}'_1 \tbf{U}'_2|\tbf{M}_3) = P(\tbf{U}'_2|\tbf{M}_3).$ 
\\
\tit{Proof:} For convenience, let us label the variables $V_i \in \tbf{U}$ such that $\tbf{U}_1 = \{ V_i : i=1,\dots,K_1 \}$ for some $K_1 \leq K$. Then: 
\eqn{
\zum{\tbf{u}_1 \tbf{u}'_1}{} \, \Delta_{\tbf{u}_1 \tbf{u}'_1} P(\tbf{u}'_1 \tbf{U}'_2|\tbf{M}_3) &=& \zum{\tbf{u}_1 \tbf{u}'_1}{} \, \left( \prod^{K_1}_{n} \,  \left[ (d_n+1)\delta_{v'_n v_n}-\frac{1}{d_n} \right] \right) P(\tbf{u}'_1 \tbf{U}'_2|\tbf{M}_3)  \nonumber \\
&=& \zum{\tbf{u}'_1}{} \, \left( \prod^{K_1}_{n} \, \zum{v_n}{}\, \left[ (d_n+1)\delta_{v'_n v_n}-\frac{1}{d_n} \right] \right) P(\tbf{u}'_1 \tbf{U}'_2|\tbf{M}_3)  \nonumber \\
&=& \zum{\tbf{u}'_1}{} \, \left( \prod^{K_1}_{n} \, \left[ (d_n+1)(1)-\frac{d^2_n}{d_n} \right] \right) P(\tbf{u}'_1 \tbf{U}'_2|\tbf{M}_3)  \nonumber \\
&=& \zum{\tbf{u}'_1}{} \, \left( 1 \right) P(\tbf{u}'_1 \tbf{U}'_2|\tbf{M}_3)  \nonumber \\
&=& P(\tbf{U}'_2|\tbf{M}_3) \, . \qquad \Box 
}

We are now ready to prove separately each of the cases 1-6 above.\\

\tbf{Case 1.} $R = (\tbf{L}_1 \, \tbf{W}_1 \, \perp \tbf{L}_2 \, \tbf{W}_2| \tbf{L}_3 \, \tbf{W}_3)$. \\
The proof proceeds by showing the stronger result that $P(\tbf{L} \tbf{W} | \brm{un}(\tbf{U})) = P(\tbf{L} \tbf{W})$. 
\eqn{
P(\tbf{L} \tbf{W} | \brm{un}(\tbf{U})) &=& P(\tbf{L} | \tbf{W}, \, \brm{un}(\tbf{U})) P(\tbf{W}) \nonumber \\
&=& \zum{\tbf{u}' \tbf{u}}{}  \, P(\tbf{L}|\tbf{u}\tbf{W}) \, \Delta_{\tbf{u} \tbf{u}'} \, P(\tbf{u}')P(\tbf{W}) \nonumber \\
&=& \zum{\tbf{u}}{}  \, P(\tbf{L}|\tbf{u}\tbf{W}) \, P(\tbf{u}) P(\tbf{W}) \nonumber \\
&=& P(\tbf{L} \, \tbf{W}) \, .
}
where we used \tbf{Lemma 1} in line 3. Since any $R$ with the assumed form holds in $P(\tbf{L} \tbf{W})$, this shows that it holds in $P(\tbf{L} \tbf{W} | \brm{un}(\tbf{U}))$ as well. $\Box$

\tbf{Case 2.} $R = (\tbf{M}_1 \, \tbf{W}_1 \, \perp \tbf{M}_2 \, \tbf{W}_2| \tbf{M}_3 \, \tbf{W}_3)$. \\
The proof proceeds by noting that $\tbf{M},\tbf{W}$ are non-descendants of $\tbf{U}$, hence by \tbf{CNS} we have that $P(\tbf{M} \tbf{W} | \brm{un}(\tbf{U})) = P(\tbf{M} \tbf{W})$. Again, since any $R$ of the assumed form holds in $P(\tbf{M} \tbf{W})$, this shows that it holds also in $P(\tbf{M} \tbf{W} | \brm{un}(\tbf{U}))$. $\Box$

\tbf{Case 3.} $R = (\tbf{L}_1 \, \tbf{M}_1 \, \perp \tbf{L}_2 \, \tbf{M}_2| \tbf{L}_3 \, \tbf{M}_3)$. \\
Using the \tbf{QMC}, an independence of the form $R$ can only hold in an LDAG with the following properties (where `$\rightarrow$' denotes a directed path):\\
(i) There can't be $\tbf{M}_1 \rightarrow \tbf{L}_2$;\\
(ii) There can't be $\tbf{M}_2 \rightarrow \tbf{L}_1$;\\
(iii) There can't be both $\tbf{M}_3 \rightarrow \tbf{L}_1$ \tit{and} $\tbf{M}_3 \rightarrow \tbf{L}_2$;\\
(iv) There can't be both $\tbf{M}_1 \rightarrow \tbf{L}_3$ \tit{and} $\tbf{M}_2 \rightarrow \tbf{L}_3$.\\

Since (iii) and (iv) each present two mutually exclusive possibilities, we expect there to be four possible classes of LDAG structure that can supporting the relation $R$. These are divided into two pairs, related to each other by interchanging of the labels $1 \leftrightarrow 2$. By the Symmetry property of conditional independences, any results obtained for one pair can be automatically carried over to the other pair, so we can restrict attention without loss of generality to just the following two classes of LDAGs:\\

\noindent (A). The class of LDAGs having $\tbf{M}_1 \rightarrow \tbf{L}_3$ and $\tbf{M}_3 \rightarrow \tbf{L}_1$;\\
(B). The class of LDAGs having $\tbf{M}_1 \rightarrow \tbf{L}_3$ and $\tbf{M}_3 \rightarrow \tbf{L}_2$.\\

\begin{figure}[!htb]
\centering\includegraphics[width=0.5\linewidth]{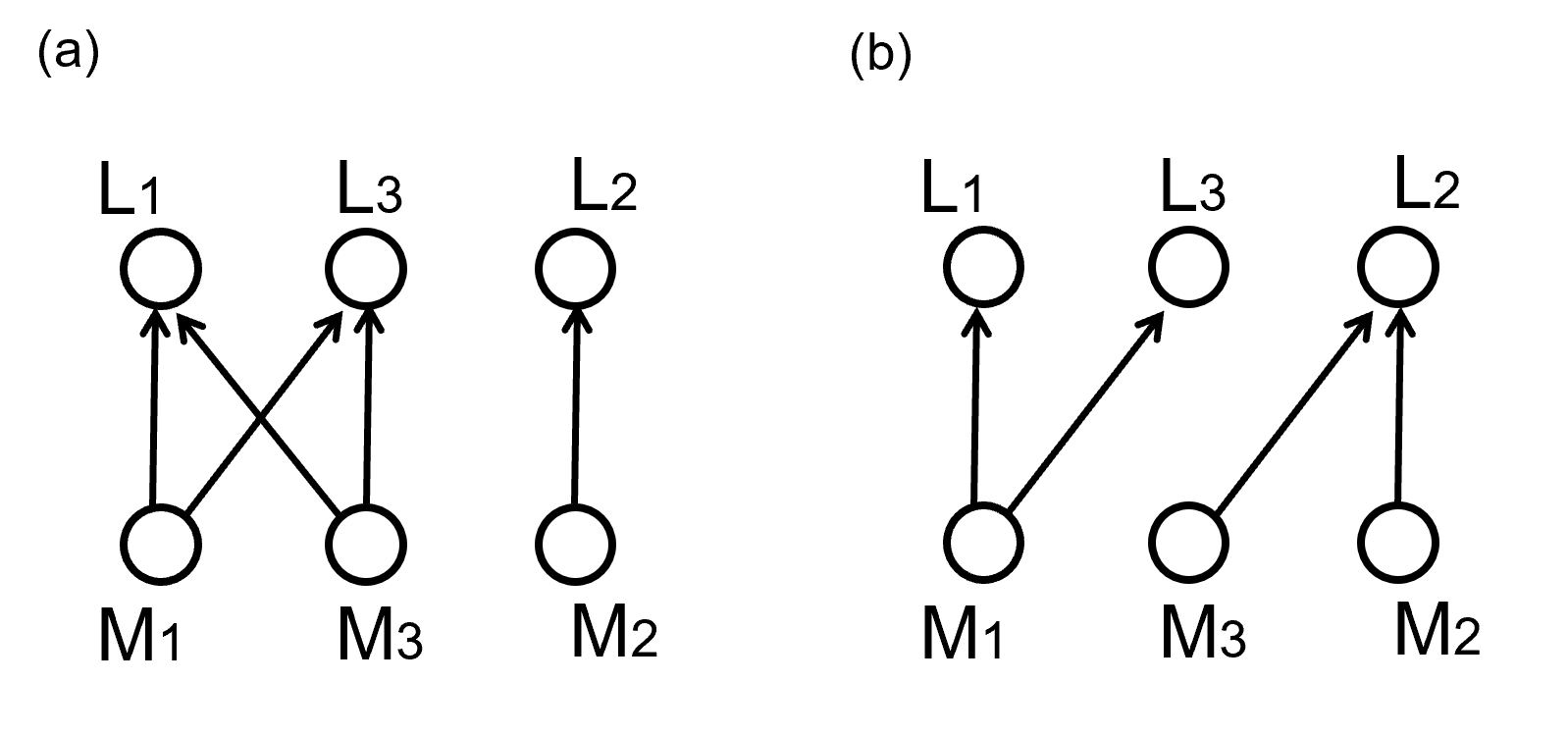}
\caption{Schematic of the two types of LDAG that can support a conditional independence of the form $R = (\tbf{L}_1 \, \tbf{M}_1 \, \perp \tbf{L}_2 \, \tbf{M}_2| \tbf{L}_3 \, \tbf{M}_3)$. Note: the arrows are indirect causes, as they are intercepted by the intervening layer \tbf{V} (not shown).}
\label{fig:classAB}
\end{figure}

Assuming that we include all paths that are not expressly forbidden by (i)-(iv) above, these classes may be characterized by the diagrams in Fig. \ref{fig:classAB}. For both classes, we begin by noting that:
\eqn{ \label{eqn:case3}
P(\tbf{L}_1 \, \tbf{L}_2 \, \tbf{L}_3 \, \tbf{M}_1 \, \tbf{M}_2 \, \tbf{M}_3 | \brm{un}(\tbf{U})) &=& \, P(\tbf{L}_1 \, \tbf{L}_2 \, \tbf{L}_3 | \tbf{M}_1 \, \tbf{M}_2 \, \tbf{M}_3 , \, \brm{un}(\tbf{U}))P(\tbf{M}_1 \, \tbf{M}_2 \, \tbf{M}_3) \nonumber \\
&=& \zum{ \tbf{u} \tbf{u}'}{} \, P(\tbf{L}_1 \, \tbf{L}_2 \, \tbf{L}_3 | \tbf{u}) \Delta_{\tbf{u} \tbf{u}'} P(\tbf{u}'|\tbf{M}_1 \, \tbf{M}_2 \, \tbf{M}_3) \, P(\tbf{M}_1 \, \tbf{M}_2 \, \tbf{M}_3) \, ,
}
where in line 2 we used \tbf{Lemma 5} with $\tbf{W}_3 = \emptyset$. We now consider each of the possibilities (A),(B) separately.\\

\tit{Possibility (A):} In this case, shown in Fig. \ref{fig:classAB}(a), we may partition $\tbf{U}=\tbf{U}_2 \tbf{U}_R$ where $\tbf{U}_2$ are all members of $\tbf{U}_2$ that are children of $\tbf{M}_2$ and parents of $\tbf{L}_2$, and where $\tbf{U}_R$ are the rest. Applying the \tbf{QMC} to this structure we see that $(\tbf{U}_2 \tbf{M}_2 \perp \tbf{U}_R \tbf{M}_1 \tbf{M}_3 )$ and also $(\tbf{L}_2 \tbf{U}_2 \perp \tbf{L}_1 \tbf{L}_3 \tbf{U}_R)$, from which it follows that $P(\tbf{U}'_2 \tbf{U}'_R|\tbf{M}_1 \, \tbf{M}_2 \, \tbf{M}_3)=P(\tbf{U}'_2|\tbf{M}_2) P(\tbf{U}'_R|\tbf{M}_1 \, \tbf{M}_3)$ and $P(\tbf{L}_1 \, \tbf{L}_2 \, \tbf{L}_3 | \tbf{U}_2 \, \tbf{U}_R )=P(\tbf{L}_2 | \tbf{U}_2) P(\tbf{L}_1 \, \tbf{L}_3 | \tbf{U}_R )$. Substituting these into \eqref{eqn:case3} and using the identities $\Delta_{\tbf{U} \tbf{U}'}=\Delta_{\tbf{U}_2 \tbf{U}'_2}\Delta_{\tbf{U}_R \tbf{U}'_R}$ and $P(\tbf{M}_1 \, \tbf{M}_2 \, \tbf{M}_3)=P(\tbf{M}_1)P( \tbf{M}_2) P(\tbf{M}_3)$, we obtain:
\eqn{
P(\tbf{L}_1 \, \tbf{L}_2 \, \tbf{L}_3 \, \tbf{M}_1 \, \tbf{M}_2 \, \tbf{M}_3 | \brm{un}(\tbf{U})) &=& \zum{ \tbf{u}_2 \tbf{u}'_2}{}\zum{ \tbf{u}_R \tbf{u}'_R}{} \, P(\tbf{L}_2 | \tbf{u}_2) P(\tbf{L}_1 \, \tbf{L}_3 | \tbf{u}_R ) \Delta_{\tbf{u}_2 \tbf{u}'_2}\Delta_{\tbf{u}_R \tbf{u}'_R} \times \nonumber \\ 
&& P(\tbf{u}'_2|\tbf{M}_2) P(\tbf{u}'_R|\tbf{M}_1 \, \tbf{M}_3) \, P(\tbf{M}_1)P( \tbf{M}_2) P(\tbf{M}_3) \nonumber \\
&=& \left( \zum{ \tbf{u}_2 \tbf{u}'_2}{} \, P(\tbf{L}_2 | \tbf{u}_2) \Delta_{\tbf{u}_2 \tbf{u}'_2} P(\tbf{u}'_2|\tbf{M}_2) P( \tbf{M}_2) \right)  \times \nonumber \\
&& \left(\zum{ \tbf{u}_R \tbf{u}'_R}{} \,  P(\tbf{L}_1 \, \tbf{L}_3 | \tbf{u}_R ) \Delta_{\tbf{u}_R \tbf{u}'_R}  P(\tbf{u}'_R|\tbf{M}_1 \, \tbf{M}_3) \, P(\tbf{M}_1) P(\tbf{M}_3) \right) \nonumber \\
&=& P(\tbf{L}_2 \, \tbf{M}_2 | \brm{un}(\tbf{U}))P(\tbf{L}_1 \, \tbf{L}_3 \, \tbf{M}_1 \, \tbf{M}_3 | \brm{un}(\tbf{U})) \, .
}
The latter distribution is guaranteed to satisfy $(\tbf{L}_2 \, \tbf{M}_2 \perp \tbf{L}_1 \, \tbf{M}_1 \, \tbf{L}_3 \, \tbf{M}_3)$ and hence, by axioms \tbf{SG1},\tbf{SG3}, also $(\tbf{L}_1 \, \tbf{M}_1 \perp \tbf{L}_2 \, \tbf{M}_2 | \tbf{L}_3 \, \tbf{M}_3)$ which is what we aimed to prove.\\

\tit{Possibility (B):} In this case, as shown in Fig. \ref{fig:classAB}(b), we may partition $\tbf{U}=\tbf{U}_1 \tbf{U}_R$ where $\tbf{U}_1$ are all members of $\tbf{U}_2$ that are children of $\tbf{M}_1$ and parents of $\tbf{L}_1 \cup \tbf{L}_3$, and where $\tbf{U}_R$ are the rest. Applying the \tbf{QMC} to this structure we see that $(\tbf{U}_1 \tbf{L}_1 \tbf{L}_3 \perp \tbf{U}_R \tbf{L}_2 )$ and also $(\tbf{U}_1 \tbf{M}_1 \perp \tbf{U}_R \tbf{M}_2 \tbf{M}_3)$, from which it follows that $P(\tbf{U}'_1 \tbf{U}'_R|\tbf{M}_1 \, \tbf{M}_2 \, \tbf{M}_3)=P(\tbf{U}'_1|\tbf{M}_1) P(\tbf{U}'_R|\tbf{M}_2 \, \tbf{M}_3)$ and $P(\tbf{L}_1 \, \tbf{L}_2 \, \tbf{L}_3 | \tbf{U}_1 \, \tbf{U}_R )=P(\tbf{L}_2 | \tbf{U}_R) P(\tbf{L}_1 \, \tbf{L}_3 | \tbf{U}_1 )$. Substituting these into \eqref{eqn:case3} and using the same tricks as before, we obtain:
\eqn{
P(\tbf{L}_1 \, \tbf{L}_2 \, \tbf{L}_3 \, \tbf{M}_1 \, \tbf{M}_2 \, \tbf{M}_3 | \brm{un}(\tbf{U})) &=& \zum{ \tbf{u}_1 \tbf{u}'_1}{}\zum{ \tbf{u}_R \tbf{u}'_R}{} \, P(\tbf{L}_2 | \tbf{u}_R) P(\tbf{L}_1 \, \tbf{L}_3 | \tbf{u}_1 ) \Delta_{\tbf{u}_1 \tbf{u}'_1}\Delta_{\tbf{u}_R \tbf{u}'_R} \times \nonumber \\
&& P(\tbf{u}'_1|\tbf{M}_1) P(\tbf{u}'_R|\tbf{M}_2 \, \tbf{M}_3) \, P(\tbf{M}_1)P( \tbf{M}_2) P(\tbf{M}_3) \nonumber \\
&=&\left( \zum{ \tbf{u}_1 \tbf{u}'_1}{} \, P(\tbf{L}_1 \, \tbf{L}_3 | \tbf{u}_1 ) \Delta_{\tbf{u}_1 \tbf{u}'_1} P(\tbf{u}'_1|\tbf{M}_1) P(\tbf{M}_1) \right) \times \nonumber \\
&& \left( \zum{ \tbf{u}_R \tbf{u}'_R}{} \, P(\tbf{L}_2 | \tbf{u}_R) \Delta_{\tbf{u}_R \tbf{u}'_R}  P(\tbf{u}'_R|\tbf{M}_2 \, \tbf{M}_3) \, P( \tbf{M}_2) P(\tbf{M}_3) \right) \nonumber \\
&=& P(\tbf{L}_1 \, \tbf{L}_3 \, \tbf{M}_1 | \brm{un}(\tbf{U}))P(\tbf{L}_2 \, \tbf{M}_2 \, \tbf{M}_3 | \brm{un}(\tbf{U})) \, ,
}

which is guaranteed to satisfy $(\tbf{L}_1 \, \tbf{L}_3 \, \tbf{M}_1 \perp \tbf{L}_2 \, \tbf{M}_2 \, \tbf{M}_3 )$ and hence, by axioms \tbf{SG1},\tbf{SG3}, also $(\tbf{L}_1 \, \tbf{M}_1 \perp \tbf{L}_2 \, \tbf{M}_2 | \tbf{L}_3 \, \tbf{M}_3)$ which is what we aimed to prove. $\Box$ \\

\tbf{Case 4.} $R = (\tbf{L}_1 \perp \tbf{W}_2 | \tbf{M}_3)$. \\
Note that $R$ can only hold for an LDAG in which there is no path $\tbf{W}_2 \rightarrow \tbf{L}_1$. Applying the \tbf{QMC} to such LDAGs then shows that $(\tbf{L}_1 \perp \tbf{W}_2 | \tbf{U})$ necessarily holds, because all undirected paths connecting $\tbf{W}_2$ to $\tbf{L}_1$ either contain a collider in $\tbf{L}$ or a fork in $\tbf{M}$, neither of which is conditioned upon in this case. Hence $P(\tbf{L}_1|\tbf{U} \, \tbf{W}_2 ) = P(\tbf{L}_1|\tbf{U})$. Therefore:  
\eqn{
P(\tbf{L}_1 | \tbf{W}_2 \, \tbf{M}_3 \, , \brm{un}(\tbf{U})) 
&=& \zum{\tbf{u}' \tbf{u}}{}  \, P(\tbf{L}_1|\tbf{u} \, \tbf{W}_2 ) \, \Delta_{\tbf{u} \tbf{u}'} \, P(\tbf{u}'|\tbf{M}_3)  \nonumber \\
&=& \zum{\tbf{u}' \tbf{u}}{}  \, P(\tbf{L}_1|\tbf{u}) \, \Delta_{\tbf{u} \tbf{u}'} \, P(\tbf{u}'|\tbf{M}_3)  \nonumber \\
&=& P(\tbf{L}_1 | \tbf{M}_3 \, , \brm{un}(\tbf{U})) \, ,
}
which is equivalent to saying $(\tbf{L}_1 \perp \tbf{W}_2 | \tbf{M}_3)$ holds in $P(\tbf{L} \, \tbf{W} \, \tbf{M} | \brm{un}(\tbf{U}))$. $\Box$ \\

\tbf{Case 5.} $R = (\tbf{M}_3 \perp \tbf{L}_1 | \tbf{W}_2)$. \\
Let us partition $\tbf{U} = \tbf{U}_3 \cup \tbf{U}_R$ where $\tbf{U}_3$ are the children of $\tbf{M}_3$ in $\tbf{U}$ and $\tbf{U}_R$ are the rest, hence $P(\tbf{U}_3 \, \tbf{U}_R |\tbf{M}_3)=P(\tbf{U}_3 |\tbf{M}_3)P(\tbf{U}_R)$ by \tbf{Lemma 3}. Note that $R$ can only hold for an LDAG in which $\tbf{M}_3$ has no path to $\tbf{L}_1$ through $\tbf{U}$, hence none of the parents of $\tbf{L}_1$ in $\tbf{U}$ can be in $\tbf{U}_3$, and so $P(\tbf{L}_1|\tbf{U}_3 \tbf{U}_R \, \tbf{W}_2 ) = P(\tbf{L}_1|\tbf{U}_R \, \tbf{W}_2 )$ by \tbf{Lemma 2}. Therefore:
\eqn{
P(\tbf{L}_1 | \tbf{W}_2 \, \tbf{M}_3 \, , \brm{un}(\tbf{U})) 
&=& \zum{\tbf{u}'_3 \tbf{u}_3 }{} \zum{\tbf{u}'_R \tbf{u}_R }{}  \, P(\tbf{L}_1|\tbf{u}_R \, \tbf{W}_2 ) \, \Delta_{\tbf{u}_3 \tbf{u}'_3} \Delta_{\tbf{u}_R \tbf{u}'_R } \, P(\tbf{u}'_3|\tbf{M}_3) P(\tbf{u}'_R)  \nonumber \\
&=& \zum{\tbf{u}'_3 \tbf{u}_3 }{} \zum{\tbf{u}_R }{}  \, P(\tbf{L}_1|\tbf{u}_R \, \tbf{W}_2 ) P(\tbf{u}_R) \, \Delta_{\tbf{u}_3 \tbf{u}'_3} \, P(\tbf{u}'_3|\tbf{M}_3)  \nonumber \\
&=&  P(\tbf{L}_1|\tbf{W}_2 ) \, ,
}
where we used \tbf{Lemma 1} in line 2; \tbf{EN} in line 3 to get $P(\tbf{U}_R)=P(\tbf{U}_R \tbf{W}_2)$; and \tbf{Lemma 6} in the second-to-last line. Since the final line is evidently independent of $\tbf{M}_3$, this establishes that $(\tbf{L}_1 \perp \tbf{M}_3 | \tbf{W}_2)$ holds in $P(\tbf{L} \, \tbf{W} \, \tbf{M} | \brm{un}(\tbf{U}))$. $\Box$ \\

\tbf{Case 6.} $R = (\tbf{W}_2 \perp \tbf{M}_3 | \tbf{L}_1)$. \\
We will establish this by proving the equivalent statement:
\eqn{ 
P(\tbf{W}_2 \, \tbf{M}_3 | \tbf{L}_1 \, , \brm{un}(\tbf{U})) = P(\tbf{W}_2 | \tbf{L}_1 \, , \brm{un}(\tbf{U}))P(\tbf{M}_3 | \tbf{L}_1 \, , \brm{un}(\tbf{U})) \, .
}
First we rearrange the expression:
\eqn{
P(\tbf{W}_2 \, \tbf{M}_3 | \tbf{L}_1 \, , \brm{un}(\tbf{U})) 
&=& P(\tbf{L}_1 | \tbf{W}_2 \, \tbf{M}_3 \, , \brm{un}(\tbf{U})) \, \frac{P(\tbf{W}_2 \, \tbf{M}_3 | \brm{un}(\tbf{U})) }{P(\tbf{L}_1 | \brm{un}(\tbf{U})) } \, .
}
Note that $R$ cannot hold in any LDAG that has $\tbf{M}_3 \rightarrow \tbf{W}_2$. This implies $(\tbf{M}_3 \perp \tbf{W}_2)$ holds in $P(\tbf{W}_2 \, \tbf{M}_3)$, and invoking \tbf{Case 2}, we conclude it holds in $P(\tbf{W}_2 \, \tbf{M}_3 | \brm{un}(\tbf{U}))$ also. Therefore:
\eqn{ \label{eqn:case6}
P(\tbf{W}_2 \, \tbf{M}_3 | \tbf{L}_1 \, , \brm{un}(\tbf{U})) 
&=& P(\tbf{L}_1 | \tbf{W}_2 \, \tbf{M}_3 \, , \brm{un}(\tbf{U})) \, \frac{P(\tbf{W}_2 | \brm{un}(\tbf{U}))P(\tbf{M}_3 | \brm{un}(\tbf{U})) }{P(\tbf{L}_1 | \brm{un}(\tbf{U})) } \, .
}
Next we note that $R$ also cannot hold in any LDAG that has both $\tbf{W}_2 \rightarrow \tbf{L}_1$ and $\tbf{M}_3 \rightarrow \tbf{L}_1$. First consider the case that $\tbf{W}_2 \rightarrow \tbf{L}_1$ is false. Then by \tbf{Lemma 2} we have $P(\tbf{L}_1|\tbf{U} \, \tbf{W}_2 ) = P(\tbf{L}_1|\tbf{U})$ and we find:
\eqn{
P(\tbf{L}_1 | \tbf{W}_2 \, \tbf{M}_3 \, , \brm{un}(\tbf{U})) 
&=& \zum{\tbf{u}' \tbf{u}}{}  \, P(\tbf{L}_1|\tbf{u} \, \tbf{W}_2 ) \, \Delta_{\tbf{u} \tbf{u}'} \, P(\tbf{u}'|\tbf{M}_3)  \nonumber \\
&=& \zum{\tbf{u}' \tbf{u}}{}  \, P(\tbf{L}_1|\tbf{u}) \, \Delta_{\tbf{u} \tbf{u}'} \, P(\tbf{u}'|\tbf{M}_3)  \nonumber \\
&=& P(\tbf{L}_1 | \tbf{M}_3 \, , \brm{un}(\tbf{U})) \, .
}
Substituting this into \eqref{eqn:case6} we get:
\eqn{
P(\tbf{W}_2 \, \tbf{M}_3 | \tbf{L}_1 \, , \brm{un}(\tbf{U})) &=&
P(\tbf{L}_1 | \tbf{M}_3 \, , \brm{un}(\tbf{U})) \, \frac{P(\tbf{W}_2 | \brm{un}(\tbf{U}))P(\tbf{M}_3 | \brm{un}(\tbf{U})) }{P(\tbf{L}_1 | \brm{un}(\tbf{U})) } \nonumber \\
&=& P(\tbf{M}_3 | \tbf{L}_1 \, , \brm{un}(\tbf{U})) P(\tbf{W}_2 | \brm{un}(\tbf{U})) \, \nonumber \\
&=& P(\tbf{M}_3 | \tbf{L}_1 \, , \brm{un}(\tbf{U})) P(\tbf{W}_2 | \tbf{L}_1 \, , \brm{un}(\tbf{U})) \, ,
}
where in the last line we made use of the fact that $(\tbf{W}_2 \perp \tbf{L}_1 )$. This establishes the result for that case. Next, let us assume that $\tbf{M}_3 \rightarrow \tbf{L}_1$ is false. This implies the children $\tbf{U}_3$ of $\tbf{M}_3$ in $\tbf{U}$ cannot be parents of $\tbf{L}_1$. Hence, using various Lemmas,
\eqn{
P(\tbf{L}_1 | \tbf{W}_2 \, \tbf{M}_3 \, , \brm{un}(\tbf{U})) 
&=& \zum{\tbf{u}'_3 \tbf{u}_3}{} \zum{\tbf{u}'_R \tbf{u}_R}{}  \, P(\tbf{L}_1|\tbf{u}_R \, \tbf{W}_2 ) \, \Delta_{\tbf{u}_3 \tbf{u}'_3}\Delta_{\tbf{u}_R \tbf{u}'_R} \, P(\tbf{u}'_3|\tbf{M}_3) P(\tbf{u}'_R) \nonumber \\
&=& \zum{\tbf{u}'_R \tbf{u}_R}{}  \, P(\tbf{L}_1|\tbf{u}_R \, \tbf{W}_2 ) \, \Delta_{\tbf{u}_R \tbf{u}'_R} P(\tbf{u}'_R) \nonumber \\
&=& \zum{\tbf{u}_R}{}  \, P(\tbf{L}_1|\tbf{u}_R \, \tbf{W}_2 ) \, P(\tbf{u}_R) \nonumber \\
&=& P(\tbf{L}_1| \tbf{W}_2 ) \, .
}
Substituting into \eqref{eqn:case6},
\eqn{
P(\tbf{W}_2 \, \tbf{M}_3 | \tbf{L}_1 \, , \brm{un}(\tbf{U})) &=&
P(\tbf{L}_1 | \tbf{W}_2 \, , \brm{un}(\tbf{U})) \, \frac{P(\tbf{W}_2 | \brm{un}(\tbf{U}))P(\tbf{M}_3 | \brm{un}(\tbf{U})) }{P(\tbf{L}_1 | \brm{un}(\tbf{U})) }  \nonumber \\
&=& P(\tbf{W}_2 | \tbf{L}_1 \, , \brm{un}(\tbf{U})) \, P(\tbf{M}_3 | \brm{un}(\tbf{U})) \nonumber \\
&=& P(\tbf{W}_2 | \tbf{L}_1 \, , \brm{un}(\tbf{U})) \, P(\tbf{M}_3 | \tbf{L}_1 \, , \brm{un}(\tbf{U})) \, ,
}
where to obtain the last line we used the fact that since there is no path $\tbf{M}_3 \rightarrow \tbf{L}_1$, the relation $(\tbf{M}_3 \perp \tbf{L}_1 )$ holds in $P(\tbf{M}_3 \, \tbf{L}_1)$, and by \tbf{Case 3} it holds in $P(\tbf{M}_3 \, \tbf{L}_1 |  \brm{un}(\tbf{U}))$ also, so $P(\tbf{M}_3 | \tbf{L}_1 \, , \brm{un}(\tbf{U})) = P(\tbf{M}_3 | \brm{un}(\tbf{U}))$. This completes the proof of the Theorem. $\Box$ \\


\begin{thebibliography}{43}%
\makeatletter
\providecommand \@ifxundefined [1]{%
 \@ifx{#1\undefined}
}%
\providecommand \@ifnum [1]{%
 \ifnum #1\expandafter \@firstoftwo
 \else \expandafter \@secondoftwo
 \fi
}%
\providecommand \@ifx [1]{%
 \ifx #1\expandafter \@firstoftwo
 \else \expandafter \@secondoftwo
 \fi
}%
\providecommand \natexlab [1]{#1}%
\providecommand \enquote  [1]{``#1''}%
\providecommand \bibnamefont  [1]{#1}%
\providecommand \bibfnamefont [1]{#1}%
\providecommand \citenamefont [1]{#1}%
\providecommand \href@noop [0]{\@secondoftwo}%
\providecommand \href [0]{\begingroup \@sanitize@url \@href}%
\providecommand \@href[1]{\@@startlink{#1}\@@href}%
\providecommand \@@href[1]{\endgroup#1\@@endlink}%
\providecommand \@sanitize@url [0]{\catcode `\\12\catcode `\$12\catcode
  `\&12\catcode `\#12\catcode `\^12\catcode `\_12\catcode `\%12\relax}%
\providecommand \@@startlink[1]{}%
\providecommand \@@endlink[0]{}%
\providecommand \url  [0]{\begingroup\@sanitize@url \@url }%
\providecommand \@url [1]{\endgroup\@href {#1}{\urlprefix }}%
\providecommand \urlprefix  [0]{URL }%
\providecommand \Eprint [0]{\href }%
\providecommand \doibase [0]{http://dx.doi.org/}%
\providecommand \selectlanguage [0]{\@gobble}%
\providecommand \bibinfo  [0]{\@secondoftwo}%
\providecommand \bibfield  [0]{\@secondoftwo}%
\providecommand \translation [1]{[#1]}%
\providecommand \BibitemOpen [0]{}%
\providecommand \bibitemStop [0]{}%
\providecommand \bibitemNoStop [0]{.\EOS\space}%
\providecommand \EOS [0]{\spacefactor3000\relax}%
\providecommand \BibitemShut  [1]{\csname bibitem#1\endcsname}%
\let\auto@bib@innerbib\@empty
\bibitem [{\citenamefont {Pearl}(2009)}]{PEARL}%
  \BibitemOpen
  \bibfield  {author} {\bibinfo {author} {\bibfnamefont {J.}~\bibnamefont
  {Pearl}},\ }\href@noop {} {\emph {\bibinfo {title} {Causality}}}\ (\bibinfo
  {publisher} {Cambridge University Press},\ \bibinfo {year}
  {2009})\BibitemShut {NoStop}%
\bibitem [{\citenamefont {Spirtes}\ \emph {et~al.}(2012)\citenamefont
  {Spirtes}, \citenamefont {Glymour},\ and\ \citenamefont {Scheines}}]{SGS}%
  \BibitemOpen
  \bibfield  {author} {\bibinfo {author} {\bibfnamefont {P.}~\bibnamefont
  {Spirtes}}, \bibinfo {author} {\bibfnamefont {C.}~\bibnamefont {Glymour}}, \
  and\ \bibinfo {author} {\bibfnamefont {R.}~\bibnamefont {Scheines}},\ }\href
  {https://books.google.com.br/books?id=oUjxBwAAQBAJ} {\emph {\bibinfo {title}
  {Causation, Prediction, and Search}}},\ Lecture Notes in Statistics\
  (\bibinfo  {publisher} {Springer New York},\ \bibinfo {year}
  {2012})\BibitemShut {NoStop}%
\bibitem [{\citenamefont {Woodward}(2003)}]{WOODW}%
  \BibitemOpen
  \bibfield  {author} {\bibinfo {author} {\bibfnamefont {J.}~\bibnamefont
  {Woodward}},\ }\href {https://books.google.com.br/books?id=LrAbrrj5te8C}
  {\emph {\bibinfo {title} {Making Things Happen: A Theory of Causal
  Explanation}}},\ Oxford scholarship online\ (\bibinfo  {publisher} {Oxford
  University Press},\ \bibinfo {year} {2003})\BibitemShut {NoStop}%
\bibitem [{\citenamefont {Costa}\ and\ \citenamefont
  {Shrapnel}(2016)}]{COSHRAP}%
  \BibitemOpen
  \bibfield  {author} {\bibinfo {author} {\bibfnamefont {F.}~\bibnamefont
  {Costa}}\ and\ \bibinfo {author} {\bibfnamefont {S.}~\bibnamefont
  {Shrapnel}},\ }\href {http://stacks.iop.org/1367-2630/18/i=6/a=063032}
  {\bibfield  {journal} {\bibinfo  {journal} {New Journal of Physics}\ }\textbf
  {\bibinfo {volume} {18}},\ \bibinfo {pages} {063032} (\bibinfo {year}
  {2016})}\BibitemShut {NoStop}%
\bibitem [{\citenamefont {Allen}\ \emph {et~al.}(2017)\citenamefont {Allen},
  \citenamefont {Barrett}, \citenamefont {Horsman}, \citenamefont {Lee},\ and\
  \citenamefont {Spekkens}}]{ALLEN}%
  \BibitemOpen
  \bibfield  {author} {\bibinfo {author} {\bibfnamefont {J.-M.~A.}\
  \bibnamefont {Allen}}, \bibinfo {author} {\bibfnamefont {J.}~\bibnamefont
  {Barrett}}, \bibinfo {author} {\bibfnamefont {D.~C.}\ \bibnamefont
  {Horsman}}, \bibinfo {author} {\bibfnamefont {C.~M.}\ \bibnamefont {Lee}}, \
  and\ \bibinfo {author} {\bibfnamefont {R.~W.}\ \bibnamefont {Spekkens}},\
  }\href {\doibase 10.1103/PhysRevX.7.031021} {\bibfield  {journal} {\bibinfo
  {journal} {Phys. Rev. X}\ }\textbf {\bibinfo {volume} {7}},\ \bibinfo {pages}
  {031021} (\bibinfo {year} {2017})}\BibitemShut {NoStop}%
\bibitem [{\citenamefont {Ried}\ \emph {et~al.}(2015)\citenamefont {Ried},
  \citenamefont {Agnew}, \citenamefont {Vermeyden}, \citenamefont {Janzing},
  \citenamefont {Spekkens},\ and\ \citenamefont {Resch}}]{RIED}%
  \BibitemOpen
  \bibfield  {author} {\bibinfo {author} {\bibfnamefont {K.}~\bibnamefont
  {Ried}}, \bibinfo {author} {\bibfnamefont {M.}~\bibnamefont {Agnew}},
  \bibinfo {author} {\bibfnamefont {L.}~\bibnamefont {Vermeyden}}, \bibinfo
  {author} {\bibfnamefont {D.}~\bibnamefont {Janzing}}, \bibinfo {author}
  {\bibfnamefont {R.~W.}\ \bibnamefont {Spekkens}}, \ and\ \bibinfo {author}
  {\bibfnamefont {K.~J.}\ \bibnamefont {Resch}},\ }\href {\doibase
  10.1038/nphys3266} {\bibfield  {journal} {\bibinfo  {journal} {Nature
  Physics}\ }\textbf {\bibinfo {volume} {11}},\ \bibinfo {pages} {414–420}
  (\bibinfo {year} {2015})}\BibitemShut {NoStop}%
\bibitem [{\citenamefont {Ried}(2016)}]{RIEDPHD}%
  \BibitemOpen
  \bibfield  {author} {\bibinfo {author} {\bibfnamefont {K.}~\bibnamefont
  {Ried}},\ }\href@noop {} {\emph {\bibinfo {title} {Causal Models for a
  Quantum World}}}\ (\bibinfo  {publisher} {University of Waterloo},\ \bibinfo
  {year} {2016})\ \bibinfo {note} {phD thesis}\BibitemShut {NoStop}%
\bibitem [{\citenamefont {Fritz}(2016)}]{FRITZ}%
  \BibitemOpen
  \bibfield  {author} {\bibinfo {author} {\bibfnamefont {T.}~\bibnamefont
  {Fritz}},\ }\href {\doibase 10.1007/s00220-015-2495-5} {\bibfield  {journal}
  {\bibinfo  {journal} {Communications in Mathematical Physics}\ }\textbf
  {\bibinfo {volume} {341}},\ \bibinfo {pages} {391} (\bibinfo {year}
  {2016})}\BibitemShut {NoStop}%
\bibitem [{\citenamefont {Henson}\ \emph {et~al.}(2014)\citenamefont {Henson},
  \citenamefont {Lal},\ and\ \citenamefont {Pusey}}]{HLP}%
  \BibitemOpen
  \bibfield  {author} {\bibinfo {author} {\bibfnamefont {J.}~\bibnamefont
  {Henson}}, \bibinfo {author} {\bibfnamefont {R.}~\bibnamefont {Lal}}, \ and\
  \bibinfo {author} {\bibfnamefont {M.~F.}\ \bibnamefont {Pusey}},\ }\href
  {http://stacks.iop.org/1367-2630/16/i=11/a=113043} {\bibfield  {journal}
  {\bibinfo  {journal} {New Journal of Physics}\ }\textbf {\bibinfo {volume}
  {16}},\ \bibinfo {pages} {113043} (\bibinfo {year} {2014})}\BibitemShut
  {NoStop}%
\bibitem [{\citenamefont {Pienaar}\ and\ \citenamefont {\v{C}aslav
  Brukner}(2015)}]{PIEBRUK}%
  \BibitemOpen
  \bibfield  {author} {\bibinfo {author} {\bibfnamefont {J.}~\bibnamefont
  {Pienaar}}\ and\ \bibinfo {author} {\bibnamefont {\v{C}aslav Brukner}},\
  }\href {http://stacks.iop.org/1367-2630/17/i=7/a=073020} {\bibfield
  {journal} {\bibinfo  {journal} {New Journal of Physics}\ }\textbf {\bibinfo
  {volume} {17}},\ \bibinfo {pages} {073020} (\bibinfo {year}
  {2015})}\BibitemShut {NoStop}%
\bibitem [{\citenamefont {Barrett}\ \emph {et~al.}(2019)\citenamefont
  {Barrett}, \citenamefont {Lorenz},\ and\ \citenamefont
  {Oreshkov}}]{BARRETTQCM}%
  \BibitemOpen
  \bibfield  {author} {\bibinfo {author} {\bibfnamefont {J.}~\bibnamefont
  {Barrett}}, \bibinfo {author} {\bibfnamefont {R.}~\bibnamefont {Lorenz}}, \
  and\ \bibinfo {author} {\bibfnamefont {O.}~\bibnamefont {Oreshkov}},\ }\href
  {https://arxiv.org/abs/1906.10726} {\  (\bibinfo {year} {2019})},\ \bibinfo
  {note} {eprint arXiv:1906.10726}\BibitemShut {NoStop}%
\bibitem [{\citenamefont {Giarmatzi}\ and\ \citenamefont
  {Costa}(2018)}]{GIARM}%
  \BibitemOpen
  \bibfield  {author} {\bibinfo {author} {\bibfnamefont {C.}~\bibnamefont
  {Giarmatzi}}\ and\ \bibinfo {author} {\bibfnamefont {F.}~\bibnamefont
  {Costa}},\ }\href {\doibase 10.1038/s41534-018-0062-6} {\bibfield  {journal}
  {\bibinfo  {journal} {npj Quantum Information}\ }\textbf {\bibinfo {volume}
  {4}},\ \bibinfo {pages} {2056} (\bibinfo {year} {2018})}\BibitemShut
  {NoStop}%
\bibitem [{\citenamefont {Leifer}\ and\ \citenamefont {Poulin}(2008)}]{LEIF08}%
  \BibitemOpen
  \bibfield  {author} {\bibinfo {author} {\bibfnamefont {M.}~\bibnamefont
  {Leifer}}\ and\ \bibinfo {author} {\bibfnamefont {D.}~\bibnamefont
  {Poulin}},\ }\href {\doibase https://doi.org/10.1016/j.aop.2007.10.001}
  {\bibfield  {journal} {\bibinfo  {journal} {Annals of Physics}\ }\textbf
  {\bibinfo {volume} {323}},\ \bibinfo {pages} {1899 } (\bibinfo {year}
  {2008})}\BibitemShut {NoStop}%
\bibitem [{\citenamefont {Milz}\ \emph {et~al.}(2017)\citenamefont {Milz},
  \citenamefont {Sakuldee}, \citenamefont {Pollock},\ and\ \citenamefont
  {Modi}}]{MILZ17}%
  \BibitemOpen
  \bibfield  {author} {\bibinfo {author} {\bibfnamefont {S.}~\bibnamefont
  {Milz}}, \bibinfo {author} {\bibfnamefont {F.}~\bibnamefont {Sakuldee}},
  \bibinfo {author} {\bibfnamefont {F.~A.}\ \bibnamefont {Pollock}}, \ and\
  \bibinfo {author} {\bibfnamefont {K.}~\bibnamefont {Modi}},\ }\href
  {https://arxiv.org/abs/1712.02589} {\  (\bibinfo {year} {2017})},\ \bibinfo
  {note} {eprint arXiv:1712.02589}\BibitemShut {NoStop}%
\bibitem [{\citenamefont {Wood}\ and\ \citenamefont {Spekkens}(2015)}]{WOOD}%
  \BibitemOpen
  \bibfield  {author} {\bibinfo {author} {\bibfnamefont {C.~J.}\ \bibnamefont
  {Wood}}\ and\ \bibinfo {author} {\bibfnamefont {R.~W.}\ \bibnamefont
  {Spekkens}},\ }\href {http://stacks.iop.org/1367-2630/17/i=3/a=033002}
  {\bibfield  {journal} {\bibinfo  {journal} {New Journal of Physics}\ }\textbf
  {\bibinfo {volume} {17}},\ \bibinfo {pages} {033002} (\bibinfo {year}
  {2015})}\BibitemShut {NoStop}%
\bibitem [{\citenamefont {Bell}(1976)}]{BELL76}%
  \BibitemOpen
  \bibfield  {author} {\bibinfo {author} {\bibfnamefont {J.}~\bibnamefont
  {Bell}},\ }\href@noop {} {\bibfield  {journal} {\bibinfo  {journal}
  {Epistemological Lett.}\ }\textbf {\bibinfo {volume} {9}} (\bibinfo {year}
  {1976})}\BibitemShut {NoStop}%
\bibitem [{\citenamefont {Colbeck}\ and\ \citenamefont
  {Renner}(2011)}]{CBRenner}%
  \BibitemOpen
  \bibfield  {author} {\bibinfo {author} {\bibfnamefont {R.}~\bibnamefont
  {Colbeck}}\ and\ \bibinfo {author} {\bibfnamefont {R.}~\bibnamefont
  {Renner}},\ }\href {http://dx.doi.org/10.1038/ncomms1416} {\bibfield
  {journal} {\bibinfo  {journal} {Nature Communications}\ }\textbf {\bibinfo
  {volume} {2}},\ \bibinfo {pages} {411 EP } (\bibinfo {year}
  {2011})}\BibitemShut {NoStop}%
\bibitem [{Note1()}]{Note1}%
  \BibitemOpen
  \bibinfo {note} {Perhaps contrary to one's first intuition, it is not
  sufficient to condition only on the set of variables that are \protect
  \textit {parents} of both $X_1,X_2$. A trivial counterexample is $X_1
  \leftarrow A_1 \leftarrow A_3 \rightarrow A_2 \rightarrow X_2$.}\BibitemShut
  {Stop}%
\bibitem [{\citenamefont {Reichenbach}(1956)}]{RPCC}%
  \BibitemOpen
  \bibfield  {author} {\bibinfo {author} {\bibfnamefont {H.}~\bibnamefont
  {Reichenbach}},\ }\href@noop {} {\emph {\bibinfo {title} {The Direction of
  Time}}}\ (\bibinfo  {publisher} {University of Los Angeles Press},\ \bibinfo
  {year} {1956})\BibitemShut {NoStop}%
\bibitem [{\citenamefont {Cavalcanti}\ and\ \citenamefont
  {Lal}(2014)}]{CAVLAL}%
  \BibitemOpen
  \bibfield  {author} {\bibinfo {author} {\bibfnamefont {E.~G.}\ \bibnamefont
  {Cavalcanti}}\ and\ \bibinfo {author} {\bibfnamefont {R.}~\bibnamefont
  {Lal}},\ }\href {http://stacks.iop.org/1751-8121/47/i=42/a=424018} {\bibfield
   {journal} {\bibinfo  {journal} {Journal of Physics A: Mathematical and
  Theoretical}\ }\textbf {\bibinfo {volume} {47}},\ \bibinfo {pages} {424018}
  (\bibinfo {year} {2014})}\BibitemShut {NoStop}%
\bibitem [{\citenamefont {Price}(1997)}]{PRICEBOOK}%
  \BibitemOpen
  \bibfield  {author} {\bibinfo {author} {\bibfnamefont {H.}~\bibnamefont
  {Price}},\ }\href {https://books.google.com.br/books?id=WxQ4QIxNuD4C} {\emph
  {\bibinfo {title} {Time's Arrow \& Archimedes' Point: New Directions for the
  Physics of Time}}},\ Oxford Paperbacks: Philosophy\ (\bibinfo  {publisher}
  {Oxford University Press},\ \bibinfo {year} {1997})\BibitemShut {NoStop}%
\bibitem [{\citenamefont {Berkson}(2014)}]{BERK}%
  \BibitemOpen
  \bibfield  {author} {\bibinfo {author} {\bibfnamefont {J.}~\bibnamefont
  {Berkson}},\ }\href {\doibase 10.1093/ije/dyu022} {\bibfield  {journal}
  {\bibinfo  {journal} {International Journal of Epidemiology}\ }\textbf
  {\bibinfo {volume} {43}},\ \bibinfo {pages} {511} (\bibinfo {year}
  {2014})}\BibitemShut {NoStop}%
\bibitem [{\citenamefont {Busch}(2009)}]{BUSCH}%
  \BibitemOpen
  \bibfield  {author} {\bibinfo {author} {\bibfnamefont {P.}~\bibnamefont
  {Busch}},\ }\enquote {\bibinfo {title} {No information without disturbance:
  Quantum limitations of measurement},}\ in\ \href {\doibase
  10.1007/978-1-4020-9107-0_13} {\emph {\bibinfo {booktitle} {Quantum Reality,
  Relativistic Causality, and Closing the Epistemic Circle: Essays in Honour of
  Abner Shimony}}}\ (\bibinfo  {publisher} {Springer Netherlands},\ \bibinfo
  {address} {Dordrecht},\ \bibinfo {year} {2009})\ pp.\ \bibinfo {pages}
  {229--256}\BibitemShut {NoStop}%
\bibitem [{\citenamefont {Holevo}(1998)}]{HOLEVO}%
  \BibitemOpen
  \bibfield  {author} {\bibinfo {author} {\bibfnamefont {A.~S.}\ \bibnamefont
  {Holevo}},\ }\href {\doibase 10.1070/rm1998v053n06abeh000091} {\bibfield
  {journal} {\bibinfo  {journal} {Russian Mathematical Surveys}\ }\textbf
  {\bibinfo {volume} {53}},\ \bibinfo {pages} {1295} (\bibinfo {year}
  {1998})}\BibitemShut {NoStop}%
\bibitem [{\citenamefont {Horodecki}\ \emph {et~al.}(2003)\citenamefont
  {Horodecki}, \citenamefont {Shor},\ and\ \citenamefont {Ruskai}}]{HORODECKI}%
  \BibitemOpen
  \bibfield  {author} {\bibinfo {author} {\bibfnamefont {M.}~\bibnamefont
  {Horodecki}}, \bibinfo {author} {\bibfnamefont {P.~W.}\ \bibnamefont {Shor}},
  \ and\ \bibinfo {author} {\bibfnamefont {M.~B.}\ \bibnamefont {Ruskai}},\
  }\href {\doibase 10.1142/S0129055X03001709} {\bibfield  {journal} {\bibinfo
  {journal} {Reviews in Mathematical Physics}\ }\textbf {\bibinfo {volume}
  {15}},\ \bibinfo {pages} {629} (\bibinfo {year} {2003})}\BibitemShut
  {NoStop}%
\bibitem [{\citenamefont {K\"{u}bler}\ and\ \citenamefont
  {Braun}(2018)}]{KUEBLER}%
  \BibitemOpen
  \bibfield  {author} {\bibinfo {author} {\bibfnamefont {J.~M.}\ \bibnamefont
  {K\"{u}bler}}\ and\ \bibinfo {author} {\bibfnamefont {D.}~\bibnamefont
  {Braun}},\ }\href {http://stacks.iop.org/1367-2630/20/i=8/a=083015}
  {\bibfield  {journal} {\bibinfo  {journal} {New Journal of Physics}\ }\textbf
  {\bibinfo {volume} {20}},\ \bibinfo {pages} {083015} (\bibinfo {year}
  {2018})}\BibitemShut {NoStop}%
\bibitem [{\citenamefont {Cavalcanti}(2018)}]{CAVNFT}%
  \BibitemOpen
  \bibfield  {author} {\bibinfo {author} {\bibfnamefont {E.~G.}\ \bibnamefont
  {Cavalcanti}},\ }\href {\doibase 10.1103/PhysRevX.8.021018} {\bibfield
  {journal} {\bibinfo  {journal} {Phys. Rev. X}\ }\textbf {\bibinfo {volume}
  {8}},\ \bibinfo {pages} {021018} (\bibinfo {year} {2018})}\BibitemShut
  {NoStop}%
\bibitem [{\citenamefont {Almada}\ \emph {et~al.}(2016)\citenamefont {Almada},
  \citenamefont {Ch'ng}, \citenamefont {Kintner}, \citenamefont {Morrison},\
  and\ \citenamefont {Wharton}}]{ALMADA}%
  \BibitemOpen
  \bibfield  {author} {\bibinfo {author} {\bibfnamefont {D.}~\bibnamefont
  {Almada}}, \bibinfo {author} {\bibfnamefont {K.}~\bibnamefont {Ch'ng}},
  \bibinfo {author} {\bibfnamefont {S.}~\bibnamefont {Kintner}}, \bibinfo
  {author} {\bibfnamefont {B.}~\bibnamefont {Morrison}}, \ and\ \bibinfo
  {author} {\bibfnamefont {K.}~\bibnamefont {Wharton}},\ }\href@noop {}
  {\bibfield  {journal} {\bibinfo  {journal} {Int. J. Quantum Found.}\ }\textbf
  {\bibinfo {volume} {2}},\ \bibinfo {pages} {1} (\bibinfo {year}
  {2016})}\BibitemShut {NoStop}%
\bibitem [{\citenamefont {Fuchs}\ and\ \citenamefont {Schack}(2013)}]{QBCOH}%
  \BibitemOpen
  \bibfield  {author} {\bibinfo {author} {\bibfnamefont {C.~A.}\ \bibnamefont
  {Fuchs}}\ and\ \bibinfo {author} {\bibfnamefont {R.}~\bibnamefont {Schack}},\
  }\href {\doibase 10.1103/RevModPhys.85.1693} {\bibfield  {journal} {\bibinfo
  {journal} {Rev. Mod. Phys.}\ }\textbf {\bibinfo {volume} {85}},\ \bibinfo
  {pages} {1693} (\bibinfo {year} {2013})}\BibitemShut {NoStop}%
\bibitem [{\citenamefont {Appleby}\ \emph {et~al.}(2017)\citenamefont
  {Appleby}, \citenamefont {Fuchs}, \citenamefont {Stacey},\ and\ \citenamefont
  {Zhu}}]{QPLEX}%
  \BibitemOpen
  \bibfield  {author} {\bibinfo {author} {\bibfnamefont {M.}~\bibnamefont
  {Appleby}}, \bibinfo {author} {\bibfnamefont {C.~A.}\ \bibnamefont {Fuchs}},
  \bibinfo {author} {\bibfnamefont {B.~C.}\ \bibnamefont {Stacey}}, \ and\
  \bibinfo {author} {\bibfnamefont {H.}~\bibnamefont {Zhu}},\ }\href {\doibase
  10.1140/epjd/e2017-80024-y} {\bibfield  {journal} {\bibinfo  {journal} {The
  European Physical Journal D}\ }\textbf {\bibinfo {volume} {71}},\ \bibinfo
  {pages} {197} (\bibinfo {year} {2017})}\BibitemShut {NoStop}%
\bibitem [{\citenamefont {Busch}(1986)}]{BUSCH2}%
  \BibitemOpen
  \bibfield  {author} {\bibinfo {author} {\bibfnamefont {P.}~\bibnamefont
  {Busch}},\ }\href {\doibase 10.1103/PhysRevD.33.2253} {\bibfield  {journal}
  {\bibinfo  {journal} {Phys. Rev. D}\ }\textbf {\bibinfo {volume} {33}},\
  \bibinfo {pages} {2253} (\bibinfo {year} {1986})}\BibitemShut {NoStop}%
\bibitem [{\citenamefont {MacLean}\ \emph {et~al.}(2017)\citenamefont
  {MacLean}, \citenamefont {Ried}, \citenamefont {Spekkens},\ and\
  \citenamefont {Resch}}]{SPEKBERK}%
  \BibitemOpen
  \bibfield  {author} {\bibinfo {author} {\bibfnamefont {J.-P.~W.}\
  \bibnamefont {MacLean}}, \bibinfo {author} {\bibfnamefont {K.}~\bibnamefont
  {Ried}}, \bibinfo {author} {\bibfnamefont {R.~W.}\ \bibnamefont {Spekkens}},
  \ and\ \bibinfo {author} {\bibfnamefont {K.~J.}\ \bibnamefont {Resch}},\
  }\href {\doibase 10.1038/ncomms15149} {\bibfield  {journal} {\bibinfo
  {journal} {Nature Communications}\ }\textbf {\bibinfo {volume} {8}},\
  \bibinfo {pages} {15149} (\bibinfo {year} {2017})}\BibitemShut {NoStop}%
\bibitem [{\citenamefont {{\.Z}yczkowski}\ \emph {et~al.}(1998)\citenamefont
  {{\.Z}yczkowski}, \citenamefont {Horodecki}, \citenamefont {Sanpera},\ and\
  \citenamefont {Lewenstein}}]{ZYCZ}%
  \BibitemOpen
  \bibfield  {author} {\bibinfo {author} {\bibfnamefont {K.}~\bibnamefont
  {{\.Z}yczkowski}}, \bibinfo {author} {\bibfnamefont {P.}~\bibnamefont
  {Horodecki}}, \bibinfo {author} {\bibfnamefont {A.}~\bibnamefont {Sanpera}},
  \ and\ \bibinfo {author} {\bibfnamefont {M.}~\bibnamefont {Lewenstein}},\
  }\href {\doibase 10.1103/PhysRevA.58.883} {\bibfield  {journal} {\bibinfo
  {journal} {Phys. Rev. A}\ }\textbf {\bibinfo {volume} {58}},\ \bibinfo
  {pages} {883} (\bibinfo {year} {1998})}\BibitemShut {NoStop}%
\bibitem [{\citenamefont {Costa}\ \emph {et~al.}(2018)\citenamefont {Costa},
  \citenamefont {Ringbauer}, \citenamefont {Goggin}, \citenamefont {White},\
  and\ \citenamefont {Fedrizzi}}]{COSTA17}%
  \BibitemOpen
  \bibfield  {author} {\bibinfo {author} {\bibfnamefont {F.}~\bibnamefont
  {Costa}}, \bibinfo {author} {\bibfnamefont {M.}~\bibnamefont {Ringbauer}},
  \bibinfo {author} {\bibfnamefont {M.~E.}\ \bibnamefont {Goggin}}, \bibinfo
  {author} {\bibfnamefont {A.~G.}\ \bibnamefont {White}}, \ and\ \bibinfo
  {author} {\bibfnamefont {A.}~\bibnamefont {Fedrizzi}},\ }\href {\doibase
  10.1103/PhysRevA.98.012328} {\bibfield  {journal} {\bibinfo  {journal} {Phys.
  Rev. A}\ }\textbf {\bibinfo {volume} {98}},\ \bibinfo {pages} {012328}
  (\bibinfo {year} {2018})}\BibitemShut {NoStop}%
\bibitem [{\citenamefont {Chiribella}\ \emph {et~al.}(2009)\citenamefont
  {Chiribella}, \citenamefont {D'Ariano},\ and\ \citenamefont
  {Perinotti}}]{COMB1}%
  \BibitemOpen
  \bibfield  {author} {\bibinfo {author} {\bibfnamefont {G.}~\bibnamefont
  {Chiribella}}, \bibinfo {author} {\bibfnamefont {G.~M.}\ \bibnamefont
  {D'Ariano}}, \ and\ \bibinfo {author} {\bibfnamefont {P.}~\bibnamefont
  {Perinotti}},\ }\href {\doibase 10.1103/PhysRevA.80.022339} {\bibfield
  {journal} {\bibinfo  {journal} {Phys. Rev. A}\ }\textbf {\bibinfo {volume}
  {80}},\ \bibinfo {pages} {022339} (\bibinfo {year} {2009})}\BibitemShut
  {NoStop}%
\bibitem [{\citenamefont {Chiribella}\ \emph {et~al.}(2008)\citenamefont
  {Chiribella}, \citenamefont {D'Ariano},\ and\ \citenamefont
  {Perinotti}}]{COMB2}%
  \BibitemOpen
  \bibfield  {author} {\bibinfo {author} {\bibfnamefont {G.}~\bibnamefont
  {Chiribella}}, \bibinfo {author} {\bibfnamefont {G.~M.}\ \bibnamefont
  {D'Ariano}}, \ and\ \bibinfo {author} {\bibfnamefont {P.}~\bibnamefont
  {Perinotti}},\ }\href {\doibase 10.1103/PhysRevLett.101.060401} {\bibfield
  {journal} {\bibinfo  {journal} {Phys. Rev. Lett.}\ }\textbf {\bibinfo
  {volume} {101}},\ \bibinfo {pages} {060401} (\bibinfo {year}
  {2008})}\BibitemShut {NoStop}%
\bibitem [{\citenamefont {Pollock}\ \emph
  {et~al.}(2018{\natexlab{a}})\citenamefont {Pollock}, \citenamefont
  {Rodr\'{\i}guez-Rosario}, \citenamefont {Frauenheim}, \citenamefont
  {Paternostro},\ and\ \citenamefont {Modi}}]{POLLOCKPRA}%
  \BibitemOpen
  \bibfield  {author} {\bibinfo {author} {\bibfnamefont {F.~A.}\ \bibnamefont
  {Pollock}}, \bibinfo {author} {\bibfnamefont {C.}~\bibnamefont
  {Rodr\'{\i}guez-Rosario}}, \bibinfo {author} {\bibfnamefont {T.}~\bibnamefont
  {Frauenheim}}, \bibinfo {author} {\bibfnamefont {M.}~\bibnamefont
  {Paternostro}}, \ and\ \bibinfo {author} {\bibfnamefont {K.}~\bibnamefont
  {Modi}},\ }\href {\doibase 10.1103/PhysRevA.97.012127} {\bibfield  {journal}
  {\bibinfo  {journal} {Phys. Rev. A}\ }\textbf {\bibinfo {volume} {97}},\
  \bibinfo {pages} {012127} (\bibinfo {year} {2018}{\natexlab{a}})}\BibitemShut
  {NoStop}%
\bibitem [{\citenamefont {Pienaar}(2019)}]{JACQUES2}%
  \BibitemOpen
  \bibfield  {author} {\bibinfo {author} {\bibfnamefont {J.}~\bibnamefont
  {Pienaar}},\ }\href {https://arxiv.org/abs/1902.00129} {\  (\bibinfo {year}
  {2019})},\ \bibinfo {note} {eprint arXiv:1902.00129}\BibitemShut {NoStop}%
\bibitem [{\citenamefont {Coecke}\ \emph {et~al.}(2017)\citenamefont {Coecke},
  \citenamefont {Gogioso},\ and\ \citenamefont {Selby}}]{COECKE}%
  \BibitemOpen
  \bibfield  {author} {\bibinfo {author} {\bibfnamefont {B.}~\bibnamefont
  {Coecke}}, \bibinfo {author} {\bibfnamefont {S.}~\bibnamefont {Gogioso}}, \
  and\ \bibinfo {author} {\bibfnamefont {J.}~\bibnamefont {Selby}},\ }\href
  {https://arxiv.org/abs/1711.05511} {\  (\bibinfo {year} {2017})},\ \bibinfo
  {note} {eprint arXiv:1711.05511.}\BibitemShut {Stop}%
\bibitem [{\citenamefont {Pollock}\ \emph
  {et~al.}(2018{\natexlab{b}})\citenamefont {Pollock}, \citenamefont
  {Rodr\'{\i}guez-Rosario}, \citenamefont {Frauenheim}, \citenamefont
  {Paternostro},\ and\ \citenamefont {Modi}}]{POLLOCKPRL}%
  \BibitemOpen
  \bibfield  {author} {\bibinfo {author} {\bibfnamefont {F.~A.}\ \bibnamefont
  {Pollock}}, \bibinfo {author} {\bibfnamefont {C.}~\bibnamefont
  {Rodr\'{\i}guez-Rosario}}, \bibinfo {author} {\bibfnamefont {T.}~\bibnamefont
  {Frauenheim}}, \bibinfo {author} {\bibfnamefont {M.}~\bibnamefont
  {Paternostro}}, \ and\ \bibinfo {author} {\bibfnamefont {K.}~\bibnamefont
  {Modi}},\ }\href {\doibase 10.1103/PhysRevLett.120.040405} {\bibfield
  {journal} {\bibinfo  {journal} {Phys. Rev. Lett.}\ }\textbf {\bibinfo
  {volume} {120}},\ \bibinfo {pages} {040405} (\bibinfo {year}
  {2018}{\natexlab{b}})}\BibitemShut {NoStop}%
\bibitem [{\citenamefont {Oreshkov}\ and\ \citenamefont {Cerf}(2015)}]{ORESH}%
  \BibitemOpen
  \bibfield  {author} {\bibinfo {author} {\bibfnamefont {O.}~\bibnamefont
  {Oreshkov}}\ and\ \bibinfo {author} {\bibfnamefont {N.~J.}\ \bibnamefont
  {Cerf}},\ }\href {\doibase 10.1038/nphys3414} {\bibfield  {journal} {\bibinfo
   {journal} {Nature Physics}\ }\textbf {\bibinfo {volume} {11}},\ \bibinfo
  {pages} {853} (\bibinfo {year} {2015})}\BibitemShut {NoStop}%
\bibitem [{\citenamefont {Gu{\'{e}}rin}\ and\ \citenamefont
  {Brukner}(2018)}]{GUERIN}%
  \BibitemOpen
  \bibfield  {author} {\bibinfo {author} {\bibfnamefont {P.~A.}\ \bibnamefont
  {Gu{\'{e}}rin}}\ and\ \bibinfo {author} {\bibfnamefont
  {{\v{C}}.}~\bibnamefont {Brukner}},\ }\href {\doibase
  10.1088/1367-2630/aae742} {\bibfield  {journal} {\bibinfo  {journal} {New
  Journal of Physics}\ }\textbf {\bibinfo {volume} {20}},\ \bibinfo {pages}
  {103031} (\bibinfo {year} {2018})}\BibitemShut {NoStop}%
\bibitem [{\citenamefont {Oreshkov}\ \emph {et~al.}(2012)\citenamefont
  {Oreshkov}, \citenamefont {Costa},\ and\ \citenamefont {Brukner}}]{OCB}%
  \BibitemOpen
  \bibfield  {author} {\bibinfo {author} {\bibfnamefont {O.}~\bibnamefont
  {Oreshkov}}, \bibinfo {author} {\bibfnamefont {F.}~\bibnamefont {Costa}}, \
  and\ \bibinfo {author} {\bibfnamefont {C.}~\bibnamefont {Brukner}},\ }\href
  {http://dx.doi.org/10.1038/ncomms2076} {\bibfield  {journal} {\bibinfo
  {journal} {Nature Communications}\ }\textbf {\bibinfo {volume} {3}},\
  \bibinfo {pages} {1092} (\bibinfo {year} {2012})}\BibitemShut {NoStop}%
\end{thebibliography}
\end{document}